\documentclass{article}
\usepackage[affil-it]{authblk}

\usepackage[english]{babel}

\usepackage[letterpaper,top=2cm,bottom=2cm,left=3cm,right=3cm,marginparwidth=1.75cm]{geometry}
\usepackage{amsmath}
\usepackage{bbm}
\usepackage{amsfonts}
\usepackage{amssymb}
\usepackage{latexsym}
\usepackage{graphicx}
\usepackage[english]{babel}
\usepackage{multirow}
\usepackage{float}
\usepackage{url}
\usepackage{slashed}
\usepackage{xcolor} 
\usepackage[utf8]{inputenc}
\usepackage{verbatim}
\usepackage{stackengine}
\usepackage{cancel}
\usepackage{minted}

\newmintedfile[pythoncode]{python}{
    bgcolor=lightgray,
    fontsize=\footnotesize,
    frame=lines
}

\def\sfrac#1#2{{\textstyle{#1\over #2}}}
\newcommand{\be}{\begin{equation}}
\newcommand{\ee}{\end{equation}}
\newcommand{\ba}{\begin{array}}
\newcommand{\ea}{\end{array}}
\newcommand{\bea}{\begin{eqnarray}}
\newcommand{\eea}{\end{eqnarray}}
\newcommand{\sss}{\scriptscriptstyle}

\newcommand{\jc}[1]{{\color{red}JC: {#1}}}
\newcommand{\tr}{{\rm tr}}

\newcommand{\nn}{\nonumber}

\newcommand{\R}{{\sss R}}
\newcommand{\T}{{\sss T}}

\newcommand{\M}{{\sss M}}
\newcommand{\K}{{\sss K}}

\newcommand{\esc}{\fbox{\tt ESC}\ }
\usepackage[utf8]{inputenc}

\usepackage{amsmath}
\usepackage{graphicx}
\usepackage[colorlinks=true, allcolors=blue]{hyperref}
\numberwithin{equation}{section}

\title{A Short Course in General Relativity}
\author{James M.\ Cline}
\affil{Physics Deptartment\ and Trottier Space Institute,\\ McGill University,\\
Montr\'eal, Qu\'ebec, Canada }

\date{}
\begin{document}
\maketitle

\begin{abstract}
These notes give a concise introduction to General Relativity at the advanced undergraduate level,
 starting from the weak
field limit and gravitational waves, then introducing curved manifolds and Riemannian
geometry.  The nonlinear gravitational action is used to derive the nonlinear field
equations, with applications to black holes and cosmology.  It is assumed that special
relativity and electromagnetic waves have been previously studied.  Some advanced topics such as
Rindler and Hawking radiation are derived, and recent developments in gravitational wave detection are briefly
covered.  Problems are included, both those suitable for homework, and simpler ones that could be
worked out by students during class sessions.

\end{abstract}

\bigskip
{\bf Preface.}  This exposition is in reverse order compared to most textbooks.  This choice is motivated in part by the great 
excitement in the physics community about the observation of gravitational waves, and their potential for revealing formerly hidden
aspects of the Universe.  In many treatments, this material gets delayed until the end, when there is typically not much time
left in the course to pursue it.  More generally, to me the weak-field limit of general relativity is a natural place to start.  It requires
less introduction of new mathematical concepts since it is a small perturbation on the already-familiar Minkowski spacetime metric, and
gravitational waves in this limit have many parallels with their electromagnetic counterparts, with which you are assumed to be already familiar.
One can proceed directly to many of the interesting phenomena present in GR such as light deflection and perihelion precession before undertaking
the full nonlinear theory.  My coverage of mathematical concepts is aimed at practical 
applications and many
results from Riemannian geometry are stated without proof; rather the emphasis is on intuitive
understanding.
I have not included any figures with these notes.  
I encourage the reader to draw your own as appropriate, as an aid to understanding.
I thank the CERN Theory Department and the Niels Bohr International Academy for their kind hospitality during the inception of these
notes.  I also thank students who took this course for thoughtful comments that helped to improve the
exposition.

\tableofcontents

\section{Introduction}
Newton's theory of gravitation provides an excellent description of nature for many purposes.  In fact one has to work rather hard to find out where it fails.  But we know that it is only an approximation to the more accurate theory known as general relativity.  In this lecture I would like for us to put ourselves in Einstein's place (with the huge extra benefits of common knowledge that has accrued since then) and try to deduce what GR should look like, how it should differ from Newton's theory, in at least some broad sense.  

A first clue is that Newton's theory is not complete in itself is that
it describes action-at-distance, which would allow signals to travel faster than light.  Imagine that a distant observer has an extremely sensitive device for measuring the gravitational force between two masses,
\be
F = {G m_1 m_2\over r^2}\,.
\ee
Suppose $m_1$ is a mass that we are shaking back and forth, 
for simplicity say along the direction pointing toward the distant 
observer, who is measuring the effect on $m_2$.  By Newton's law, the time-dependent force will be 
\be
F = {G m_1 m_2\over (r_0 + \delta r(t))^2}\,,
\label{aaad}
\ee
where $r_0$ is the mean distance and $\delta r(t)$ is the displacement we are making.  According to Eq.\ (\ref{aaad}), the distant observer would see the effect immediately, no matter how large $r_0$ is, thereby allowing for communication at superluminal speeds.  This goes against a cherished principle from relativity, that nothing can travel faster than light.

From the analogous situation in electrodynamics, where we would be shaking a charge instead of a mass, we can anticipate how this should get fixed: the distant observer sees how the force must have looked at the retarded time, $t_r = t - r/c$, where $c$ is the speed of light:
\be
F(t,r) \cong {G m_1 m_2\over (r_0 + \delta r(t-r_0/c))^2}\,.
\label{fret}
\ee
However, this is not a very satisfying theory.  We know there is much
more to electrodynamics than Coulomb's law, and the retardation of
signals is something that should follow from the theory, not have to
be put in by hand.  Furthermore, we want the theory to be manifestly
Lorentz invariant, like Maxwell's equations.  A first step is to
reformulate  it not in terms of a force between two distant objects,
but instead between a single object and a field at the same point. 
Then it is the field, which exists throughout spacetime, that is
responsible for generating the force between distant objects.  And the
field will respond to changes in the sources that create it in a
causal way, consistent with the speed of light.

Therefore we need a gravitational {\it field}.  Well, Newton's theory already had that; it is the gravitational potential that is generated by a point mass.  Let's call it $\phi$,
\be
    \phi(r) = -{G m_1\over r}\,.
    \label{phieq}
\ee
The sign is negative because gravity is attractive: the bigger $m_1/r$
is, the deeper the potential.  
The force exerted by the field on a mass $m_2$,
now including its direction, is
\be
    \vec F = -m_2 \vec\nabla \phi  = - {G m_1 m_2\over r^2}\,\hat r\,.
    \label{Feq}
\ee
This looks just like the electric force equation $\vec F = q\vec E$, with mass playing the role of charge, 
and $\vec\nabla\phi$ that of the electric field.  
Could the relativistic theory of gravity be some version of Maxwell's equations?  

Two immediate problems tell us that this won't work. First,
electrodynamics requires there to be magnetic as well as electric
fields.  There seems to be no analog in gravitation.  Second,
like-sign charges repel, but masses (whose sign is always positive)
gravitationally attract each other.   A way around this is to
hypothesize that $\phi$ itself rather than a vector field is the
fundamental gravitational field. (Recall that in electrodynamics,  the
scalar potential is really $A^0$, the zeroth component of a
four-vector).  It is relatively easy to write a relativistic equation of motion
for a scalar field $\phi$ that couples to point masses $m_i$ following
trajectories $\vec x_i(t)$:\footnote{This scalar field is normalized
differently from the one in Eq.\ (\ref{Feq}), but otherwise they are
the same thing.  We use the Einstein summation convention: repeated
indices are implicitly summed over.  The Minkowski metric tensor is
$\eta^{\mu\nu} = {\rm diag}(1,-1,-1,-1)$.}
\be
    \partial_\mu\partial^\mu\phi \equiv
\eta^{\mu\nu}{\partial\over
	\partial x^\mu}{\partial\over\partial x^\nu}\phi = 
    \left({1\over c^2}\partial_t^2 -\nabla^2\right)\phi = 
    {4\pi G}\,\sum_i {(m_i c)^2\over E_i}\, \delta^{(3)}(\vec x-\vec x_i(t))
    \label{eom1}
\ee
where $E_i$ is the total energy of the particle,  including its
mass-energy.  In the limit of a stationary particle, $E_i = m_i c^2$
and the coefficient $(m_i c)^2/E_i \to m_i$.  You will show
why it must have  this peculiar form to be Lorentz invariant when you
solve problem 1.  The short answer is that even though the Dirac delta
function and $E_i$ are not invariant under Lorentz transformations,
their ratio is.\\

\noindent\fbox{%
    \parbox{\textwidth}{%
{\bf Metric sign convention.}  In this text I adopt the $(+,-,-,-)$ sign convention for the metric:
$\eta^{\mu\nu} = {\rm diag}(1,-1,-,1-,1)$.  Other authors, including Carroll, take the opposite.  There is no standard choice:
you should choose the one you like, and be sure to let people know what you decided so there is no ambiguity.  My choice has the 
advantage that $p^2 = +m^2$ for a massive particle with four-momentum $p^\mu$, whereas the other choice is more convenient for going between
$(3+1)$ dimensions (3 space plus 1 time) and situations where only spatial components of vectors  and tensors are needed.
}}
\medskip

Consider the case of a single mass $m_1$ that is at rest.  Then $\phi$ is time-independent, and Eq.\ (\ref{eom1}) becomes the 
Poisson equation (with $Gm_1$ playing the role of $q/\epsilon_0$ in the analogy to electrostatics), whose solution is Eq.\ (\ref{phieq}).
Now if we ask what the potential energy of another mass $m_2$ at $\vec x_2$ in this field is, it turns out to be
\be
    V = m_2\int d^{\,3}x\, \phi(x)\, \delta^{(3)}(\vec x - \vec x_2) =         -{G\, m_1 m_2\over |\vec x_1-\vec x_2|}\,,
\ee
which matches the Newtonian result, with the correct sign.  Moreover, if we solve the equation of motion
(\ref{eom1}) for sources that are moving, the solution will automatically incorporate the desired time retardation as in Eq.\ (\ref{fret}).

Alas, this theory, while it seems to solve our problems for 
explaining causal gravitational attraction between massive objects, 
it is
invalidated by the observations of the deflection of  
{\it massless} particles in gravitational fields.
If we imagine the photon has a tiny mass $m_2$ and take
the limit of its equation of motion $m_2\ddot x_2 = - G m_1 m_2/|\vec
x_1-\vec x_2|$ as $m_2\to 0$, we see that even
Newtonian gravity predicts that light will be deflected by massive
objects.  However it turns out that the predicted deflection angle 
(published by Johann von Soldner in 1801 \cite{Soldner}) is
smaller than what is observed, by a factor of 2! 
 This was first tested in 1919 by
Eddington and collaborators during a total eclipse of the sun
\cite{Dyson:1920cwa}.   Our scalar theory of gravity seems to be even
worse than Newtonian gravity in this respect, since it predicts that
massless particles should not interact with $\phi$.  This can be seen
from the vanishing of the source term in Eq.\ (\ref{eom1}) when $m_i=0$.
It says that massless particles cannot act as a source of the field.  
Therefore they cannot be influenced by it either.  To get light deflection
we would need some modification of the source term. 

We are out of historical order here since
Einstein had already invented GR and predicted the angle by which
light rays should
be bent by massive objects when Eddington's observation was made, precisely
in order to
test Einstein's prediction, but let's take it as experimental guidance
for drawing our own conclusions.   
What if 
we replace $(m_i c)^2$ in Eq.\ (\ref{eom1}) by $(E_i/c)^2$, 
the particle's energy?\footnote{I may omit factors of $c$ in subsequent equations,
using natural units $c=1$.  They can be restored when needed using dimensional analysis.}  This would have the right limit for a slowly moving
massive particle, and it would allow light to be deflected, but it
would spoil the Lorentz invariance of our theory since $E$ transforms
as the zeroth component of a four-vector (and so does the delta
function).  To maintain Lorentz
invariance, we would need to replace the right-hand-side of Eq.\
(\ref{eom1}) with an expression involving the four-momenta $p_i^\mu$
of each source particle.  To make it invariant, we would need to find
some other 4-vector to contract with the Lorentz index $\mu$. (Contracting
$p_i^\mu$ with itself is no good since that gives $m_i^2$.) But
there is no such natural 4-vector relevant to our setup.  If we simply invented a
4-vector $n^\mu=(1,0,0,0)$ and wrote $E_i = p_i^\mu n_\mu$, we would be fooling ourselves;
it superficially looks Lorentz-invariant, but $n^\mu$ turns into some other vector under
a Lorentz transformation.  It only has the desired form in some special frame.

We have thus determined that the gravitational field cannot be a
scalar field (since the source term would not be Lorentz invariant, or
it would predict the wrong deflection angle for light),
nor a vector field (which would lead to repulsive gravity).  The next
simplest possibility is that it is a {\it two}-index tensor, call it
$h^{\mu\nu}$.  This turns out to be correct.  A reasonable guess for
its relativistic equation of motion could take the form
\be
    \partial_\mu\partial^\mu h^{\alpha\beta} \sim 
    \kappa\, T^{\alpha\beta}
        \label{eom2}
\ee
where $T^{\alpha\beta}$ is some 2-index Lorentz tensor constructed out
of the 4-momenta of the particles that are sourcing the gravitational
field; we could expect it to contain terms like 
$p_i^\alpha p_i^\beta$ for each particle $i$ contributing to it.  Notice that this new equation is not Lorentz-invariant {\it
per se}, since it has uncontracted indices.  Instead it must be
Lorentz {\it co}variant, meaning that the left- and right-hand sides
transform in the same way under Lorentz transformations (notice that
the 4D Laplacian $\partial_\mu\partial^\mu$ is {\it in}variant since
upper and lower indices are contracted together). A
word about the ``$\sim$'': I did not write ``='' in Eq.\ (\ref{eom2})
because of a technicality.  Consider the quantities
$h=\eta_{\mu\nu}h^{\mu\nu}$ and $T = \eta_{\mu\nu}T^{\mu\nu}$ which
are the traces of the respective tensors.  By imagining all the ways
of consistently contracting indices, we could more generally
anticipate an equation of the form
\be
    \partial_\mu\partial^\mu (h^{\alpha\beta}
    + a_1\,h\eta^{\alpha\beta}) + a_2\,\partial^\alpha\partial^\beta h + a_3\,(\partial_\mu\partial^\alpha h^{\mu\beta} +\partial_\mu\partial^\beta h^{\mu\alpha} ) 
    + a_4\, \partial_\mu\partial_\nu h^{\mu\nu} \eta^{\alpha\beta}= 
    \kappa\, (a_6\,T^{\alpha\beta} + a_5\,T \eta^{\alpha\beta})
        \label{eom3}
\ee
with some constants $a_i$.  This more general form is second order in
derivatives and has the right Lorentz transformation properties.  The
argument I am making here cannot determine what the correct values of
the constants are; that has to await a detailed derivation.  The
right answer turns out to be $a_1=-\sfrac12$, $a_{2,3,4,5}=0$, when we
define $T^{\alpha\beta}$ in the conventional way, immediately below, 
and choose a convenient coordinate system.   The value $a_1=-\sfrac12$ has
the following significance: if one defines
\be
    \bar h^{\mu\nu} = h^{\mu\nu} - \sfrac12 h \eta^{\mu\nu}
\label{barhdef}
\ee
we notice that its trace is $\bar h = h - 2h = -h$, that is, the negative of the trace of $h^{\mu\nu}$ (using the fact that the
trace of $\eta^{\mu\nu} = 4$, the number of space-time dimensions). 
Hence $\bar h^{\mu\nu}$ is known as the {\it trace-reversed} metric perturbation, and its equation of motion has the elegant form 
$-\partial^2 \bar h^{\mu\nu} = 2\kappa T^{\mu\nu}$, with an important caveat to be discussed in the next chapter.  
That is, we have chosen a special coordinate system in which $\partial_\mu \bar h^{\mu\nu}=0$, known as Lorenz 
gauge,\footnote{Named after Danish physicist and mathematician Ludwig Lorenz,
not the Dutch physicist Hendrik Antoon Lorentz, after whom the transformation is named} to obtain this simple form.  
To reiterate, the field equation in this special coordinate system is
\be
    -\partial^2\bar h^{\mu\nu} = 2\kappa T^{\mu\nu}
\label{eom4}
\ee
and the proportionality constant is conventionally normalized as 
$\kappa = 8\pi G$, where $G$ is Newton's
constant.\\

\noindent\fbox{%
    \parbox{\textwidth}{%
{\bf Upper and lower indices.}  Any vector or tensor can be expressed with upper (contravariant) or lower (covariant) indices.
The metric tensor is used to switch between one form and the other.  In Minkowski space, the metric tensor is $\eta_{\mu\nu}$ for lowering
indices, as in $a_\mu = \eta_{\mu\nu}a^\nu$, and $\eta^{\mu\nu}$ for raising them.  This rule applies to the metric tensor itself, implying
\be
	\eta_{\mu\nu} = \eta_{\mu\alpha}\eta_{\nu\beta}\eta^{\alpha\beta}\,.
\ee
This relation is consistent with the fact that $\eta_{\mu\nu}$ and $\eta^{\mu\nu}$ are inverse to each other.
}}

\medskip
\noindent\fbox{%
    \parbox{\textwidth}{%
{\bf Lorentz transformation of tensors.}
Let us recall what we mean by the Lorentz transformation of a 
tensor, which in general could have any number of indices.  
A scalar is a zero-index tensor (which does not transform), a vector is a one-index tensor, {\it etc.}.  Denote the Lorentz transformation matrix by $\Lambda^\mu_{\ \nu}$.
A 4-vector such as the 4-momentum transforms as
\be
    p^\mu \to \Lambda^\mu_{\ \nu} p^\nu\,,
\ee
being careful to contract contravariant (upper) indices only with covariant (lower) ones.  Violating this rule would lead to a Lorentz-noninvariant expression.  One could also write
\be
    p^\mu \to \Lambda^{\mu\alpha}\eta_{\alpha\nu} p^\nu\,, 
\ee
where $\eta_{\mu\nu} = {\rm diag}(1,-1,-1-,1)$
is the Minkowski metric tensor.  This is the general way of converting between covariant and contravariant indices.  Then a tensor transforms as
\be
    T^{\alpha\beta} \to \Lambda^{\alpha}_{\ \mu}\Lambda^{\beta}_{\ \nu} T^{\mu\nu},
    \label{Ttrans}
\ee
just like the outer product\footnotemark\ of two four vectors,
$p^{\alpha}q^{\beta}$ (which  is also a tensor).  The generalization
to a higher-index tensor is obvious.  For the special case of a
two-index tensor, it is possible to rewrite Eq.\ (\ref{Ttrans})  as
matrix multiplication by transposing the second Lorentz matrix,
\be
    T^{\alpha\beta} \to \left(\Lambda T\Lambda^T\right)^{\alpha\beta}
    \label{TLtrans}
\ee    
but this is not useful for higher-index tensors, where we have to just
keep all the indices like in Eq.\ (\ref{Ttrans}).
}}
\footnotetext{To distinguish from the inner
product, where we would contract the two indices, in the outer product
we simply leave them uncontracted}
\medskip

Coming back to our proposed equation of motion for the gravitational field, there is a natural candidate for $T^{\alpha\beta}$, known as the {\it
stress-energy} tensor.  For a single particle of 4-momentum $p^\mu$, following the trajectory $x_1(t)$,  it can be written as
\be
    T^{\mu\nu} = {p^\mu p^\nu \over p^0}
        \,\delta^{(3)}(\vec x- \vec x_1(t))\,.
        \label{Teq}
\ee
To prove it behaves as a tensor under Lorentz transformation, one must
show that the delta function divided by the energy is invariant, as we mentioned above (see problem
1), so that only the $p^\mu p^\nu$ factors transform. For a particle
at rest, only the $T^{00}$ component is nonvanishing, and in this
case, the source term matches our guess in Eq.\ (\ref{eom1}), assuming the
appropriate choice of the constant $\kappa$.  For a collection of
particles, such as a solid or a fluid, $T^{\mu\nu}$ can be obtained by
summing over all the point particles.  We will find out that for systems that are rotationally invariant (in some preferred frame), the off-diagonal entries cancel between particles moving in different directions; then $T^{\mu\nu}$ is diagonal, with entries $T^{\mu\nu} = {\rm
diag}(\rho,p,p,p)$ where $\rho$ is the energy density and $p$ is the
pressure.  In the special case of massless particles, $p = \rho/3$ so
that $T^{\mu\nu}$ is traceless.\footnote{It is interesting to notice that energy density and pressure have the same units}   

We now have a candidate for a viable equation of gravitation, but what is the meaning of the tensor $h^{\mu\nu}$?  
Notice that from the form of $T^{\mu\nu}$, it must be symmetric under interchange of the indices.  
The only symmetric two-index tensor we already know about is the Minkowski metric,\footnote{The reader might object, ``what about
the Kronecker delta?"  But this is just the mixed-index version of the Minkowski tensor: $\eta^{\mu}_{\ \nu}\equiv
\delta^\mu_\nu$.} that appears in the line element of special relativity,
\be
    ds^2 = \eta_{\mu\nu} dx^\mu dx^\nu\,.
  \label{linel}
\ee
It is tempting to think that $h^{\mu\nu}$ might be somehow related, but since $\eta^{\mu\nu}$ is a constant, it is not immediately obvious.  However, the constancy of $\eta^{\mu\nu}$ depends on a choice of coordinates.  
For example, suppose we wanted to use polar coordinates for the $x$-$y$ plane.
Then the line element takes the form
\be
    ds^2 = dt^2 - dz^2 - d\rho^2 - \rho^2 d\theta^2 \,.
\ee
In this coordinate system, $\eta^{\theta\theta} = -\rho^2$ is no longer constant.  More generally, we could let $x^\mu\to x^\mu(x')$ so that $dx^\mu = ({\partial x^\mu\over\partial x'^\nu}) dx'^\nu$, 
and then in the new coordinate system, $ds^2 =
\eta'_{\alpha\beta}dx'^\alpha dx'^\beta$, with 
\be
    \eta'_{\alpha\beta}= {\partial x^\mu\over\partial x'^\alpha} 
    {\partial x^\nu\over\partial x'^\beta} \,
        \eta_{\mu\nu}\,,
            \label{diffeo}
\ee
which follows from Eq.\ (\ref{linel}) using the chain rule.  The transformation (\ref{diffeo}) is an example of a {\it diffeomorphism} or {\it general coordinate transformation}.  Notice the similarity to 
Eq.\ (\ref{Ttrans}).  In fact, Lorentz transformations are a special case of diffeomorphisms, but unlike Eq.\ (\ref{diffeo}), which generally transforms
$\eta^{\mu\nu}$ into some complicated form,
under a Lorentz transformation $\eta^{\mu\nu}$ remains invariant.
In fact, one can use the latter observation as a defining property\footnote{The other defining property is that the determinant
of $\Lambda^\mu_{\ \nu}$ should be $+1$ rather than $-1$}
of Lorentz transformations: they leave the Minkowski metric tensor
invariant, just like rotations leave the unit matrix invariant.

Of course, Minkowski space should exist in the absence of any gravitational sources,
regardless of what coordinate system we choose,  so it would not make
sense to identify $h^{\mu\nu}$ with $\eta^{\mu\nu}$. Instead,  the
spacetime metric gets {\it perturbed} in the presence of gravitational
sources: 
\be
    ds^2 =  g_{\mu\nu}dx^\mu dx^\nu \cong (\eta_{\mu\nu} + h_{\mu\nu})dx^\mu dx^\nu\,,
\ee
where we have ignored terms of $O(h^2)$. 
Indices should now be raised and lowered with the metric tensor, rather than $\eta_{\mu\nu}$.  The inverse
metric is
\be
	g^{\mu\nu} \cong \eta^{\mu\nu} - h^{\mu\nu}
\ee
(again ignoring terms of $O(h^2)$), where for  $h^{\mu\nu}$
it is permissible to use the Minkowski metric to raise indices,
$h^{\mu\nu} = \eta^{\mu\alpha}\eta^{\nu\beta}h_{\alpha\beta}$,
since the error is higher order in $h^{\mu\nu}$.  This
gives a physical interpretation of the gravitational field: it
describes a distortion in the fabric of
spacetime, caused by gravitational sources.  Spacetime itself has
become a dynamical quantity in general relativity.  This must have
been a shocking proposition to physicists after centuries of thinking
of space as a static, rigid nonentity.

The equation of motion (\ref{eom4}) is still only an approximation. 
It is valid in the {\it weak field} limit, $|h^{\mu\nu}| \ll 1$.  The
full equations of general relativity (Einstein equations) are
nonlinear in $h^{\mu\nu}$, and Eq.\ (\ref{eom4}) is only the leading
term in a small-$h^{\mu\nu}$ expansion. We will not be able to describe black
holes using this equation, for example.  But it provides the correct
generalization of Newton's theory for relativistic systems in which
the gravitational effect is weak; for example binary neutron stars
that are rapidly orbiting each other before they merge.  And it
accurately describes systems like the gravitational field of the
Earth.  The exact correspondence between the Newtonian potential and
the metric perturbation in the nonrelativistic limit turns out to be
(see for example Ref.\ \cite{Hirata})\footnote{The sign in this equation,
relative to that reference, comes from the different choice of metric sign convention. 
 It leads to the opposite sign relation between $h^{\mu\nu}$ and the Newtonian gravitational potential. It will become apparent
as we go along that this is just one of several 
sign conventions that one has to keep track of in general relativity.  The textbook \cite{Misner:1973prb} has an extensive table of sign
conventions for classic texts written before 1974.}
\be
    h^{\mu\nu} = +2\phi \times  {\rm diag}(1,1,1,1)\,.
\label{weak-h-sol}
\ee
The gravitational source thus not only warps space but also 
dilates time: clocks run slower in a gravitational field,
by the factor $\sqrt{1 + 2\phi}$ (recall that $\phi<0$).
Notice that the tensor in Eq.\ (\ref{weak-h-sol}) is not Lorentz invariant.  That is because the
source (e.g., Earth) prefers a special frame, the one where it is at rest.

\subsection{Problems}
% 2026: x->y, look for x in solutions.
 1.1 (a) Consider a particle of mass $m_1$ initially at rest at the 
origin.  Show that $\delta^{(3)}(\vec x - \vec x_1)/E$
    is invariant under a boost in the $y$ direction.  That is, in the
new frame it takes the form $\delta^{(3)}(\vec x' - \vec
x_1'(t'))/E'$.
Recall that a boost along one direction is given by
\be
        \Lambda = \gamma\left({1\atop \beta}{\beta\atop 1}\right)\nn
\ee
acting on the relevant components of the 4-vector,
    where $\beta = v/c$ and $\gamma = 1/\sqrt{1-\beta^2}$.\\
% 2026: v_0 -> v_1.  Look for v_0 in solutions.
(b) The problem is more tricky if the particle was not initially at
rest.  Then $\vec x_1 = \vec x_1(t)$.  Suppose $y_1(t) = v_1 t$,
moving along the $y$ axis, and do the boost along the same axis.
By transforming both $y$ and $t$ (but not $v_1$, consider it to be a
fixed parameter), find the more general transformation of the delta
function and of the energy, and show that their ratio does not change
under the Lorentz transformation.  Hint: how is $v_1$ related to the
original energy and momentum of the particle?\\
(c) Show that the argument of the transformed delta function tells
you how the originally velocity $v_1$ gets transformed when going to
the new frame.  Correct any missing factors of $c$ in case you have
been using $c=1$ in your 4-vectors. 
 This is why you do not transform $v_1$ inside
the delta function: it gets done for you by transforming $y$ and
$t$.\\

\noindent 1.2. (a) Compute the stress-energy tensor of the particle in
problem 1(a), before and after the boost, using Eq.\ (\ref{Teq}).  Take its boosted velocity to
be $-v$.  Express your answer after the boost in terms of $m$, $\beta$
and $\gamma$.  You can ignore the dimensions in which the particle is not moving so that it appears as a $2\times 2$ matrix.\\
(b) Show that you get the same result for the boosted stress tensor by
doing the matrix Lorentz transformation on the unboosted one, using Eq.\ (\ref{TLtrans}).\\

\noindent 1.3. Consider the stress-energy tensor from a very large
collection of particles moving in random directions, i.e., a perfect
fluid.  (a) Show that by summing over the directions of motion,
the off diagonal elements average to zero.\\
(b) If the particles are so dense that their individual positions are
not kept track of, show that $T^{00}$ becomes the energy density 
$\rho$ of the
fluid.  Integrate over some small volume $\Delta V$ to show it.\\
(c) The $T^{ii}$ elements are similarly related to the ``momentum times velocity
density'' of the fluid (ignoring a factor of $c$).  Show that this has the same units as pressure.
Argue that the rate of momentum transport across a unit area is the same as pressure.
Can you derive the relation $p=\rho/(3c)$ for massless particles?  Hint:
consider $\sum_i T^{ii}$.\\

% 2026: x->y, look for x.
\noindent 1.4. (a) Show that $\eta^{\mu\nu}$ remains invariant under a
boost in the $y$-direction.\\
(b) In (1+1) dimensions (one space plus one time), the Levi-Civita
tensor $\epsilon^{\mu\nu}$ is antisymmetric, with off-diagonal 
entries $\pm 1$.  Show that this also remains invariant under a
such a Lorentz transformation.\\
(c) The Levi-Civita tensor in (3+1) dimensions is
$\epsilon^{\alpha\beta\mu\nu}$, again totally antisymmetric with
nonzero entries of $\pm 1$, and it is invariant under cyclic permutations of its indices.
 Show that it is invariant under a boost
in the $y$ direction.  Hint: using (b) you can verify that
$\epsilon'^{0123} = \epsilon^{0123}$.  Further show that 
$\epsilon'^{\alpha\beta\mu\nu}=0$ if any two of the indices are the
same.\\

%2026: Jupiter-> Neptune
\noindent 1.5. (a) Compute the value of the Newtonian potential on the surface of Neptune,
and make it dimensionless using appropriate powers of $c$.
Show that the corresponding metric perturbation is indeed $\ll 1$.\\
%2026: 1.4->1.7, 10->12.
(b) Repeat the calculation for a neutron star whose mass is 1.7 solar
masses and radius is 12\,km. \\
%2026: 10->15
(c) Find the factor by which a clock ticks faster at
an altitude of 15\,km relative to one at the Earth's surface.\\
(d) How
much slower do clocks run on the surface of the neutron star than on Neptune?

\section{General coordinate transformations}
In the previous chapter, we briefly introduced the concept of general coordinate transformations, 
\be
    x^\mu\to x^\mu(x')\,.
    \label{gcteq}
\ee
They turn out to be of central importance to the theory of GR, like Lorentz transformation are to special
relativity (SR).  Lorentz transformations are a special kind of coordinate transformation, with the
defining property that $ds^2 = dt^2 - dx^2$ remains invariant, in the sense that it has the same form
$ds^2 = dt'^2 - dx'^2$ in the new coordinate system.  Equivalently, the Minkowski metric tensor remains
invariant under Lorentz transformations.  Recall that SR could be derived from a few basic principles, including that the laws of
physics should be invariant under a change of reference frames, parametrized by Lorentz transformations.  GR
carries this logic a step further and asserts that the laws of physics should be invariant under {\it any}
change of the coordinate system, linear or not.  Physics must not depend upon how we label the points in
spacetime.  This will provide an essential clue in the coming quest to derive Einstein's field equations.
However the metric tensor is not invariant under general coordinate transformations, despite that
the physics it describes is unchanged.

Our linearized theory of gravitation does not have the property of complete invariance under general
coordinate transformations.  Instead, it will have an approximate version of this symmetry, valid in the case
of small transformations.  Suppose that the exact spacetime metric is
\be
    g_{\mu\nu} = \eta_{\mu\nu} + h_{\mu\nu}\,\quad (\hbox{hence\ }g^{\mu\nu} \cong \eta^{\mu\nu} - h^{\mu\nu})
\ee
and the full theory of GR is invariant under 
coordinate changes of the metric,
\be
    g_{\alpha\beta} \to g'_{\alpha\beta} = 
{\partial x^\mu\over\partial x'^\alpha} 
    {\partial x^\nu\over\partial x'^\beta} \,
        g_{\mu\nu}\,
,\quad  g^{\alpha\beta} \to  g'^{\alpha\beta} = 
{\partial x'^\alpha\over\partial x^\mu} 
    {\partial x'^\beta\over\partial x^\nu} \,
        g^{\mu\nu}
            \label{diffeo2}
\ee
as in Eq.\ (\ref{diffeo}).  (The second form can be obtained from the
first using the fact that $g^{\mu\nu}$ is inverse to $g_{\mu\nu}$,
and $\partial x'^\alpha/\partial x^\mu$ is inverse to 
$\partial x^\alpha/\partial x'^\mu$.)  
But here we want to consider $|h^{\mu\nu}| \ll 1$, which 
suggests to look at small diffeomorphisms of the form
\be
    x^\mu = x'^\mu + \xi^\mu(x')\,,%\quad x_\mu = x'_\mu + \xi_\mu(x')\,,
\ee
where $|\xi^\mu(x')|\ll |x'^\mu|$ in some formal sense.  The idea is to keep 
only terms that are either linear in $h_{\mu\nu}$ or $\xi^{\mu}$ and consider 
them to be of the same order.  In this way, we can find  coordinate 
transformations that will simplify $h^{\mu\nu}$ in some 
convenient way.  Keeping just these terms, Eq.\ (\ref{diffeo2}) takes the form
\bea
    \eta_{\mu\nu} + h_{\mu\nu}&\to&
        \eta_{\mu\nu} +  {\partial\, \xi^\alpha\over\partial x'^\mu}\eta_{\alpha\nu} + 
        {\partial\, \xi^\beta\over\partial x'^\nu}
        \eta_{\mu\beta} + h_{\mu\nu}\,,\nn\\
        &\equiv& \eta_{\mu\nu}  + \xi_{\nu,\mu} + \xi_{\mu,\nu} +h_{\mu\nu}
        \label{diffeo3}
\eea
Then it makes sense to identify the new terms with the transformation of $h^{\mu\nu}$, which can be more simply written as 
\be
    h_{\mu\nu}\to h'_{\mu\nu} =  h_{\mu\nu} +
    \xi_{\mu,\nu} + \xi_{\nu,\mu}
\ee
Given some initial $h_{\mu\nu}$, it is always generically possible to find a coordinate system such that, for example
\bea
    \partial'^\mu h'_{\mu\nu} &=& \partial'^\mu h_{\mu\nu} + 
    {\partial^2\,\xi_\mu\over\partial x'^\nu\partial x'^\alpha}
	\eta^{\alpha\mu} + 
        {\partial^2\,\xi_\nu\over\partial x'^\mu\partial x'^\alpha}
        \eta^{\alpha\mu}\nn\\
    &\equiv&  \partial'^\mu h_{\mu\nu} + 
    \xi_{\mu,\nu}^{\ ,\mu} + \xi_{\nu,\mu}^{\ ,\mu} 
    =0\,.
        \label{lorentz}
\eea
This is because Eq.\ (\ref{lorentz}) is a set of four differential equations (labeled by $\nu$) that depend on four independent functions $\delta x^\mu(x')$, which is a well-posed problem.

Of course, the choice $\partial^\mu h'_{\mu\nu} \equiv {h'}_{\!\mu\nu}^{\ \ ,\mu} = 0$ is just one possibility.  As another example, we could have
imposed the conditions $h'_{0\mu}=0$ instead.  The choice is a matter of convenience, depending on how it might simplify calculations.  It is
perfectly analogous to a similar procedure in electrodynamics, known as choice of gauge condition.  Because of gauge symmetry, we can impose
$\partial_\mu A^\mu = 0$ on the vector potential.  Or we could choose axial gauge,  $A^0 = 0$.  In E\&M, the gauge transformation depends on only
one function, $A^\mu\to  A^\mu + \partial^\mu\theta(x)$, so we can set only one quantity to zero, whereas in GR we have four functions, hence four
gauge conditions.

This partly explains how  Eq.\ (\ref{eom3}) can be simplified to the form (\ref{eom4}).
First notice that all the same terms that are allowed by Lorentz invariance for
$h^{\mu\nu}$ are also allowed for $\bar h^{\mu\nu}$ since they are both
two-index tensors.  By imposing $\bar h^{\mu\nu}_{\ ,\nu}=0$, we get rid of
the unwanted terms that do not involve the traces $h$ or $T$.  A more detailed
derivation is needed to take care of those terms.  
In analogy to EM, $\bar h^{\mu\nu}_{\ ,\nu}=0$ is known as the 
{\it Lorenz gauge condition} in linearized GR.  
The EOM would look different if we had chosen
some other gauge.  In general one should always know what the choice of gauge was
and use it as a constraint on the solutions that arise from the EOM (such as Eq.\ (\ref{eom3}) in Lorentz
gauge) in order to find a consistent solution.

As an example,  consider  the possible gauge choice (which might be called ``transverse gauge''), $h^{\mu\nu}_{\ \ ,\nu}=0$.
Suppose we are given that  $h^{\mu\nu} = \epsilon x^\mu x^\nu$ for some small $\epsilon$, and we consider
only $x$ such that $|h^{\mu\nu}|$ remains small, so that it remains a small perturbation on Minkowski space. 
After transforming to the new coordinate $x'$, $h$ will have the same form up to higher order corrections,
since we take $\delta x \sim h$.  The divergence is 
$\partial'_\mu h^{\mu\nu} = \epsilon x'^\mu_{\
,\mu}x'^\nu  + \epsilon x'^\mu\delta^\nu_\mu =  5\epsilon x'^\nu$.  From Eq.\ (\ref{lorentz}) we see that $\delta x^\mu$ must therefore be
cubic in $x'$.  Further, it should only have one uncontracted Lorentz index.  One can easily guess that
$\delta x^\mu = \alpha x'^2 x'^\mu$.  I leave it for you to determine 
the value of $\alpha$ in problem 2.1.  
%(Answer: $\alpha=-\epsilon/6$.)

We derived the tensor transformations for the metric and its inverse,
but similar transformation laws apply to other tensors, with arbitrary
numbers of indices.  One can say that this is the defining property of
a tensor: its lower or upper indices transform by the Jacobian, or its
inverse, respectively, of the coordinate transformation.  For the
contravariant components of a vector $A_\alpha$,
\be
   A'_\alpha = {\partial x^\mu\over \partial x'^\alpha}A_\mu\,,
\label{contrans}
\ee
while for the covariant components $A^\alpha$, 
\be
	A'^\alpha = {\partial x'^\alpha\over\partial x^\mu}A^\mu\,,
	\label{covtrans}
\ee
similarly to Eq.\ (\ref{diffeo2}).  An easy way to recover these
formulas is to demand that geometrical objects remain invariant 
under coordinate transformations, in particular the differential $A_\mu dx^\mu = A'_\mu
dx'^\mu$ and the
directional derivative $A^\mu\partial_\mu = A'^\mu\partial'_\mu$.   The
transformation laws (\ref{contrans},\ref{covtrans}) then follow
directly from the chain rule when one expresses $dx^\alpha$ in 
terms of $dx'^\alpha$
or $\partial_\alpha$ in terms of $\partial'_\alpha$.

It should be noted that with a general metric $g_{\mu\nu}$ and
its inverse $g^{\mu\nu}$, indices should be raised and lowered 
using the full metric (or its inverse).  In the linearized theory however, we can use
the Minkowski tensor to do the job, just like we did in Minkowski
space.  The reason is that the error one makes is at quadratic order
in the perturbation $h_{\mu\nu}$, or the small coordinate
transformation $\xi^\mu$, or their product, and we are neglecting such corrections.
The one place where it matters is when raising  indices of
$g_{\mu\nu}$ or lowering indices of $g^{\mu\nu}$ themselves.   But this process is equivalent
to using the fact that they are inverse to each other, as you can
easily verify.

\subsection{Rindler space}
A very interesting example of a general coordinate transformation is the one that describes the reference frame
of a uniformly accelerating observer moving in the $x$ direction:
\bea
\label{rindler-def}
	t &=& a^{-1}e^{a\xi}\sinh a\eta\,,\nn\\
	x &=& a^{-1}e^{a\xi}\cosh a\eta\,,
\eea 
with $y$ and $z$ untransformed.  The inverse transformation is
\bea
	\eta &=& {1\over 2a}\ln\left({x+t\over x-t}\right)\,,\nn\\
	\xi &=& {1\over 2a}\ln a^2(x^2-t^2)\,.
\eea
It is only well defined in two patches of the full Minkowski space, since we need $x>|t|$ or $x < -|t|$
for the logarithms to be real-valued.  Call them $R$ and $L$ for ``right'' and ``left,'' respectively.
 In fact, only one of these patches is valid for a given observer, since $R$ corresponds to the case
$a>0$ and $L$ represents $a < 0$.  The regions $t>|x|$ and $t<-|x|$ not covered by the Rindler coordinates are labeled as
$F$ for ``future'' and $P$ for ``past.''\ 
Notice that the timelike coordinate $\eta$ runs backwards in $L$.
One can readily show that the transformed Minkowski line element is given by
\be
	ds^2 = e^{2a\xi}\left(d\eta^2 - d\xi^2\right) - dy^2 - dz^2\,.
\ee
%Henceforth I will ignore $y$ and $z$ and treat the space as being two-dimensional.

The boundaries of the $R$ and $L$ patches are the null surfaces $|x| = |t|$, which are the lightlike
trajectories the observer approaches as $t\to\pm\infty$.  They play the role of {\it horizons} for the
observer, meaning that they can never receive signals emanating from $F$ (if they are in $R$)
or $P$ (if they are in $L$).  They are moving too fast for light rays from those regions to catch up to them.
On the other hand, they are able to send signals into these regions.  This turns out to provide a close analogy
to the causal structure of spacetime around black holes, that will be useful in chapter \ref{BHsect}.  

Rindler
space provides a nice example showing that a given coordinate system does not always cover the entire manifold.
In such cases one has to find other coordinates to describe the missing regions, and patch them together
with the original coordinates in a continuous fashion.  In the present case, one could define supplementary
coordinates $\bar\eta$ and $\bar\xi$ such that
\bea
\label{rindler-sup}
	t &=& a^{-1}e^{a\bar\xi}\cosh a\bar\eta\,,\nn\\
	x &=& a^{-1}e^{a\bar\xi}\sinh a\bar\eta\,,
\eea
whose inverse is  
\bea
	\bar\eta &=& {1\over 2a}\ln\left({t+x\over t-x}\right)\,,\nn\\
	\bar\xi &=& {1\over 2a}\ln a^2(t^2-x^2)\,.
\label{rindler-inv}
\eea
These are suitable to describe the $F$ and $P$ regions.  The metric is 
\be
	ds^2 = e^{2a\bar\xi}\left(d\bar\xi^2-d\bar\eta^2 \right) - dy^2 - dz^2\,,
\ee
where now $\bar\xi$ plays the role of the timelike coordinate.  The same interchange of coordinate roles when
passing through the horizon will feature in the Schwarzschild metric in Chapter \ref{BHsect}, further
strengthening the analogy to black holes.

\subsection{Radiation seen by accelerated observers}
\label{rindler-rad}
Readers may wish to skip this section and come back to it later.  It serves as a model for Hawking radiation, to
be discussed in Section \ref{hawking-sect}.\footnote{It was discovered by Davies \cite{Davies:1974th} soon after Hawking's famous paper
\cite{Hawking:1975vcx}
appeared.}  A surprising prediction of quantum field theory (QFT) is that
accelerated observers will find themselves in a bath of thermal radiation, with temperature proportional to 
$a$ \cite{Davies:1974th}, which appears to be emitted and absorbed by the horizons $x = \pm t$.  We start by 
introducing
the elements of QFT needed to understand the result.
A massless quantum field, such as the electromagnetic field,
 can be thought of as an infinite collection of harmonic oscillators, one for each wave number $\vec k$, 
with frequency $\omega_k = |\vec k|$.  The excitation number of the oscillator is the number of photons carrying
that wave number.  For simplicity, we will consider a massless scalar field, which can represent a single
polarization of the electromagnetic field.  We will also ignore the transverse dimensions $y,z$ and consider it
as a $1+1$ dimensional system.  The generalization to $1+3$ dimensions is straightforward.

The massless scalar field satisfies the wave equation $\eta^{\mu\nu}\partial_\mu\partial_\mu\phi = (\partial_t^2 - \partial_x^2)\phi = 0$, which has the general 
solution
\be
	\phi = \int {dk\over \sqrt{2\pi(2\omega_k)}}\left(a_k e^{-i(\omega_k t - kx)} + 
a^*_k e^{i(\omega_k t - kx)}\right) \equiv 
\int dk\left(a_k u_k(t,x) + 
a^*_k u_k^*(t,x)\right)
\,,
\label{phiM}
\ee
where the normalization of the mode functions is chosen so that they are orthonormal under the inner product
 \cite{Bjorken:1965zz}
\bea
\label{innerprod}
	(u_k,u_{k'}) &=& -i\int dx\, \left(u_k \stackrel{\leftrightarrow}{\partial_t} u^*_{k'}\right) = 
-i\int dx\, u_k \left(\stackrel{\rightarrow}{\partial_t} - \stackrel{\leftarrow}{\partial_t}\right) u^*_{k'}
	= \delta(k-k')\,.\\
	(u_k,u^*_{k'}) &=& 0\,.\nn
\eea
In QFT, the Fourier coefficients
get promoted to creation and annihilation operators (with $a^*_k\to a^\dagger_k$) that act like raising and 
lowering operators of the harmonic oscillator.  There is a vacuum state containing no particles, defined by
\be
	a_k|0\rangle_\M = 0 \hbox{\ for all\ }k\,.
\ee
I included the subscript $M$ for Minkowski, since we will see that the vacuum state is not unique, but can
depend on the observer's frame of reference, if it is noninertial.  The number operator for Minkowski observers
$N^\M_k = a^\dagger_k a_k$,
that counts the number of photons in the state $k$, is obviously zero acting on $|0\rangle_\M$.

We can repeat the exercise in Rindler coordinates, where $g^{\mu\nu}\partial_\mu\partial_\mu\phi = 
e^{-2a\xi}(\partial_\eta^2-\partial_\xi^2)\phi = 0$.  The solutions have the same form as in Minkowski space,
\be
	\phi = \int {dk\over \sqrt{2\pi(2\omega_k)}}\left(b_k e^{-i(\omega_k \eta - k\xi)} + 
b^*_k e^{i(\omega_k \eta - k\xi)}\right) \equiv 
\int {dk}\,\Big(b_k v_k(\eta,\xi) + 
b^*_k v^*_k(\eta,\xi)\Big)\,,
\label{phiR}
\ee
but the coefficients $b_k$ must be different from $a_k$.  They correspond to another vacuum state
\be
		b_k|0\rangle_\R = 0 \hbox{\ for all\ }k\,.
\ee
It is important to notice that since the Rindler coordinates cover only one fourth of  Minkowski space,
the $R$ region, the equivalence of $\phi$ between Eqs.\ (\ref{phiM}) and (\ref{phiR}) is only valid in this
patch.  It is possible to extend the definition of $\phi$ in Eq.\ (\ref{phiR}) to remedy this \cite{Birrell:1982ix}, by defining an
additional set of creation and annihilation operators for the $L$ region (and analytically continuing the
mode functions into the $F$ and $P$ regions), 
but for our
purposes it will not be necessary.

One can show that there is a linear relation between the creation and annihilation operators in the two
systems, known as
a Bogoliubov transformation, and its inverse,
\be
\label{bogo}
	\left(a_k\atop a^\dagger_k\right) = \int dk'\left({\alpha_{k'k}\atop \beta^*_{k'k}}\ {\beta_{k'k}\atop 
\alpha_{k'k}^*}\right)\left(b_{k'}\atop
b^\dagger_{k'}\right)\,,\quad
\left(b_k\atop b^\dagger_k\right) = \int dk'\left({\alpha^*_{kk'}\atop -\beta_{kk'}}\ {-\beta^*_{kk'}\atop 
\alpha_{kk'}}\right)\left(a_{k'}\atop
a^\dagger_{k'}\right)\,.
\ee
The Bogoliubov coefficients satisfy orthogonality  relations that can be deduced by demanding
that the two equations (\ref{bogo}) are compatible:
\be
	\int dk\left(\alpha_{kp}\alpha^*_{kq} - \beta_{kp}\beta^*_{kq}\right) = \delta(p-q),\quad
	\int dk\left(-\alpha_{kp}\beta^*_{kq} + \beta_{kp}\alpha^*_{kq}\right) = 0\,.
\ee
Similar relations with all subscripts transposed also hold, since Eqs.\ (\ref{bogo}) can be substituted into
each other in either order.

 Suppose that the true state of the spacetime is the
Minkowski vacuum.  Then if $\beta_{kk'}$ is nonzero, the number operator $N^\R_{k} = b_k^\dagger b_k$ is nonzero
when acting on $|0\rangle_\M$, since $b_k$ contains some admixture of $a^\dagger_{k'}$s.  The interpretation is
that an accelerating observer in the $|0\rangle_\M$ vacuum will see a spectrum of particles given by
\be
	{dN\over dk} = \phantom{x}_\M\langle 0|N^R_k|0\rangle_\M = \int dk'|\beta_{k'k}|^2\,.
\label{Nkint}
\ee	
Thus we need to solve for $\beta_{kk'}$.  This can be done by taking the inner product (\ref{innerprod})
between $u_k$ and both versions of $\phi$, Eqs.\ (\ref{phiM},\,\ref{phiR}):
\be
	(\phi,u_k) = a_k = \int dk'\left(b_{k'}(v_{k'},u_k) + b^\dagger_{k'}(v^*_{k'},u_k)\right)\,.
\ee
Comparing this with Eq.\ (\ref{bogo}), we see that\footnote{The restriction to $t=0$ is a convenient choice, but any spacelike ray passing through the origin of $(t,x)$
provides a valid Cauchy surface for defining the inner product. One can show by Gauss's theorem that the result
is independent of this choice. The operator
$\stackrel{\leftrightarrow}{\partial_t}$ should be replaced by derivatives with respect to the direction normal
to the surface.  Notice that we have restricted $x\ge 0$ because the Rindler mode functions are defined
only in the $R$ region.}
\bea
	\beta_{k'k} &=& (v^*_{k'},u_k)\nn\\
 &=& {1\over 4\pi\sqrt{\omega_k\omega_{k'}}}
	\int_0^\infty dx\,\left(\omega_{k'}-\omega_{k}\partial_t\eta + k'\partial_t\xi\right) 
e^{i(\omega_k\eta-k\xi) + i(\omega_{k'}t - k'x) }\Big|_{t=0}\nn\\
	&=& {-k\over 4\pi\sqrt{|k k'|}}\left(i{k'\over a}\right)^{i{k\over a}}\left({\rm sgn}\,k' + {\rm sgn}\,k\right)
	\,\Gamma(-ik/a)
\label{rindler-beta}
\eea
We used Eq.\ (\ref{rindler-inv}) to set $\partial_t\eta=1/(ax)$ and $\partial_t\xi=0$ at $t=0$.  Taking
$(i)^{ik/a} = e^{i\pi/2 \cdot ik/a}$ and $|\Gamma(it)|^2 = , \pi/(t\sinh\pi t)$, this gives
\be
	|\beta_{k'k}|^2 = {1\over 2\pi a |k'|}\left(1\over e^{2\pi|k|/a} -1\right)
\ee
when $k$ and $k'$ have the same sign, 
which is a Bose-Einstein distribution with temperature $T = a/(2\pi)$.\footnote{$dN/dk$ diverges logarithmically
when integrated in Eq.\ (\ref{Nkint}).  This is an artifact from the assumption that the observer has been
accelerating since the infinite past, so that radiation emitted from the horizon has been accumulating
during an infinite time \cite{Davies:1974th}.}  Remarkably, we will see in Section \ref{hawking-sect} that the
same conclusion holds for the spacetime surrounding a black hole; quasi-thermal radiation is emitted from the
event horizon, with a temperature going as the acceleration associated with the surface gravity at the horizon.

This radiation seems to belie Einstein's equivalence principle, which
states that physics in a uniformly accelerating frame is
indistinguishable from that in a uniform gravitational field.  We
do not expect to see radiation in the vacuum of the latter
situation, unless a horizon is present.  If the stress-energy comes
from an infinite slab in the $x$-$y$ plane, giving rise to a uniform
field $\phi = gz$ along the $z$ direction, is there a horizon at some
finite value of $z$?  We must await the full nonlinear formulation of
GR to find out.\footnote{In fact, the exact solution of Einstein's
equations for an infinite thin slab, which is the gravitational analog
of Rindler space, has a horizon \cite{Vilenkin:1983ykn}.  I am not aware of any paper
that has derived the radiation emanating from the horizon for
observers at rest in this spacetime.}

\subsection{Problems}

\noindent 2.1. Let the  inverse metric be $g^{\mu\nu} = \eta^{\mu\nu} - h^{\mu\nu} = \eta^{\mu\nu} + \sfrac12\beta x^\mu x^\nu$, 
considering $\beta$
to be small. Suppose that a right triangle 
with sides of length $\Delta y$, $\Delta z$ is located with its bottom left corner 
at position $x^\mu_0 = (0,0,y_0,z_0)$ in the $y$-$z$ plane.  Its other vertices are at $x^\mu_0 + \Delta y\, 
\hat y$
and $x_0^\mu + \Delta z\, \hat z$, where $\hat y$ and $\hat z$ are the unit vectors.  Assume the sides are infinitesimal, so that the metric is approximately constant
over the triangle. \\  
(a) Find the extra contribution to the length $\ell$ of the hypotenuse, to leading order in $\beta$, relative to its value
$\ell_0$ in Minkowski space. \\
(b) As suggested in the text, perform a coordinate transformation with $\xi^\mu = \alpha\, x'^\mu x'^2$.  Determine $\alpha$ 
in terms of  $\beta$ such that $h'^{\mu\nu}$  satisfies the Lorenz gauge condition, and solve for $h'^{\mu\nu}$.\\
(c) Determine $\Delta x'$ and $\Delta y'$ in the new coordinate system in terms of 
$x_0^\mu$, $\Delta x$ and $\Delta y$.\\
(d) Using the results from parts (b) and (c), show that the length of the hypotenuse in the new coordinate system
 remains the same as in the original one.
\begin{comment}
I changed $-\epsilon$ to $\sfrac12\beta$ in 2026, and $x\to y$, $y\to z$ to make it a bit different from previous year.
 Answers:\\
(a)  $-(\epsilon/2)(x_0\Delta x + y_0\Delta y)^2(\Delta x^2 + \Delta y^2)$.\\
(b) $h'^{\mu\nu} = h^{\mu\nu} - \sfrac23\epsilon x^\mu x^\nu - \sfrac13 \epsilon x^2\eta^{\mu\nu}$. \\ (c) $\Delta x' = (1+\alpha x_0^2)\Delta x + 2 \alpha x_0^\beta \Delta x_\beta x_0$, \quad
$\Delta y' = (1+\alpha x_0^2)\Delta y + 2 \alpha x_0^\beta \Delta x_\beta y_0$.
\end{comment}

\medskip
\begin{comment}
In 2026 I changed sin -> cos and cos -> -sin relative from 2025, also B->A.
\end{comment}
\noindent 2.2.  A metric has the perturbation $h^{tx} = h^{xt} = A\cos(k(y-t))$, $h^{ty} = h^{yt} =
A\sin(k(x-t))$, $h^{xy} = h^{yx} = A(\sin(k(x-t)) + \cos(k(y-t)))$, $h^{xx} = A\cos(k(z-t))= - 
h^{yy}$, with all other elements vanishing. \\
(a) Find the coordinate transformation that makes $h'^{0\mu}=0$ for all $\mu$ and display the transformed 
perturbation $h'^{\mu\nu}$ in this gauge.\\
(b) Do the same for the transverse gauge $h^{\mu\nu}_{\ \ ,\nu}=0$. 
Hint: the coordinate transformation has the form $\xi^x = f(z, t-y)$,
$\xi^y = g(z, t-x)$.  For simplicity, set any integration constants to
zero.

\medskip
\noindent 2.3.  (a) Construct a diagram of 2D Minkowski space $(t,x)$ that shows the various regions $R,L,P,F$ corresponding to 
Rindler coordinates. Draw the lines of constant $\eta$ and $\xi$ in the $R,L$ regions, and $\bar\eta$ and
$\bar\xi$ in the $P,F$ regions.\\
(b) Construct the transition functions that give the relationship between the $\eta,\xi$ and $\bar\eta,\bar\xi$
coordinates along the horizons.

\
\section{Gravitational waves in vacuum}

Before we try to solve the EOM (\ref{eom4}) for the metric in the presence of a source (for example the
Earth, or black holes orbiting around each other), an easier task is to look for vacuum solutions, that is,
where no matter or energy is present, and therefore the stress-energy tensor $T^{\mu\nu}$ vanishes.  You
might object that this should also cause  the solution for $h^{\mu\nu}$ to vanish.  While it is true that
this is one solution, it is not unique.   Even if matter was present to initially create the gravitational
wave, it will continue to propagate through empty space once it gets started.  For example, it is easy to see
that 
\be
    h^{\mu\nu} = \epsilon^{\mu\nu}\cos(k(t-z)) = \epsilon^{\mu\nu}\cos(k_\mu x^\mu)
    \label{gwsol}
\ee
is a solution to the EOM, for any fixed polarization tensor $\epsilon^{\mu\nu}$,  and any value of the
wavenumber $k$.  It is a gravitational wave traveling in the $z$ direction, moving at the speed of light, in
our units of $c=1$.  The 4-vector has components $k^\mu = (k,0,0,k)$ in this example.

An important question is how many independent polarizations can be present in the solution (\ref{gwsol}).  A $4\times 4$ symmetric tensor has
$(4\times 5)/2 = 10$ independent components.  This would naively imply that gravitons have 10 polarizations,
analogous to the two possible polarizations of electromagnetic waves.  However, we need to impose the 
Lorenz gauge condition on the solution, because this was used to put the more general initial
EOM (\ref{eom3}) into the simpler form (\ref{eom4}). Since $\partial_\mu \cos(k_\nu x^\nu) = -k_\mu
\sin(k_\nu x^\nu)$,  this implies
\be
    k_\mu (\epsilon^{\mu\nu} - \sfrac12\epsilon^\alpha_\alpha \eta^{\mu\nu}) = 0\,,
\ee
(For general plane waves, one could take $h^{\mu\nu}$ to be the real part of 
$\epsilon^{\mu\nu} e^{-ik^\mu
x^\mu}$ with $\epsilon^{\mu\nu}$ possibly being complex.)   These are four conditions that restrict the
possible allowed forms of $\epsilon^{\mu\nu}$, so we might think that
this leaves six graviton polarizations, but that is not correct: in fact there are only two.

A similar counting problem happened in electrodynamics: we started with 4-component solutions
$A^\mu = \epsilon^\mu e^{-ik\cdot x}$ and imposed one gauge condition, for example $k\cdot\epsilon = 0$, which would seem to leave three independent solutions, but we know that only the transverse polarizations are physical; the longitudinal one with
$\vec\epsilon \propto \vec k$ for the spatial parts of the 4-vectors is unphysical.  The problem is that the solutions left after imposing $k\cdot\epsilon = 0$ still admit a residual gauge symmetry: to any such solution we can add a term of the form
\be
    \epsilon^\mu \to \epsilon^\mu + a k^\mu
\ee
for arbitrary $a$, and this still respects the gauge condition $\epsilon\cdot k=0$, since $k^2 = 0$ for any solution to the equation of motion.  We can use this residual gauge symmetry to fix the longitudinal polarization to zero, leaving just two physical, transverse, polarizations.  

The analogous thing happens in gravity.  Since we are interested in transformations that simplify plane wave solutions, we will consider
gauge
(coordinate) transformations that are also plane waves, $\delta x^\mu \sim {\rm Re}(\xi^\mu e^{ik\cdot x})$, 
so that the gauge transformation of the metric
gives rise to a shift in the polarization tensor of the form
\be
    \epsilon^{\mu\nu}\to \epsilon^{\mu\nu}
    + k^\mu \xi^\nu + \xi^\mu k^\nu\,.
\ee
The corresponding shift in the trace-reversed perturbation is 
\be
    \bar\epsilon^{\mu\nu}\to \bar\epsilon^{\mu\nu}
    + k^\mu \xi^\nu + \xi^\mu k^\nu - \eta^{\mu\nu} k\cdot \xi\,,
\ee
and the Lorenz gauge condition transforms into
\be
	\bar\epsilon^{\mu\nu} k_\nu \to \bar\epsilon^{\mu\nu} k_\mu + k^2\xi^\mu + (k\cdot\xi)k^\mu - (k\cdot\xi)k^\mu \,.
\ee
Since $k^2=0$ for a gravitational plane wave, the extra terms vanish: the Lorenz condition is preserved for any choice of
$\xi^\mu$.  Since there are four linearly independent choices, these extra four gauge transformations reduce 
the number of physical degrees of freedom from six to two.

A convenient choice of basis for the physical polarizations is the transverse, traceless one, where the two independent polarization tensors take the 
form
\be
    \epsilon_{+}^{\mu\nu} = \left(\begin{array}{rrrr} 0 & 0 & 0 & 0\\
           0 & 1 & 0 & 0\\
           0 & 0 & -1 & 0\\
           0 & 0 & 0 & 0\end{array} \right)\,,\qquad
      \epsilon_{\times}^{\mu\nu} = \left(\begin{array}{cccc} 0 & 0 & 0 & 0\\
           0 & 0 & 1 & 0\\
           0 & 1 & 0 & 0\\
           0 & 0 & 0 & 0\end{array} \right)  \,. 
           \label{GWpol}
\ee
This is for a plane wave traveling in the $z$ direction.  For a wave traveling in some other direction, we can find the appropriate basis tensors by doing the required Lorentz transformation.  (Recall that pure rotations are a subgroup of the Lorentz group.)
The shape of these tensors has an intuitive meaning, we will later confirm.  Imagine a gravitational wave passing through an array of particles arranged in a circle, in the $x$-$y$ plane.  The shape of the circle will be deformed by the wave, first being squashed in one direction, and then in the other.
For the $+$ polarization, the squashing is along the $x$ and $y$ 
axes.  For the $\times$ polarization, it is along the diagonals.  
To prove this, we need to learn how particles move in a 
gravitational field.

However one case is simple enough to discuss immediately: the motion
of light (photons).  Light has $ds^2=0$, so for light traveling in the
$x$ direction, say, we can immediately write
$dt = \pm dx(1 - h_{xx})^{1/2}$.  The arrival times of pulses along
the $x$ direction will be distorted in a time-dependent way by the 
gravitational wave.  For example, pulsar arrival times are expected to
be extremely regular, and these provide a way for searching for
stochastic GWs, that is superpositions of GWs of different
polarizations and frequencies that are permeating the
Universe.  Long wavelength GWs in particular would tend to distort the
arrival times of pulsars from the same region of the sky in a
correlated manner, providing a distinctive dependence on the angular
separation of two pulsars.

\medskip
\noindent\fbox{%
    \parbox{\textwidth}{%
{\bf Tensors as geometrical objects.}
Writing out all the components of the sparse matrices (\ref{GWpol}) is a cumbersome and inefficient notation.
They are more nicely expressed as outer products of vectors,
\be
	 {\boldsymbol\epsilon_{+}} = \hat x\hat x - \hat y \hat y\,,\quad {\boldsymbol\epsilon_\times} = \hat x \hat y + \hat
y\hat x
\ee
In this notation, $\epsilon_+$ and $\epsilon_\times$ are geometric objects, analogous to vectors, that exist
independently of the coordinate system that determines their components.  A tensor in this sense can be
regarded as an object that is designed to be contracted with other vectors or tensors, to produce a number in
the end.  For example a 3-index tensor can be thought of as a function ${\mathbf T}[\_,\_,\_]$ with three
slots waiting to accept vectors (or a vector and a two-index tensor, or a 3-index tensor) to produce a number such as
\be
	{\mathbf T}[a,b,c] = T^{\alpha\beta\gamma}a_\alpha b_\beta c_\gamma\,.
\ee
This point of view is discussed at length in the textbook \cite{MTW}.

Any tensor can be expressed in a similar way.  For example, the Levi-Civita tensor is
\be
	{\boldsymbol\epsilon} = (\hat t \hat x \hat y \hat z - \hat x \hat t \hat y \hat z) + \dots
\ee
where $\dots$ stands for all the cyclic permutations.  Notice that two vectors do not in general commute in 
an outer product: $\hat t\hat x \neq \hat x\hat t$; the order matters.  In components this is because
\be
	\left({0\atop 0}{1\atop 0}\right) \neq \left({0\atop 1}{0\atop 0}\right)\,.
\ee
}}\\

It should be noticed that the plane wave solution (\ref{gwsol}) does not display the effects of the wave
getting weaker as it spreads away from its source.  A more realistic solution would be a plane wave in
spherical coordinates, whose amplitude decreases with $r$, and whose polarization tensor depends on the
angular direction in order to be transverse along all rays coming from the source.  We will need  to rewrite the field equations in such a coordinate system to properly study this more
realistic case.

\subsection{Problems}
\noindent 3.1.  (a)  Reconsider the metric of problem 2.2 but with the
opposite sign of $h^{xy}$ and $h^{yx}$.  Find a coordinate
transformation that removes the $h^{0i}$ and $h^{i0}$ elements,
and show that $h'^{\mu\nu}$ also satisfies the $h'^{\mu\nu}_{\ \
,\nu} = 0$ and 
Lorenz gauge condition for this gauge choice.  Further show that 
the resulting $h'^{\mu\nu}$ is in the standard form of a gravitational
wave solution.  Hint: the required coordinate
transformation is simpler now than in problem 2.2; 
it does not depend on $z$.\\
(b) Show that a 45$^\circ$ rotation (which is a subset of the Lorentz group) transforms the given
polarization tensor into the other allowed one, by doing the appropriate matrix transformation.\\
(c) Construct a plane wave solution that displays right circular polarization: the polarization rotates
periodically between $\epsilon_+$ and $\epsilon_\times$, with the rotation following the right hand rule with respect to the
wave vector.  (One way to do this is by taking the real
part of complex exponentials, with the appropriate superposition of polarization tensors, like in
electrodynamics.)  To get the relative sign right, suppose the polarization at $z=0$ starts as $\epsilon_+$ when $t=0$,
and find out how $\epsilon_+$ changes under a rotation by a positive angle $\theta$ to see which linear combination
of $\epsilon_+$ and $\epsilon_\times$ this corresponds to at a later time.  Notice that the polarization tensors rotate 
twice as fast as a vector would, under the same rotation angle.  This is a sign that the graviton has spin 2, unlike a photon
that has spin 1.\\

\noindent 3.2.  We would like to show that a spherical wave of the form
\be
	h^{\mu\nu}(t,r,\theta,\phi) = \epsilon^{\mu\nu}(\theta,\phi)\,{f(r)\over r}\,e^{ikt}
\ee
is a solution to the wave equation, with transverse, traceless $\epsilon^{\mu\nu}$, and some $f(r)$ to be
determined.  It is convenient
to keep the vector indices $\mu\nu$ (which will have only nonvanishing
spatial components) in Cartesian coordinates even though we want to express the spatial
positions in spherical coordinates; nothing prevents us from  doing that.  We expect that
$\epsilon^{0\mu} = \epsilon^{\mu 0}=0$, so we can focus on the spatial components $\epsilon^{ij}$.
 The natural generalization of
Eq.\ (\ref{GWpol}) is
\bea
	\epsilon_+^{ij} &=& \hat\theta^i\hat\theta^j - \hat\phi^i\hat\phi^j,\qquad
	\epsilon_\times^{ij} = \hat\theta^i\hat\phi^j + \hat\phi^i\hat\theta^j\,,\nn\\
	\boldsymbol{\epsilon}_+ &=& \hat\theta\hat\theta - \hat\phi\hat\phi, \qquad\quad\quad
	\boldsymbol{\epsilon}_\times = \hat\theta\hat\phi + \hat\phi\hat\theta\,,
	\label{sphPT}
\eea
the first version being in terms of the Cartesian components of the unit vectors $\hat\theta$ and $\hat\phi$.  The spatial Laplacian
appearing in the wave operator $\partial_t^2 -\nabla^2$ is
\be
	\nabla^2  = {1\over r^2}\partial_r\left(r^2\partial_r\right) + {1\over
r^2\sin^2\theta}\partial_\phi^2 + {1\over r^2\sin\theta}\partial_\theta\left(
\sin\theta\partial_\theta\right) \equiv \nabla^2_r + {1\over r^2}\nabla^2_\Omega\,.
\ee 
(a) We want to calculate $\nabla^2_\Omega \boldsymbol{\epsilon}_+$ and $\nabla^2_\Omega \boldsymbol{\epsilon}_+$.
It will turn out that $\boldsymbol{\epsilon}_+$ and $\boldsymbol{\epsilon}_\times$ are eigenfunctions of the angular part of the Laplacian.
As a first step, argue that $\hat\theta =
\partial_\theta\hat r$.  Then using
$\hat r = (\sin\theta\cos\phi,\,\sin\theta\sin\phi,$ $\,\cos\theta)$, obtain the explicit expressions for
$\hat\theta^i$.  Similarly, show that $\hat\phi = (1/\sin\theta)\partial_\phi\hat r$.\\
(b) Because of spherical symmetry, it should be sufficient to evaluate $\nabla^2_\Omega \boldsymbol{\epsilon}_+$ at
some convenient point on the sphere, which we will take to be on the equator, $\theta=\pi/2$ and $\phi=0$.
Show that $\nabla^2_\Omega$ simplifies at this point.
Compute all the first and second derivatives with respect to $\theta$ and $\phi$ of $\hat\theta$ and $\hat\phi$,
except the mixed partial which is not needed, and tabulate their values at the point of interest in terms of the 
Cartesian unit vectors.  For example, $\hat\theta =
-\hat z$, $\hat\phi = \hat y$, $\partial_\phi\hat\phi =-\hat x$.  Then use your results and the chain rule to show that 
$\boldsymbol{\epsilon}_+$ and $\boldsymbol{\epsilon}_\times$ are eigenfunctions of $\nabla^2_\Omega$, and 
determine the eigenvalue. Notice that there is no need for components at this stage of the calculation.
 The tensors (\ref{sphPT}) are examples of generalizations
of spherical harmonics (0-index tensors), known as tensor spherical harmonics.  
Both are eigenfunctions of 
$\nabla^2_\Omega$.  Spherical harmonic tensors exist with any number of indices.\\
(c)  From the wave equation $(\partial_t^2 - \nabla^2) h^{\mu\nu} = 0$ and your previous result, show that the
ordinary differential equation for $f(r)$ is $f'' + k^2 f = 2 f/r^2$.  Show that the general solution can be written
as (an example of spherical Bessel functions)
\be
	f = Ae^{ikr}\left({1\over kr} -i\right) + B e^{-ikr}\left({1\over kr} +i\right)
\label{sph-wave}
\ee
Write down the real part of an outgoing wave solution, that has the property of being well-defined at $t=r=0$.
Give a physical argument for why it makes sense that the amplitude 
decreases as $1/r$ at large $r$.\\

\noindent 3.3.  Simplified version of problem 3.2: Instead of doing parts (a)
and (b), assume that the polarization tensors as given are
eigenfunctions of the angular part of the Laplacian, and determine
what the eigenvalue must be in order to get the o.d.e.\ in part (c).
Then complete part (c). \\

\noindent 3.4.  (a) Consider a gravitational wave (GW) detector consisting
of a laser source at the origin of coordinates, pointing in the $x$
direction, and a detector at $x=L$.  The laser and detector are
suspended by fine wires along the $z$ direction (which is pointing upward from
the Earth), and suppose a GW is approaching along the
$z$ direction, with metric perturbation $h^{\mu\nu} =
\eta\epsilon^{\mu\nu}_+\cos(k(t-z))$.  Assume that the frequency
$\omega_{gw} = k$ (in $c=1$ units) is much smaller than $1/L$, the
inverse of the light-crossing time to the detector, so that the phase of the GW
doesn't change appreciably during the transit of a laser pulse. 
The pulses are emitted at regular intervals $\Delta t$ at times
$t_n = n\Delta t$, and the arrival time of the $n$th pulse is denoted
by $t_{a,n} = t_n + L + \delta t_n$, where the last term represents
the deviation caused by the GW.    
Assuming that the laser and the detector remain fixed at $x=0,L$
respectively, solve for $t_{a,n}$ by integrating the equation of
motion  that arises from $ds^2 = dt^2-(1-h_{xx})dx^2 =
0$ for a light ray moving along the $x$ direction.  
Determine $\delta t_n$, and show that it varies
sinusoidally with $t_n$.\\
(b) A more realistic experiment has two perpendicular arms making an 
L shape, each with equal length $L$, and four mirrors, 
one at each end, and two at right angles to each other 
at the vertex of the L.  These latter are partially-silvered mirrors,
which allow a fraction of light bouncing between the two pairs of
mirrors to be transmitted and detected near the vertex.
A laser of frequency $f_l\gg\omega_{gw}$
simultaneously emits two pulses from the central mirrors along each
arm.  These pulses are initially in phase with each other, but they get out of
phase (as measured at the origin) 
if a GW passes through the experiment, because of the distortion in 
the distances traveled along the arms.
An analyzer
determines the phase difference between the two beams, which
accumulates with the number $N$ of bounces between mirrors.  This
effectively increases the arm lengths by a factor of $2N$ relative to 
the thought experiment of part (a).  Using insight gained from part
(a), calculate the phase difference $\Delta\phi$
that accumulates between
the laser pulses during $N$
bounces, assuming that the phase of the GW remains constant during
the measurement.  Show how by measuring it as a 
function of time,
one can determine the strength and the frequency of the GW.\\
(c).  The maximum $N$ is limited by the beam losses from the partial
transmission at the vertex mirror, and also by the
requirement of the GW phase
not changing too much during a single measurement of the phase
difference, which could wash out the effect to be measured.  (One
must be able to observe the sinusoidal variation of the phase
differences.)
For given values of $N$ and $L$, find the limitation on the observable
GW frequency $\omega_{gw}$ coming from the
requirement that the GW phase changes by no more than 10\% during the
measurement of the dephasing of the pulse.  The LIGO experiment has
arms of length $L = 4\,$km and is sensitive to frequencies up to
$f_{gw}=10\,$kHz.  Assuming this limitation arises from the present
consideration, how many bounces $N$ are the laser beams undergoing?
What distance do the corresponding photons travel before being 
detected?\\
(d) LIGO is able to detect gravitational strains as low as $\eta \sim
10^{-22}$.  What is the corresponding deviation $\Delta L$
in the distance $L$ between the mirrors?  How does it compare to the
size of a proton, $\sim 10^{-15}$\,m?

\section{Geodesic motion}
So far, we have assumed there is some spacetime metric given, but we do not yet know how particles will respond to it.  Imagine that a particle 
follows some path $x^\mu({\lambda})$
in the spacetime, parametrized by $\lambda$, which will be known as an {\it affine} parameter.  
We might impose some boundary conditions such as $x^\mu(0) = x_0^\mu$, and
$x^\mu(1) = x_1^\mu$.  If the particle were living in Minkowski space, there is a unique straight-line path through spacetime that describes its force-free motion,
\be
    x^\mu(\lambda) = x_0^\mu + \lambda(x_1^\mu-x_0^\mu)\,,
	\label{lintraj}
\ee
with a constant velocity.  But if the metric is perturbed, gravity is present and it will bend the path.  What is the correct trajectory in that case?

In relativity, the line element $ds^2 = g_{\mu\nu}dx^\mu dx^\nu$ encapsulates the basic meaning of the 
metric tensor.  Consider the invariant squared distance traveled by the particle during some interval $d\lambda$ along the path:
\be
    ds^2 = g_{\mu\nu}{dx^\mu\over d\lambda}d\lambda\, {dx^\nu\over d\lambda}d\lambda \implies
    |ds| = \left(g_{\mu\nu}\dot x^\mu\dot x^\nu\right)^{1/2} d\lambda 
\ee
where I defined $\dot x^\mu = dx^\mu/d\lambda$.   We can integrate this to find the total invariant distance along the path,
\be
    |s| = \int_{\lambda_1}^{\lambda_2} d\lambda\, \left(g_{\mu\nu}\dot x^\mu\dot x^\nu\right)^{1/2}
	\equiv \int_{\lambda_1}^{\lambda_2} d\lambda\, \sqrt{I}\,.
\ee
Apart from the square root and the nontrivial metric tensor, this expression is reminiscent of the action in classical mechanics, for a particle with only kinetic energy.  In fact, we could choose $\lambda = t$
so that $\dot x^i$ is the normal velocity, $\dot x^0=1$, and consider Minkowski space where $g_{\mu\nu} = \eta_{\mu\nu}$.  Taylor expanding to first order in the small velocity squared, $|s|$ is proportional to the action for a free particle, 
up to an irrelevant constant.  In fact, $m|s|$ is precisely the relativistic form of the action for a massive particle.
This strongly suggests that we must find the path that makes $|s|$ stationary in the general case.  That is,
we should vary $x^\mu\to x^\mu + \delta x^\mu$ with the endpoints fixed and set the variation to zero, to find the equation of motion for the particle.

The variation of $|s|$ is given by
\bea
    \delta|s| &=& \int_{\lambda_1}^{\lambda_2} d\lambda\, 
    {1\over 2\sqrt{I}} \left({\partial g_{\mu\nu}\over \partial x^\alpha}\dot x^\mu\dot x^\nu \delta x^\alpha(\lambda)  + 
    g_{\alpha\nu}{d\delta x^\alpha\over d\lambda}\dot x^\nu +g_{\mu\alpha} \dot x^\mu {d\delta x^\alpha\over d\lambda}\right)
\eea
In the next step, we want to integrate by parts to turn ${d\delta x^\alpha/d\lambda}$ into $-\delta x^\alpha d/d\lambda$, but the problem is
that the factor of $1/\sqrt{I}$ will generate a complicated mess when it gets differentiated.  There is a special choice of parametrization
that avoids this complication \cite{wiki-geodesic}:
change variables such that  $d\lambda = d\tau/\sqrt{I}$.  Then the factor of $I$
will be absorbed by the change of variables and we can integrate by parts
to get
\bea
  \delta|s|   &=& \int_{\tau_1}^{\tau_2} d\tau\, 
     \left({\partial g_{\mu\nu}\over \partial x^\alpha}\dot x^\mu\dot x^\nu  - 
    {d\over d\tau} (g_{\alpha\nu}\dot x^\nu)-{d\over d\tau}(g_{\mu\alpha} \dot x^\mu) \right)\delta x^\alpha(\tau) = 0\,,
     \label{variation}
\eea  
where now $\dot x^\mu = dx^\mu/d\tau$.
This should vanish for any choice of $\delta x^\alpha(\tau)$ by stationarity, so the term in large parentheses must vanish.  
Carrying out the $d/d\tau$ derivatives using the chain rule, it is straightforward to arrive at the {\it geodesic equation} of motion,
\be
    \ddot x^\mu + \Gamma^\mu_{\alpha\beta}\dot x^\alpha \dot x^\beta = 0,\qquad 
\Gamma^\mu_{\alpha\beta} = 
\sfrac12 g^{\mu\nu}\left(- g_{\alpha\beta,\nu} + g_{\nu\alpha,\beta}+g_{\beta\nu,\alpha}\right)\,.
    \label{geodesic}
\ee
Notice that both the metric and its inverse appears in $\Gamma^\mu_{\alpha\beta}$,
which is known as the  Christoffel symbol.   It is obviously symmetric under
interchange of $\alpha$ and $\beta$.   It is important to know that, despite
appearances, $\Gamma^\mu_{\alpha\beta}$ is not a tensor.  Under general coordinate
transformations $x^\mu\to x^\mu(y)$, it does {\it not} obey
\be
	\Gamma^\mu_{\alpha\beta} \to \Lambda^\mu_\nu \Lambda^\gamma_\alpha
\Lambda^\delta_\beta\, \Gamma^\nu_{\gamma\delta} 
\ee
where $\Lambda^\mu_\nu = \partial x^\mu/\partial y^\nu$.  Instead, extra terms get
generated, involving derivatives of the $\Lambda$ matrices.  This will be useful to us
in a later chapter.

I emphasize that we had to make a special choice of parametrization of
the path, such that $I\to 1$, in order to get the simple form of the
geodesic equation.  Suppose we transform back to some arbitrary
parameter $\tau \to \lambda$. The left and right sides of Eq.\
(\ref{geodesic}) do not in general change by the same factor
$(d\lambda/d\tau)^2$, because of the second derivative, that
introduces an extra term involving $d^2\lambda/d\tau^2$.  The geodesic
equation only takes its simple form for this special choice of affine
parameter.\footnote{Or a constant multiple of it.  This kind of
rescaling allows us to use the geodesic equation also for lightlike
trajectories (massless particles), since we can rescale $\tau$ by
$\gamma$ and take the limit $\gamma\to\infty$ after $\gamma$ has
dropped out from the equations.}
 This underscores the fact that $\tau$ has a special
significance in the geodesic equation: it is chosen so that
$g_{\alpha\beta}\dot x^\alpha \dot x^\beta = 1$.  The 4-velocities are
normalized to be unit vectors.  This means that $\tau$ measures the
{\it proper time} experienced by a particle moving along the
geodesic---the time measured on a clock that is freely falling  along
the worldline.

Now we can return to the problem of how a gravitational wave would affect free particles through which it is moving.  Since it is a weak field, we can work to linear order in $h^{\mu\nu}$,
hence replace $g^{\mu\nu}$ by $\eta^{\mu\nu}$ in the Christoffel symbol, since the derivatives of the metric are already linear in $h^{\mu\nu}$.  We can again set $\tau = t$ and use $\dot x^0=1$, $\dot x^i = v^i$.  
Since the velocities are small, the leading terms in the geodesic equation will be
\be
    -\ddot x^i = \Gamma^i_{00} + \left(\Gamma^i_{0j} + \Gamma^i_{j0}\right) v^j = 2\Gamma^i_{0j} v^j 
    = \dot h_{ij} \dot x^j \,.
    \label{wfgeo}
\ee
This result doesn't seem to match the claim of the last section about how particles arranged along a circle would get deformed by the gravitational
wave.  If the particles started at rest, then Eq.\ (\ref{wfgeo}) predicts that they will remain at rest.  But notice, they are only at rest {\it
within the chosen coordinate system}.  The distance between two particles will change with time, because space itself is oscillating!  The effect
becomes easy to understand once we realize this.

To measure the effect, one could use mirrors that are suspended in such a way that they could freely move along the $x$ direction, say.
A gravitational wave along the $z$ axis would cause the distance between two mirrors alont the $x$ axis to change with time.  Photons
being reflected between the mirrors would experience oscillating travel times between the mirrors as the gravitational wave passed.  By counting
the number of reflections per unit time, one would find that the distance between the mirrors is oscillating with a frequency given by
that of the gravitational wave.  In principle, one could also measure the distances using a ruler, since 
 a rigid object like a ruler will not get stretched and compressed by these undulations of space: gravity is too weak
to do that in the weak-field limit we are working in.  But in practice one needs to measure displacements much smaller than the size 
of an atom, hence more sophisticated methods involving reflection and inteference of light beams are needed.

Suppose the two mirrors are placed at $x = \pm L/2$ in the lab frame, before the gravitation wave (GW) passes.  The physical distance between 
them is given by the spatial part of the metric,
\be
	D = \int_{L/2}^{L/2} dx \sqrt{|g_{xx}|} \cong \int_{L/2}^{L/2} dx \left(1 - \sfrac12 h_{xx}\right)
\ee
Supposing the polarization is $\epsilon_+$ and the GW is traveling in the $z$ direction, then $h_{xx} = A\cos(k(z-t))$, where $A$ is some
(very small) amplitude related to the source.  The separation $D$ thus oscillates with amplitude $AL/2$.  Since the strain $A$ from astrophysical
sources is extremely small, it is advantageous to make the baseline $L$ as large as possible, to make the displacement measurable.
The LIGO observatories use baselines of 4\,km and are sensitive to displacements of $10^{-18}$\,m, much smaller than the size of a nucleon.
The GW must be oscillating more slowly than the light travel time between the mirrors in order for the signal to not get washed out; LIGO is therefore sensitive to
frequencies $\omega \lesssim c/L \sim 100$\,kHz.  This includes the interesting frequency range of GWs produced by the merger of inspiralling
binary systems consisting of neutron stars or black holes.  LIGO (USA) operates in collaboration similar observatories Virgo (Italy) and KAGRA (Japan)
to combine data and thereby increase sensitivity.  Nearly 100 mergers have been observed so far \cite{Virgo}.

More recently, pulsar timing arrays have found strong evidence for GWs at much lower frequencies, using the fact that a GW causes undulations
in the arrival times of the pulses, which otherwise would be extremely regular, like those of an atomic clock \cite{NANOGrav:2023gor}.  In this case, the baseline is the distance
between pulsars, leading to much smaller frequencies or order 1/y.  These GWs are believed to come from mergers of supermassive black holes at the
centers of galaxies, which are orbiting each other much more slowly than the $\sim 50$ solar-mass mergers seen by LIGO.

\medskip
\noindent\fbox{%
    \parbox{\textwidth}{%
{\bf Spatial geodesic curves.}
The same equations we have derived above can be used in any number of dimensions, including spatial manifolds where time is absent.
Then the affine parameter no longer has the interpretation of proper time; rather it is the proper (physical) distance traveled along the path.
The geodesic equation thus determines the trajectory between two points on the manifold that has the minimum distance. Note: if we have found a
good affine parameter that makes $ds^2 = d\lambda^2$, it can always be rescaled by a constant without spoiling the derivation of the 
geodesic equations.
}}

\medskip
\noindent\fbox{%
    \parbox{\textwidth}{%
{\bf A nice trick for computing Christoffel symbols.}  It can be tedious to compute Christoffel
symbols from the general formula (\ref{geodesic}).  Typically one spends a lot of time just
trying to figure out which ones are zero and which are nontrivial.  It can often be more
efficient to start from the explicit expression for the distance between two points in a given
metric.  As an example, take the unit sphere with $ds^2 = d\theta^2 + \sin^2\!\theta\, d\phi^2$.
If $\lambda$ parametrizes the distance along a curve $[\theta(\lambda), \phi(\lambda)]$, then
one can write the distance between two points as
\be
	s = \int d\lambda\sqrt{\dot\theta^2 + \sin^2\!\theta\,\dot\phi^2}\,,
\ee
where the dot denotes $d/d\lambda$.  By demanding that $s$ is stationary under variations of
$\theta$ and $\phi$, we get the two geodesic equations
\bea
	{\delta s\over \delta\theta} &=& -\ddot\theta + \sin\theta\cos\theta\,\dot\phi^2 = 0\nn\\
	{\delta s\over \delta\phi} &=&-{d\over d\lambda}(\sin^2\!\theta\,\dot\phi) = 0\,.
	\label{two-sphere}
\eea
The choice of $\lambda$ is such that $\sqrt{I} = \sqrt{g_{ab}\dot x^a\dot x^b}=1$, so we avoid having to
include the factor $1/\sqrt{I}$ in the integration by parts, like in our original derivation of
the geodesic equation.  One easily sees that Eqs.\ (\ref{two-sphere}) are in the form $-\ddot x^a =
\Gamma^a_{\ b c}\dot x^b\dot x^c$, so one can immediately read off the nonzero Christoffel
symbols, and not have to determine which ones vanish.
}}\\

\subsection{Problems}

\noindent 4.1.  (a) Suppose a particle follows the trajectory $t = \lambda$, $x = A\sqrt{\lambda}$ in Minkowski space.  (This is not a solution
to its force-free equation of motion; rather it is a possible trajectory in the variational problem before finding the stationary
solution.)  Find the proper time parameter $\tau$ as a function of $\lambda$ that makes $\eta_{\mu\nu}(dx^\mu/d\tau)(dx^\nu/d\tau) = 1$.\\
% answer: $d\tau = d\lambda X$ where $X = \sqrt{1- A/(4\lambda)}$, integrates to $X - (A/8)\ln((1+X)/(1-X))$.\\
(b) Suppose the particle follows the trajectory $t = \gamma \lambda$, $x = \gamma v \lambda$ in Minkowski space, where $\gamma$ is the usual
Lorentz factor.  What is the relation between $\tau$ and $\lambda$?\\
% answer: they are the same
(c) Consider the same trajectory as in (b), but now the metric is perturbed by the gravitational wave $\boldsymbol{h} = 
A(\hat x\hat x - \hat y\hat y)\cos(k(z-t))$.  Find $\tau$ as a function of $\lambda$ to leading order in $\boldsymbol{h}$.  You can treat
$z$ as being fixed.\\
% answer: $\tau = \lambda + (v^2\gamma A/2k)\sin(k(z-\gamma\lambda))$.

\noindent 4.2. (a) Write the general expression for the Christoffel symbol $\Gamma^\mu_{\ \alpha\beta}$ in terms of a metric perturbation
$h_{\mu\nu}$, to leading order in the perturbation around Minkowski space.\\
\noindent (b) For the GW in problem 4.1(c), which components of $\Gamma^\mu_{\ \alpha\beta}$ will be nonzero?\\
% answer: must have one component being $0$ or $z$, and two being $xx$ or $yy$.
(c) Compute the components of $\Gamma^\mu_{\ \alpha\beta}$ that will be relevant for a particle traveling in the $x$ direction,
with the same GW as in 4.1(c), and write the ensuing geodesic equations for $\ddot x$ and $\ddot t$.\\
% answer: $\ddot x = \dot h_{xx}\dot x\dot t$, $\ddot t = \sfrac12 \dot h_{xx}\dot t^2$.  Hence
% $\Gamma^x_{xt} = -\sfrac12\dot h_{xx} = -\Gamma^0_{xx}$. 
(d) We would like to solve the equations in the case of light, where 
$\dot t \cong \dot x\cong 1$ up to $O(h)$ corrections.  
Integrate the equations twice to find the trajectory for light, assuming it starts at the 
origin at $t=0$.\\
% answer: $x = \tau + A/k \sin k\tau$, $t = \tau + A/2k \sin k\tau$.
(e) Light rays are supposed to obey $ds^2 = 0$ along their path.  Verify that your trajectory satisfies this, including the $O(h)$ corrections.\\
% answer: $ds^2/d\tau^2 = (1 + h/2)^2 - (1-h)(1+h)^2 = 0$.

\noindent 4.3. Consider a two-dimensional hyperbolic space with line element $ds^2 = R^2(d\theta^2 + \sinh^2\theta\, d\phi^2)$, where $\phi$ is
periodic in the interval $[0,2\pi]$ and $\theta \in (-\infty,\infty)$.\\
(a) Qualitatively sketch a two-dimensional surface embedded in three dimensions that might have such a metric.  Hint: the circles of constant
$\theta$ have radii given by $R\sinh\theta$.\\
(b) Compute the nonvanishing Christoffel symbols.\\
(c) Write the geodesic equations.\\
(d) Suppose we want to find the geodesic that starts at $(\theta,\phi) = (5,0)$ and ends at $(1,\pi)$.  We have to guess an initial direction
for the trajectory, 
$\dot\theta/\dot\phi$, and vary it until the geodesic path (found by integrating the equations) passes through the desired endpoint.  If the space was flat,
we would guess the straight line direction $d\theta/d\phi = \dot\theta/\dot\phi = (1-5)/(\pi-0)$.  Prove that the actual solution requires $\dot\theta/\dot\phi$
to be more negative than this guess.  Illustrate what is going on with a qualitative plot of the expected shape of the geodesic on the 
$\phi$-$\theta$ plane.

\section{Static fields}
\label{sect:static}

Let's now solve the field equations in the case where there is a gravitational source,
$T^{\mu\nu}\neq 0$.  Consider the Earth, approximated as a sphere of constant density $\rho$.  Recall that we are using $c=1$ units, which means that $\rho$ could stand for mass density or energy density; we need not make the distinction in these units.  Further recall the statement that $T^{\mu\nu}= {\rm diag}(\rho,p,p,p)$ for such a system.  The pressure is negligible compared to the density
near the surface of the Earth
(see problem 5.1), so only $\bar h^{00}$ has a nonvanishing source.  Since there is no time-dependence, the field equation becomes
\be
    +\nabla^2 \bar h^{00} = 2\kappa\rho = 16\pi G\rho(r)
\ee
where $\rho(r)$ is taken to be constant,
$\rho_e$, within the Earth and vanishes beyond its radius $r_e$.   This is exactly the same as the equation for the Newtonian gravitational 
potential $\phi$ if we make the identification $\bar h^{00} = 4\phi$.  Hence we know the solution,
\be
    \bar h^{00} = -4\left\{\begin{array}{ll}
    (G\,M_e/r_e)(3 r_e^2/2- r^2/2),& r < r_e\\
    G\,M_e/r,&  r \ge r_e\end{array}\right.\,,
	\label{Earthpot}
\ee
where $M_e$ is the Earth mass.  From the definition of $\bar h^{\mu\nu}$ it is straightforward to show that $h^{\mu\nu} = h_{\mu\nu} = 2\phi\,{\rm diag}(1,1,1,1)$ (Problem 5.2).

Now we can verify that the geodesic equation 
(\ref{geodesic}) correctly predicts the motion of a test particle in this field.  Unlike for 
the gravitational wave metric perturbation, 
$\Gamma^i_{00}$ has a nonvanishing contribution, so the acceleration  
has a leading term of the form $\ddot x^i \cong -\Gamma^i_{00}$, since $\dot x^0\cong 1$ 
for nonrelativistic motion.  (Recall that $dx^\mu/d\tau = \gamma(1, \vec v).$)
From the definition of the Christoffel symbol
we find $\Gamma^i_{00} = -\sfrac12 h_{00,i}=
-\phi_{,i}$.  In particular, we could choose spherical coordinates and $\Gamma^r_{00} =
-\phi_{,r} = GM_e/r^2$.  This of course gives the expected result for the acceleration of a test particle toward the Earth.

So far we have only reproduced Newtonian gravity.  But now we have the  means
to discover how relativistic particles, including photons, get deflected in a
gravitational field.  In this case, the velocities are no longer negligible. 
Let's consider a relativistic particle initially moving along the 
line $x = t$, $y=b$, $z=0$
for $t \to -\infty$, with mass $M$ at the origin. We expect the particle to be
deflected in the $x$-$y$ plane, hence the equations for $dt/d\tau$, $dx/d\tau$
and $dy/d\tau$ will be nontrivial.  For a relativistic particle it is no
longer true that $dt/d\tau\gg dx/d\tau$; we cannot ignore $\Gamma^i_{\mu
x}\dot x^\mu \dot x$ or  $\Gamma^i_{x x}\dot x^2$.  However if the deflection
is small, then $\dot y\ll \dot x$ and at least terms involving $\Gamma^i_{\mu
y}$ will be subdominant.  For the deflection, we are most interested in the
evolution of $\dot y$,  hence focus on 
\bea
    -\ddot y &=& \Gamma^y_{\mu\nu}\dot x^\mu \dot x^\nu\nn\\ &=&  
    \Gamma^y_{00}\dot t^2 + 2  \Gamma^y_{0x}\dot t \dot x +  
\Gamma^y_{xx}\dot x^2 +O(\dot y)  \nn\\
 &\cong& 
    \gamma^2\left(\Gamma^y_{00}+   2\Gamma^y_{0x} +  \Gamma^y_{xx}\right) =
2\gamma^2{GM y\over r^3}\,,
    \label{light-deflect}
\eea
where we used $\dot t \cong \dot x\cong \gamma$ for 
a particle moving at the speed of light, and the final result you will 
derive in problem 5.3.
  The radial distance is $r^2 \cong t^2 + y^2$ in the denominator.

To deal with the factor of $\gamma^2$, let's put it on the other side of the 
equation and consider
\be
	\gamma^{-2}\ddot y = \left({d\tau\over dt}\right)^2 {d^2y\over
d\tau^2} =  {d^2y\over dt^2} + {d\ln\gamma\over dt }{dy\over dt}\,.
	\label{d2yconv}
\ee
In a weak field, $\gamma$ is nearly constant, up to corrections of order
the small Newtonian potential, so the second term is quadratic in the
perturbation and we can ignore it.  Then Eq.\ (\ref{light-deflect})
becomes
\be
	{d^2y\over dt^2} = -2{GM y\over r^3}\,.
\label{light-deflect2}
\ee
This will be true for any highly relativistic particle, regardless of whether
it is exactly massless, hence including the photon.  
Notice the factor of 2!  In Newtonian gravity, if one gives the photon
a small fictitious mass $m$, which cancels out of both sides of the $F =
ma$ equation, the predicted acceleration is half as large.  Hence GR predicts
twice as large a deflection angle for light.

We don't know how to solve Eq.\ (\ref{light-deflect2}) analytically, but we can estimate the angle of deflection, $\theta\sim \dot y/\dot x\cong \dot y \cong \ddot y \Delta t$, where $\Delta t \sim b$ is the time during which $\ddot y$ is relatively large.   This gives $\theta \sim GM/b$, independently of the energy of the photon.  
In the limit $b\to 0$ the angle would diverge, but we must remember that for any realistic system, $b$ can be no smaller than the radius of the 
massive object causing the deflection. More precisely, for smaller $b$, we have to use the interior solution for $\phi$ in Eq.\ (\ref{Earthpot}), which 
does not diverge as $r\to 0$.

\medskip
\noindent\fbox{%
    \parbox{\textwidth}{%
{\bf Natural units.}
As you know, you should always check your answer to make sure the dimensions are correct before proceeding any further.  
We have already set $c=1$, but to further simplify dimensional analysis, it is convenient to also take $\hbar =1$, called ``natural units.''  
With this choice, there is only one kind of dimensionful quantity, which can be considered as mass or energy.  
Since $\hbar c = 197.3$\,MeV\,fm has units of energy times length (fm = femtometer $= 10^{-15}$\,m), it means that distance or time has the same dimensions as inverse mass (or energy).  Newton's constant has units of inverse mass squared,
\be
    G = {1\over M_p^2}
\label{PlanckMass}
\ee
where $M_p = 1.22\times 10^{19}$\,GeV is the Planck mass.  Velocity is dimensionless, acceleration has units of mass.  
In this system, we can immediately see that the dimensions in Eq.\ (\ref{light-deflect2}) are consistent.

We could alternatively express $G$ as a distance squared, $G = \ell_p^2$, using $(\hbar c)^2$ to convert.
This gives $\ell_p = 197.3{\,\rm GeV\,fm}/M_p = 1.61\times 10^{-32}\,$m, a remarkably short distance.
It is the length scale at which quantum mechanics should start to
 modify the behavior of gravity.

\qquad In natural units, the wave 4-vector $k^\mu = (\omega, \vec k)$ that
appears in plane wave solutions via expressions like $\cos k\cdot x$ has
a nice interpretation in terms of the associated massless particle,
the photon or the graviton.  It is the same as the momentum 4-vector
of the particle,
since $E = \hbar \omega$ and $\vec p = \hbar \vec k$.
}}
\medskip

\subsection{Conservation of energy}
What about the zeroth component of the geodesic equation,
\be
\label{dotx0eq}
	-\ddot x^0 = \Gamma^0_{\ \alpha\beta}\dot x^\alpha\dot x^\beta\,;
\ee
what is its signifcance in the present context?  If the metric perturbation is
time-independent, as we are assuming, then $\Gamma^0_{\ 00}=0$, and we have
\be
	-\ddot x^0 = - \dot\gamma = 2\Gamma^0_{\ 0i}\gamma \dot x^i + \Gamma^0_{\ ij}
\dot x^i\dot x^j
\label{coe1}
\ee
One finds that $\Gamma^0_{\ 0i}=\phi_{,i}$ and  $\Gamma^0_{\ ij}=0$ for the potential around 
a point mass.  Then Eq.\ (\ref{coe1}) implies
\be
	-{d\over dt}\ln\gamma \ {\stackrel ?=}\  2\phi_{,i}v^i = 2{d\over dt}\phi\,,
\label{coe2}
\ee
where we used the chain rule in the last step.  For a massive particle, the kinetic
plus mass energy is given by $E = \gamma m$, and since $m$ is constant, Eq.\ (\ref{coe2}) says that
$d(\ln E+2\phi)/dt = 0$, which looks like a peculiar form of energy conservation. 
Consider the nonrelativistic limit, where $\ln(E) \cong \ln(m(1+ v^2/2)) \cong
\ln m + \sfrac12 v^2$.  We can ignore $\ln m$ since its derivative vanishes.
Eq.\ (\ref{coe2}) implies that $\sfrac12 m v^2 + 2m\phi$ is constant,
which is not what we expect: kinetic plus gravitational potential energy must be conserved,
but we seem to be getting twice the gravitational potential.  What is going wrong?

Recall that in the derivation of the geodesic equation, we set 
\be
	g_{\mu\nu}\dot x^\mu\dot x^\nu=1\,;
\label{proptimedef}
\ee
this is the definition of proper time in the nontrivial metric.  Multiply both
sides by $m^2$ and define $P^\mu = m \dot x^\mu$.  Using the explicit form of the perturbed
metric, Eq.\ (\ref{proptimedef}) becomes
\be
	(1+2\phi) (P^0)^2 - (1-2\phi) \vec P^2 = m^2\,.
\ee
For simplicity, consider a particle at rest, $\vec P=0$.  Then
\be
	P^0 = {m\over \sqrt{1+2\phi}}\cong m(1 - \phi)
\ee
Hence what we mistakenly thought was $\gamma$ in Eq.\ (\ref{coe2}) was really 
$P^0/m=\gamma(1 - \phi)$, and
correcting for this gives the expected result
\be
	-{d\over dt}\ln\gamma = \phi_{,i}v^i = {d\over dt}\phi\,.
\label{coe3}
\ee
It might seem strange that the zeroth component of the 4-momentum gives the mass plus kinetic energy
{\it minus} the gravitational potential energy, but at least this leads to a result consistent
with our usual understanding of energy conservation.

On further reflection, perhaps it is not so strange after all.  Consider the covariant
component, $P_0 = g_{00}P^0 = m\gamma(1+\phi)$.  It meets our expectations of adding kinetic
and potential energy.  In Minkowski space we never needed to worry about the distinction, but
here we do.   Is there a reason that $P_0$ should be the more natural choice for the energy
than $P^0$?  In quantum mechanics, the momentum operator is $-i\hbar {\partial\over \partial x^\mu}$.  Therefore it
may be natural that the 4-momentum should have a lower index. The wave function of a particle
depends upon $P_\mu x^\mu$, and its group velocity will be $-dP_i/dP_0$. 

One might at first be surprised that the potential energy of a relativistic particle is not just
$m\phi$ as in Newtonian gravity, but $m\gamma\phi$.  This is another difference between GR
and Newtonian gravity.  Its interpretation is that not only mass, but also kinetic energy
feels the effect of gravity.  This makes sense when we realize that photons get deflected
by the gravitational field, and they are pure energy, no mass. 

Notice that the distinction between $\gamma$ and $\gamma(1-\phi)$ is not important for our
interpretation of the spatial components of the geodesic equation.   This is because the
correction would only appear on the right-hand side of the equation, where it is multiplied by
a Christoffel symbol that is already of order $\phi_{,i}$.  Hence the correction is higher
order in the perturbation.

For a highly relativistic particle, like a photon, we can integrate Eq.\ (\ref{coe3})
to find that the energy varies in the gravitational field according to 
\be
	E = E_0 e^{-\phi}\cong E_0(1-\phi)\,,
\ee
where we  interpret $E_0$ as the unperturbed energy in the absence of $\phi$, and $E$
is the energy after the influence of $\phi$.  Since $\phi$ is negative, the
energy is increasing: the photon gains energy if it falls into the gravitational potential
well (or loses it if it is climbing out of the potential).  Of course the velocity is not
changing in this case; instead the frequency and wavelength are changing.  This argument is
valid for arbitrarily small values of the particle mass, compared to its energy, so there is
nothing to keep us from taking the limit as $m\to 0$ and applying it to the photon.   	

It should be emphasized that in practice, there is no need to solve
for $\dot t$ starting from its geodesic equation (\ref{dotx0eq}). 
From the definition (\ref{proptimedef}), we have a first integral of
the motion.  For the weakly perturbed metric, one can solve for $\dot
t$,
\bea
	\dot t &=& \sqrt{(1+2\phi) - (1-2\phi)\vec v^{\,2}}\nn\\
	&\cong& 1 + \phi - \sfrac12(1-2\phi)\vec v^{\,2}\,.
\eea 
Unless the object is relativistic, one can typically ignore the
correction of order $\phi v^2$.

\subsection{Gravitational time dilation}
The $r$-dependence of $h^{00}$ implies that time will run at different speeds
depending on distance from the gravitational source.   Of course, clocks at different
altitudes tick at the same speed according to observers who are next to the
clocks.  But someone timing a distant clock will perceive it as running at a different
speed, similarly to the case in special relativity where the clock is moving with
respect to the observer.  Consider the line element in the $t$-$r$ slice of the 4D
spacetime:
\bea
	ds^2 &=& (1+h_{00})(dx^0)^2 - (1 - h_{rr})(dx^r)^2\nn\\
	 &=& (1 + 2\phi)(dx^0)^2 -(1 - 2\phi)(dx^r)^2 \,.
\eea
The physical time interval at a distance $r$ is 
\be
	d\tau = (1+h_{00})^{1/2}dt = (1+2\phi)^{1/2}dt = \left(1-{2MG\over r}\right)^{1/2}dt\,.
\ee 
Suppose that an observer at $r=\infty$ counts ticks of his own clock for some interval $\Delta t$, as well
as those from a clock at radius $r$.  The time interval $\delta\tau$ elapsed on the clock at $r$ is 
is smaller by a factor $(1-MG/r)$; it is running more slowly, and registers fewer ticks.  The observer at
$r$ has aged less than the one at $\infty$ during this interval.
Near the Earth, this is a small deviation, but much more dramatic results will be found in nonlinear gravity,
when we study black holes.  There, time comes to a standstill at the horizon.

Suppose we would like to know the time elapsed on a clock that moves through a weak 
gravitational field with some trajectory $\vec x(t)$.  It is given by
\be
	\tau = \int dt \sqrt{(1 + 2\phi) - (1-2\phi)(d\vec x/dt)^2}\,.
\ee
If the clock is moving slowly, with $v^2 \lesssim \phi$, we can Taylor expand to get
\be
	\tau \cong \int dt \left(1 +\phi -\sfrac12 v^2\right)\,,
\ee
revealing that a weak gravitational field has a similar effect to the special relativistic
time dilation of a moving clock.  Since $\phi <0$, both act to make the clock tick more slowly.
It is interesting to recall that $-m\tau$ is equal to the {\it action} associated with the
path, where $m$ is the particle mass, if we count mass energy as part of the potential
energy.  We used this fact to justify that the geodesic equations correspond to the equations
of motion for the particle in the previous chapter.  Now we see that it holds also including the
gravitational potential energy.

Analogously to time dilation, spatial distances are contracted relative to each other, depending upon the
altitude.  Once again, the sign of the effect goes in the same direction as the special
relativistic Lorentz contraction.  
However this effect is more difficult to observe experimentally than time dilation.  With
atomic clocks it is possible to measure extremely small deviations in clock rates.  On
the other hand, the distortion of space is a quite weak effect that is resisted by 
rigid objects like rulers.  Gravity is so much weaker than the electromagnetic force 
that one would have to design a clever experiment to detect the 
effect.  See problem 5.6.

\subsection{Orbital motion and perihelion precession}
We would like to check that circular orbits exist as expected, using the geodesic equations.  It is convenient to go to polar coordinates
\be
	ds^2 = (1+2\phi)dt^2 - (1-2\phi)(dr^2 + r^2 d\theta^2 + dz^2) 
\label{lin-metric}
\ee
and set $z=0$ for orbits in that plane.  Then we can continue to write $\phi = \phi(r)$ since its $z$ dependence will play no role.
In problem 5.5 you will compute the Christoffel symbols
\bea
\label{Xseq}
	\Gamma^0_{r0} &=& \Gamma^0_{0r} = \phi_{,r}\,,\nn\\
	\Gamma^r_{00} &=& \phi_{,r}\,,\quad\Gamma^r_{\ rr} = -\phi_{,r}\,,\quad \Gamma^r_{\ \theta\theta} = r(r\phi_{,r} -1)\cong -r\,.\\
	\Gamma^\theta_{r\theta} &=& \Gamma^\theta_{\theta r} = -\phi_{,r} + {1\over r} \cong {1\over r}\,.\nn
\eea
The geodesic equation for $r$ is
\be
	-\ddot r = \Gamma^r_{00}\dot t^2 + \Gamma^r_{\ rr}\dot r^2 + \Gamma^r_{\ \theta\theta}\dot\theta^2\,.
\ee
For a nonrelativistic circular orbit, we set $\dot r=\ddot r = 0$ and $\dot t = 1$, to find $v^2 = (r\dot\theta)^2 = GM/r$, which is 
the familiar result.  A more interesting case is elliptical orbits.  In Newtonian gravity, they close, but in GR they do not.  Instead, the
point of closest approach (perihelion) precesses with each orbit.  This effect was first verified for the orbit of Mercury as one of the
early tests of GR.  We need in addition the equations for $\theta$ and $t$,
\be
	\ddot\theta = -2\Gamma^\theta_{r\theta} \dot r \dot\theta\,,\quad \ddot t =  - 2\Gamma^0_{\ r0}\dot r\dot t\,.
\ee
Suppose the ellipticity is small so that we can linearize in the deviation from circularity, $r = r_0 + \delta r$, and $\theta = \dot\theta_0 \tau
+ \delta\theta$, where $\dot\theta_0 = \sqrt{GM/r_0}$.  It will turn out that $\delta\dot t\sim \phi$ and for the present argument we can ignore it
in $t = \tau + \delta t$, hence take 
$t=\tau$.  One subtlety is that the correction $\phi_{,r} = \phi_{,r}|_{r=r_0} + \phi_{,rr}\delta r = \phi_{,r}(1 - 2\delta r/r_0)$ is relevant in the $\Gamma^r_{\ 00}$ contribution to $\ddot r$,
since $\dot t^2 = 1$.  On the other hand, we can ignore the $\Gamma^r_{\ rr}$ contribution since $\dot r^2 = (\delta\dot r)^2$ which is higher
order in the deviation from circularity.  When we keep all the terms that are linear in the deviation, and leading in $\phi$, we get
\bea
	-\delta\ddot r &\cong&  \phi_{,r}(1 - 2\delta r/r_0) -(r_0+\delta r)(\dot\theta_0^2 + 2\dot\theta_0\delta\dot\theta)\nn\\
	-\delta\ddot\theta &\cong& {2\over r_0}\delta\dot r\, \dot\theta_0
\eea
The second of these can be integrated to give $\delta\dot\theta = (2/r_0)\dot\theta_0 \delta r$, and substituted back into the first
to get an equation for $\delta r$ alone.  When we remove the zeroth order terms that cancel for the circular orbit, it becomes
\be
	-\delta\ddot r = \left[-2\phi_{,r}/r_0 - \dot\theta_0^2  + 4\dot\theta_0^2\right]\delta r = \dot\theta_0^2 \delta r\,,
	\label{ddotreq1}
\ee
which has solution $\delta r = \epsilon\cos(\dot\theta_0 t)$, for example.  Then one can integrate the equation for $\theta$ to find
$\theta = \dot\theta_0 t - 2(\epsilon/r_0)\sin\dot\theta_0 t$.  This orbit closes after one period, without any precession of the perihelion.  

However we neglected corrections of higher order in $\phi$.    Taking account of them, the period of $\delta r(t)$ gets shifted by a fractional amount of order
$\phi$.  It requires more work to keep track of all the corrections at this order.\footnote{
\label{Xsfn} Namely, the $\Gamma^0_{\mu\nu}$ symbols should be 
divided by $1+2\phi$, and all the $\phi_{,r}$ factors in the  other Christoffel symbols get divided by  $1-2\phi$.}  But even without computing
them exactly, we can estimate the expected magnitude of the perihelion shift.  For Mercury, 
$|\phi| = 2.55\times 10^{-8}$ (taking the semi-major axis for the orbital radius), and the number of orbits in a century
is 415.2, giving an accumulated precession angle $\Delta\theta$ of order $415.2|\phi|\times 360^\circ \sim 13.7$ arcseconds per century.  
The measured number is approximately 3 times bigger.  The exact prediction from GR (see Eq.\ (7.128) of 
\cite{Ohanian:1995uu}) is
\be
	{\Delta\theta\over 2\pi} = 3{GM\over a(1-\epsilon^2)}
\ee
where $a$ is the semi-major axis of the ellipse and $\epsilon$ is its eccentricity.  For Mercury, $\epsilon = 0.2056$
and $a = 57.91\times 10^6$\,km.     Most textbooks derive the perihelion precession from the exact orbits of the Schwarzschild 
solution in full GR.  It is not clear that the metric from linearized GR can yield the correct answer, since the full theory could
further perturb the metric by higher powers of $\phi$, that would enter into the post-Newtonian correction.  
To see whether 
one can derive the factor of 3 within the linearized theory, do Problem 5.5.  
%This seems to have not been guaranteed, 
%since the effect arises at $O(\phi^2)$, which should be beyond the validity of linearized gravity, but there may be a
%deep reason for it to be valid to one order higher than naively expected.

\subsection{Problems}
5.1.  Show that atmospheric pressure, when converted to units of energy density using the appropriate powers of $c$, is negligible compared to the average mass-energy density of the Earth.\\

\noindent 5.2.  Assuming that only $\bar h^{00}$ is nonzero, determine the nonvanishing components of $h^{\mu\nu}$ in terms of it.\\

\noindent 5.3.  (a)  Derive the small correction term in Eq.\ (\ref{d2yconv}).\\
(b) Compute the Christoffel symbols needed for Eq.\ (\ref{light-deflect2}) 
to leading order in the gravitational perturbation and verify the final result.\\

\noindent 5.4.  Convert the equation of motion (\ref{light-deflect2}) for light deflection into a dimensionless equation by choosing $b$ as the unit of distance (and time):  $\hat y = y/b$, $\hat t = t/b$.
Then the initial condition is $\hat y = 1$ in the far past.  What is the dimensionless parameter, restoring factors of $c$, 
that distinguishes different solutions?  Call it $\lambda$.
Numerically solve the o.d.e.\ for several values of $\lambda$ and sketch the solutions using some suitably large initial and final
values of $\hat t$.  Sketch the deflection angle as a function of $\lambda$ and show that there is some value where it starts to diverge:
light would start getting deflected by more than $\pi/2$, and then our small-angle approximation for the equation is no longer valid.
What is the approximate relation between the deflection angle and $\lambda$ for small $\lambda$?\\

\noindent 5.5. (a) By pretending that the linearized GR metric (\ref{lin-metric}) is exact, and keeping appropriate corrections to the Christoffel symbols
that are higher order in $\phi$,  and further using that fact that $g_{\mu\nu}\dot x^\mu \dot x^\nu = 1$ to eliminate $\dot t$, show that
the geodesic equations for an orbit in the $z=0$ plane can be written as
\bea
\ddot r &=& -\phi_{,r} + (1 + 2\phi) r\dot\theta^2 + O(\dot r^2, \phi^3)
\label{ddotreq1}\\
\ddot\theta &=& 2\left(-{1\over r} + {\phi_{,r}\over 1 - 2\phi}\right)\dot r\dot\theta
\eea
Why is it unnecessary to keep the $\dot r^2$ terms in Eq.\ (\ref{ddotreq1}), when considering a small perturbation to a nearly circular orbit?\\
(b) Considering a circular orbit with $r = r_0$ and $\dot\theta = \dot\theta_0$, find the relation between the parameters
needed to satisfy the geodesic equations.\\
(c) Now perturb the orbit as $r = r_0+\delta r$ and $\dot\theta=\dot\theta_0 + \delta\dot\theta$, and find the equations
for $\delta\ddot r$ and $\delta\ddot\theta$, which are linearized in the small perturbations.  Show that the equation for
$\delta\ddot\theta$ can be integrated to relate $\delta\dot\theta$ to $\delta r$.  Use this to eliminate $\delta\dot\theta$
in the $\delta\ddot r$ equation, and solve it.  How does the frequency of oscillations of $\delta r$ compare to the known result derived
from full general relativity?  Hint: to get the frequency into the form $\dot\theta_0(1 + a\phi)$ for some constant $a$, remember
that $\phi_{,r}$ is related to $r_0\dot\theta_0^2$.\\

\noindent 5.6. A possible experiment that could test the gravitational contraction of space would be to
put two satellites in geosynchronous orbit close to each other, subtending a small angle $\Delta\theta$,
and time laser pulses traveling between them.  Assume that the angle is small enough so that one can
ignore the difference between the arc of constant radius connecting the two satellites and the
straight-line path followed by the light.  The satellites are carrying synchronized atomic clocks and
a pulse is sent at an agreed-upon time.  Its arrival time at the other satellite is carefully measured
to be $\Delta\tau$ later.\\  (a) Compute $\Delta\tau$ in terms of the radius $r$ and
$\Delta\theta$.  Show that it deviates from the expected value 
(in Newtonian gravity) by some amount $\delta\tau \ll
\Delta\tau$.\\
(b) Compute $\delta\tau/\Delta\tau$ numerically.  How does it compare to the relative accuracy of an
atomic clock?\\
(c) One should also do the experiment on the surface of the Earth to verify that the effect has the expected
$r$ dependence.  What is the value of $\delta\tau/\Delta\tau$ there?

\section{Dynamical fields: gravitational waves}
We now return to the earlier problem of gravitational wave generation.  Previously we assumed the GW had already been created and we considered its free propagation.  Here we want to see how to predict its amplitude and frequency from a moving source.  A classic example is two massive objects orbiting around each other.  They have a time-dependent stress-energy tensor, 
hence they are a good candidate to produce a time-dependent metric perturbation, 
which is a GW.  We will see that not all time-dependent sources produce GWs.  There needs to be
a time-dependent quadrupole in the mass distribution.

Let's consider two equal point masses $M$ separated by a distance $2R$ in the $x$-$y$ plane, in a circular orbit with angular frequency $\omega$. 
Each mass moves with linear velocity $v = \omega R$ along the circle.  We will assume $v\ll 1$ for simplicity. The first step is to determine
$T^{\mu\nu}$ for the combined system, using Eq.\ (\ref{Teq}).  It is the sum of the contributions from each mass.  The largest component is
\be
    T^{00} = M\sum_i\delta(\vec x - \vec x_i(t))
	\label{T00eq}
\ee
where $\vec x_1(t) = R(\cos\omega t,\, \sin\omega t,\,0)$ and  $\vec x_2(t) = - \vec x_1(t)$.  However, we know that this component is not the one
responsible for creating the gravitational wave,  since the GW polarization tensors do not have a $00$ component, Eq.\ (\ref{GWpol}).  Rather, this
component generates the average gravitational potential corresponding to the static source from the two masses.   We are interested in the
components of $T^{\mu\nu}$ that  source the nonvanishing ones in Eq.\ (\ref{GWpol}).   Notice that the trace $T^{ii}$ over the spatial diagonal
components of $T^{\mu\nu}$ is not vanishing; in fact it is just $Mv^2\delta(\dots)$, which also does not contribute to the GW.  It corresponds to
the static pressure $p$ of the source, that we neglected when computing the metric perturbation in the previous chapter.  We can neglect it here
too, so long as $v\ll 1$.   (For binary systems that merge into a black hole, $v\to 1$ just before the merger; then full GR is needed.)
The relevant components of $T^{jk}$ are the ones with transverse spatial indices,  and trace subtracted,
\be
    T^{jk}_{\cancel{tr}} = M\sum_i \left(v_i^j v_i^k-\sfrac12 \delta^\perp_{jk} v_i^2\right)\delta(\vec x - \vec x_i(t))
\label{Tjkeq}
\ee
where $\vec v_1 = d\vec x_1/dt$ and $\vec v_2 = - \vec v_1$.  The symbol 
$\delta^\perp_{jk}$ means a Kronecker delta for the indices in the orbital plane,
that is, the transverse indices: $\delta^\perp_{jk}={\rm diag}(0,1,1,0)$ in the present example.  The procedure of subtracting the trace ahead of
time is a shortcut.  In problem 6.4 you will show that it amounts to doing the gauge transformation to go to TT (transverse traceless) gauge.

\subsection{Green's function for d'Alembertian}

The field equation for $\bar h^{\mu\nu}$ is a second-order p.d.e.\ with a complicated source term, which would be difficult to solve in general,
but we have seen the same kind of equation in electrodynamics, in the theory of electromagnetic waves.  We will construct a Green's function
to solve the equation  in terms of an integral over the source term.  We define $G(x,y)$ as the inverse of the 4D Laplacian,
\be
    \partial^2_x G(x,y) = \delta^{(4)}(x^\mu-y^\mu)\,.
\ee
Then formally the solution can be written as
\be
    \bar h^{\mu\nu}(x) = -2\kappa \int d^{\,4}y\, G(x,y)\, T^{\mu\nu}(y)\,.
\label{GFsol}
\ee
Also formally, we can write the Green's function in the form
\be
    G(x,y) \cong -\int {d^{\,4}k\over (2\pi)^4} {e^{-i k\cdot( x-y)}\over k^2}
\label{Gxyeq}
\ee
since $\partial_x^2 e^{-i k\cdot x} = -k^2 e^{-i k\cdot x}$, 
and we get the integral representation for a 4D delta function.  

However the integral (\ref{Gxyeq}) is not well-defined, because $k^2 = \omega^2 - \vec k^2$ (denoting $k^0=\omega$) vanishes on the light cone, and the integrand diverges.  
The problem is that the integration contour for $\omega$ passes right over the poles
at $\omega = \pm|\vec k|$, leading to an ambiguous result.  In general, the ambiguity
can be resolved by extending the $\omega$ integral into the complex plane, and deforming the contour so that it avoids the
poles.  But should one go above or below a given pole?  The choice corresponds to
different boundary conditions.  After all, the solution to a second order p.d.e.\ is
not unique until appropriate boundary conditions are specified, so it understandable
that the result (\ref{Gxyeq}) is ambiguous; we did not impose any b.c.'s.  In the present case, we would like
b.c.'s that respect causality: the solution should only be nontrivial
after the source has turned on, not before.  The correct choice of contour can be 
specified  by adding a small term $2i\epsilon k^0$ to the denominator,
\be
    G(x,y) \to -\int {d^{\,4}k\over (2\pi)^4} {e^{-i k\cdot( x-y)}\over k^2 + 2i\epsilon
k^0} 
\ee
where $\epsilon \to 0$ from above; this allows us to avoid the divergence and make the integral well-defined.  Instead of deforming the contour
to go around the poles, we leave it on the real $\omega$ axis and shift the positions of the poles slightly below it; this is equivalent to
leaving the poles on the real axis and deforming the contour to go above them, in the limit $\epsilon\to 0$.

One can now do the integral over  $\omega$ unambiguously. The denominator can be written as 
$(\omega +E+i\epsilon)(\omega-E+i\epsilon)$ with $E =
|\vec k|$, neglecting $O(\epsilon^2)$.  If $x^0>y^0$, we can complete the contour in the complex $\omega$ plane by adding the semicircle in the
lower half plane with infinite radius, since it makes a vanishing contribution to the integral (the complex exponential is a decaying real
exponential in this region).  Then the two poles determine the value of the integral by the theory of residues.  If $x^0<y^0$, we can complete the
contour with the upper semicircle, and no poles are enclosed, so the integral vanishes.  In this way, the choice of $\epsilon>0$ gives rise to the
``retarded'' Green's function $G_r$,  whereas taking $\epsilon < 0$ would give the ``advanced'' one (whose physical  significance is that it
corresponds to a GW being absorbed by the source rather  than emitted).\footnote{This covers
the cases when the contour goes above or below both poles.  What about the cases where it goes
above one and below the other?  These would cause positive and negative frequency solutions,
written as complex exponentials, to propagate in opposite directions with respect to time. 
For classical, real-valued solutions, we do not distinguish between positive and negative
frequencies, but in quantum gravity the distinction would become relevant.}\ \ 
  After doing the $\omega$ contour integral we are left with 
\bea G_r(x,y)
&=& \Theta(x^0-y^0) \int {d^{\,3}k\over (2\pi)^3}\,  e^{i\vec k\cdot (\vec x-\vec y)}\,{\sin[k(x^0-y^0)]\over k} \nn\\ 
&=& {\Theta(x^0-y^0) \over
2\pi^2}\int_{-\infty}^\infty {dk}\, { \sin[k(x^0-y^0)]\sin k|\vec x-\vec y|\over |\vec x-\vec y|}\nn\\ 
&=&  {\Theta(x^0-y^0) \over 4\pi |\vec
x-\vec y|}\,\delta[|\vec x-\vec y| - (x^0-y^0)] \label{Gret} 
\eea 
where $\Theta$ is the Heaviside function.  This result has the satisfying feature
of vanishing except on the light cone  connecting the source to the observer.  This makes it manifestly consistent  with the expectation that the
signal travels at the speed of light.

\subsection{GW solution in far-field limit}
\label{gwffsect}

We can put these results together to find an explicit expression for $\bar
h^{\mu\nu}$, but it is still quite complicated.  A significant simplification can be
made if the observer is far from the source, so that $|\vec x|\gg|\vec y|$.  We can
then set $\vec y\to 0$ in the Green's function.  The integral over $\vec y$ is trivial
thanks to the delta functions in Eq.\ (\ref{Tjkeq}).  The remaining integral over $y^0$ is
also simple because of the delta function in Eq.\ (\ref{Gret}).   The result is\footnote{This is a factor of
2 smaller than Carroll's result (7.149)}
\be
	\bar h^{jk} = {\kappa M\over \pi |\vec x|} \left(v_1^j v_1^k
-\sfrac12\delta^\perp_{jk}v_1^2\right)_{t=t_r} = {\kappa M (R\omega)^2\over 2\pi |\vec x|}\left({-\cos 2\theta\atop\phantom{-}\sin 2\theta}
	\ {\sin2\theta\atop\cos2\theta}\right)\,, \\
\label{GWsol}
\ee
where $v_1^j$ is evaluated at the retarded time $t_r = x^0-|\vec x|$,
\be
	\vec v_1 = -R\omega\,(\sin\omega t_r,\,\cos\omega t_r,\, 0)\,,
\ee 
and $\theta = \omega t_r$.
We see that it is a spherical wave moving radially outward from the source, and
falling off linearly with the distance traveled.  The absence of the $1/r^2$ correction
found in the exact solution of Problem 3.2.\ can be understood from our adoption of the
far-field limit, which neglects such contributions.
The frequency of the GW
oscillations is twice that of the orbital frequency, since we have the product of
two velocities. 

One might at first be disturbed that the solution (\ref{GWsol}) is
only transverse for waves traveling in the $z$ direction.  Recall that we have only 
imposed the Lorenz gauge condition so far, not the residual gauge transformations that
were used to make (\ref{GWpol}) traceless and transverse.  We did this for the plane wave to 
convince ourselves that there were only two independent graviton polarizations propagating
in the GW, but it is not obligatory.  We have a circularly polarized wave in (\ref{GWsol}),
which is natural given that it comes from a rotating source.

If the orbiting masses are gravitationally bound, $R$ is not independent of $\omega$.
For a circular orbit, $F = MR\omega^2 = GM^2/(2R)^2$, hence $\omega^2 = GM/(4R^3)$.  The
amplitude of the GW is of order
\be
	|h^{jk}|\sim {(G M)^2 \over R|\vec x|}\,. 
	\label{GWhest}
\ee

In 2015, the LIGO collaboration
\cite{LIGOScientific:2016aoc} observed the first GW signal from the merger of two
orbiting black holes, of mass $\sim 30\,M_\odot$ each, at a distance of $\sim
400\,$Mpc.  The black holes merge because of their energy loss from emitting GWs.
There is a maximum amplitude, because black holes are not point masses; they have a
finite radius (the Schwarzschild radius), given by $r_s = 2 G M$.  When they reach
this separation, they merge into a single black hole.  The magnitude of $|h^{jk}|$,
known as the strain, is around $10^{-21}$.
To detect it, the LIGO
experiment must measure fractional changes in distance between two bodies at rest
to one part in $10^{21}$.  Even if the two bodies are 4\,km apart, the change in
distance is much smaller than an atomic nucleus, requiring extraordinary technology.

The displacements at the detector occur within the same plane as the orbit of the
source masses, so the maximum signal is achieved if the detector is on the $z$ axis,
assuming its arms are oriented along the Earth's surface.  Both polarizations of the GW
are sourced in this configuration.  If we view the orbit edge-on, only one of the
polarizations is sourced, giving a somewhat smaller signal.

It is interesting to ask what other kinds of time-dependent sources could give rise to
gravitational waves.  What about a supernova, which is an exploding star?  We can
model it as an outgoing shell of material, moving at some radial speed $v$.  The
stress-energy tensor takes the form
\bea
	T^{00} &=& \rho(t) = {M\delta(r-vt)\over 4\pi r^2},\quad T^{0i} = \rho(t) v^i\,,\nn\\
	T^{ij} &=& \rho(t)\, v_i v_j \,,%-\sfrac12\delta_{ij}^\perp v_\perp^2
	\label{sph-exp}
\eea
where $v_i = v(\sin\theta\,\cos\phi,\, \sin\theta\,\sin\phi,\,\cos\theta)$.
%, and $v_\perp = v\sin\theta$.  
Here we
have used spherical coordinates to label the space, and Cartesian ones for the
velocity components, to match the coordinate system in which the GW polarization
tensor (\ref{GWpol}) was written (assuming the $z$ direction is toward the observer
from the source).  It is not hard to see that the integral of $T^{ij}$
in (\ref{GFsol}) cannot match the form of the GW polarization tensor: when we integrate
over the sphere it gives a result proportional to $\delta_{ij}$.  
The problem is that this kind of 
matter distribution only contributes only to the trace of $T^{ij}$, hence it cannot
generate a GW, which can always be expressed in TT gauge.  However a system that
breaks spherical symmetry, such as a collision between two stars, will generate
gravitational waves.  Moreover real supernovae are not spherically symmetric explosions, hence
they will also create GWs.

If the stress-energy tensor (\ref{sph-exp}) doesn't produce a GW, then what kind of solution
is it?  Let the observer be along the $z$ axis with the explosion at the origin.  Initially,
only signals emanating from the origin can reach the observer; later the spherical shell will
expand and signals coming from a circle on the shell will contribute to the total observed
perturbation.  However, when integrating over the azimuthal angle, all the $v_x$ and $v_y$
components will average to zero, except those on the diagonals $\bar h^{xx}$ and $\bar h^{yy}$.
Hence we see that initially $\bar h^{00}$, $\bar h^{0z}$, $\bar h^{z0}$ and $\bar h^{zz}$ will
be nonzero, and then $\bar h^{xx}$ and $\bar h^{yy}$ will slowly grow as the spherical shell
becomes larger.  The $\bar h^{00}$ component is no different from the static solution
that already existed before the star exploded: its solution is the same as for a point mass
until the shell passes the observer.  The $\bar h^{zz}$ component is the same as $\bar h^{00}$
except for the extra factor $v^2$.  It is a pressure correction, similar to what you will 
study in problem 6.1.  The only qualitatively new elements are the off-diagonal ones, 
$\bar h^{z0}=\bar h^{zz}\cong v \bar h^{00} \sim GM/|z|$, but this can be removed by a
coordinate transformation.  Hence there is no dramatic signal coming from the explosion, just
a modification to the static potential, which would be very difficult to detect. 

\subsection{Multipole expansion}

These results can be phrased in terms of the multipole expansion of the mass
distribution.  The exploding star with spherical symmetry has a 
monopole mass distribution.  In GR, the dipole moment has a trivial effect, since it
can be made to vanish by going to the center of mass of the distribution.  
The minimal configuration needed to produce GWs
is a time-dependent quadrupole distribution.  You have seen the multipole expansion in the context
of electromagnetic waves, where it characterizes a distribution of charges, rather than masses, and
gives corrections to the leading $1/r$ behavior of EM waves from a source.  
We could imagine trying to do the analogous thing
for the Green's function solution for the metric.  
The multipole expansion arises from the
Taylor expansion of $1/|\vec x-\vec y|$ in the integral 
\bea
	\bar h^{\mu\nu}(x) &=& {\kappa \over 2\pi}\int d^{\,3}y 
{T^{\mu\nu}(t_r,\vec y)\over |\vec x-\vec y|}\nn\\
 &\cong&  {\kappa \over 2\pi |\vec x|}\int d^{\,3}y\, 
\left[1  +{\vec y\cdot \hat x\over |\vec x|} + {3(\vec y\cdot\hat x)^2 -y^2\over 2|\vec
x|^2}+ \dots \right]T^{\mu\nu}(t_r,\vec y)\,.
\eea
However, as previous examples showed, the sources of observable GWs are at astronomical 
distances, so the corrections higher order in $1/|\vec x|$ are hopelessly out of 
experimental reach.  Nevertheless, there is a connection between the leading GW solutions
and the quadrupole moment of the mass distribution, which we will prove in this section.

Consider the multipole expansion of $T^{00}$, the mass density.  It is just like the
familiar one from electrostatics, with the difference that $\rho$ is always positive,
while the charge distribution can be negative.  If we compute the multipole moments of
two point masses $M$ orbiting in the $x$-$y$ plane, the monopole moment is just $2M$, the dipole moment
vanishes because we are in the center of mass system, and the nonreduced quadrupole moment\footnote{The nonreduced quadrupole is the one where the
trace is not subtracted. The trace should not contribute to the GW, but explicitly subtracting it turns out
to be unnecessary for the following, since it does not contribute to the time derivatives of
$I_{ij}$. } is
\bea
	I_{ij} &=& \int d^{\,3}y \left(y_i y_j \right)\rho(y)\nn\\
	&=& 2MR^2\left(\begin{array}{ccc} c^2 & cs & 0\\
	cs & s^2  & 0\\ 0 & 0 & 0\end{array}\right)
\eea
where $c=\cos\omega t_r$ and $s=\sin\omega t_r$.   We are interested in the oscillating 
part of this expression, which can be obtained by subtracting its time-averaged value.
Using the identities $\cos^2\theta = (1 + \cos 2\theta)/2$ and
$\sin^2\theta = (1 - \cos 2\theta)/2$, the oscillatory part can be written as
\be
	\overline I_{ij} = MR^2\left({\cos2\theta \atop \sin2\theta}{\sin2\theta\atop \cos2\theta}\right)
\ee
in the nonvanishing components, with $\theta = \omega t_r$.
Moreover, the second time derivative is given by $\ddot{\overline T}_{ij} = -(2\omega)^2
\overline I_{ij}$.  If we compare to Eq.\ (\ref{GWsol}) we see that $\bar h_{ij}= 
2G\ddot{\overline I}_{ij}/r$  (in fact  $2G\ddot{I}_{ij}/r$, since the nonoscillatory part is
constant in time): the GW is proportional to the second time derivative of
the quadrupole moment.  

This is supposed to be a general statement.  To check it,
let's write $\rho(t,\vec y) = \sum_k m_k\, \delta(\vec y-\vec y_k(t))$ for a collection
of point particles (which could be continuous) in the definition of the quadrupole
moment, and then take two time derivatives.  Using the chain rule,
\bea
	\ddot I_{ij} &=& \int d^{\,3}y\, \left(y_i y_j \right)\sum_k m_k\,\left[\ddot{\vec y}_k(t)
\cdot\nabla_y + (\dot{\vec y}_k\cdot\nabla_y)^2\right]
 \delta(\vec y-\vec y_k(t))\nn\\
&=& \sum_k m_k\,  \left[2\dot y_{k,i} \dot y_{k,j} -(\ddot y_{k,i}y_{k,j} 
+ \ddot y_{k,j}y_{k,i})\right] = 2T_{ij} - (\dots) \,,
\label{ddquad}
\eea
where we integrated by parts to get the second line.  The terms denoted by $(\dots)$ only contribute to the
trace part of $I_{ij}$ for particles that are on a circular orbit, since the 
acceleration is always directed oppositely to the position of the particle in this case.  
However it is not obvious why we should ignore them for particles oscillating linearly, 
as we explore in problem 6.4.  The extra terms
can be recognized as having only half the frequency of the ones that are associated with the 
gravitational wave.

\subsection{Energy carried by GWs}
Although perfectly spherical explosions do not produce GWs, no explosion is ever perfectly symmetrical.  Supernovae are known to be quite
asymmetrical in fact, and then one expects them to produce GWs.  
It is interesting to know what fraction of the energy of the explosion
comes out in GWs.  
Similarly, we could wonder how much energy is lost to GW emission in the merger of
black holes or neutron stars.
In analogy to EM waves, we expect GWs to carry energy.  

As motivation, let's review energy density of EM waves.  In convenient units,
they have an energy density $\rho = \sfrac12(\vec E^2 + \vec B^2)$.  This can be written in terms of the vector potential $\vec A$ in the
gauge where $A^0=0$ and $\vec\nabla\cdot\vec A=0$ as 
\be
	\rho = \sfrac12\left(\dot{\vec A}^2 + (\vec\nabla\times\vec A)^2\right) = \sfrac12\left(\dot{\vec A}^2 + (\nabla_i A_j)^2\right)\,,
\ee 
where we used a vector identity to simplify the last term.  It is interesting to compare to the Lagrangian density, which is ${\cal L} = \sfrac12(\vec E^2 -
\vec B^2)$, corresponding to $\sfrac12\left(\dot{\vec A}^2 - (\nabla_i A_j)^2\right)$.  We see that $\vec E^2$ plays the role of kinetic energy
density, and $\vec B^2$ is like potential energy density.  An EM wave $A_0\cos\omega (t-x)$ will have energy density $\sfrac12 A_0^2 \omega^2$.
According to our simplified system of units where distance and time go as inverse mass, energy density has units of $M^4$, and $\omega$ has units
of $M$.  This is consistent if $A_0$ also has units of $M$, implying that $|\vec E|$ and $|\vec B|$ go as $M^2$.

For gravitational waves, we expect the energy density should go like $|h^{\mu\nu}|^2\omega^2$, but this doesn't have the right units, since
$h^{\mu\nu}$ is dimensionless.  The problem is that Newton's constant, which has dimensions $1/M^2$, appears in the Lagrangian for gravity.
By dimensional analysis, we could then guess that $\rho \sim (1/G)|h^{\mu\nu}|^2\omega^2$, which is correct up to a constant factor.  To find the
constant, we need to reverse engineer the Lagrangian that leads to the linearized Einstein field equation (\ref{eom4}).  That is, we want to
construct the action whose variation leads to the EOM when we demand it is stationary.  The action constructed in this way is not unique; it can
be multiplied by an arbitrary factor, call it $F$, but we may be able to more easily guess the factor in this context.  It is not hard to see that
the action giving rise to Eq.\ (\ref{eom4}) is
\bea
	S &=& F \int d^4 x\left[ \sfrac12\left( \dot{\bar h}^{\mu\nu}\dot{\bar h}_{\mu\nu} -\nabla_i\bar h^{\mu\nu}\nabla_i\bar h_{\mu\nu}\right)
	- 2\kappa \bar h_{\mu\nu}T^{\mu\nu}\right]\nn\\
	&=& F \int d^4 x\left[ \sfrac12\left({\bar h}^{\mu\nu}(-\partial^2)\bar h_{\mu\nu}\right)
	- 2\kappa \bar h_{\mu\nu}T^{\mu\nu}\right]
\eea
where we integrated by parts to get the second line.  Varying with respect to $\bar h^{\mu\nu}$ immediately gives the desired EOM, but what about
dimensions?  $T^{\mu\nu}$ already has dimensions of $M^4$, suggesting that the factor $\kappa$
 should be canceled by choosing $F \sim 1/\kappa$.  To get the constant of proportionality,
we can use the fact that $2\kappa F\bar h_{\mu\nu}T^{\mu\nu}$ must be the gravitational potential energy of a point
mass $M$ when $T^{00} = M\delta(\vec x)$ and $\bar h_{00} = 4\phi$.  This fixes $F =
1/(8\kappa)$.  In analogy to the electromagnetic wave, the energy density of the GW
is then given by
\be
	\rho = {1\over 8\kappa}\times\frac12\left((\dot{\bar h}^{\mu\nu})^2 + (\nabla_i \bar
h^{\mu\nu})^2\right) = {\omega^2 |{\bar h}^{\mu\nu}|^2\over 64\pi G}\,.
\label{GWeden0}
\ee
If we decompose the wave as $h^{\mu\nu} = (h_+ \epsilon_+^{\mu\nu} + h_\times 
\epsilon_\times^{\mu\nu})\cos(\vec k\cdot \vec x - \omega t)$ and define 
$\langle h\rangle^2 = h_+^2 + h_\times^2$, then
\be
	\rho = {\langle h\rangle^2\omega^2\over 64\pi G}\,.
\label{GWeden}
\ee
The factor of 2 coming from tracing over the $\epsilon_+^2$
or $\epsilon_\times^2$ tensors is canceled by the time average of $\cos^2(\dots)$, which gives $1/2$.  
{\it Notice that $\omega$ denotes the frequency of the GW here,
which is twice the frequency of the source (the $\omega$ used in Section \ref{gwffsect}).}

This tells us the energy density in the GW, but our original question was about the energy
transport.  The relation is simple: the energy transported through a surface per unit time and
area is just $\rho c$ since the wave is moving at the speed of light.  If it is not obvious to
you, draw a small box of length $c\Delta t$ and transverse area $A$, and compute the total
energy passing through the area in time $\Delta t$.  It is $\rho A c\Delta t$, assuming the
wave is traveling perpendicularly to the area.  It is reduced by a factor of $\cos\theta$ if
it is not perpendicular.  We can apply this to the problem of two orbiting masses.  Leaving
aside some factors of order unity, the power in GWs passing through a sphere in the far-field
region is given by $4\pi |\vec x|^2 \rho \sim (GM/R)^5/(8G)$.  This has the interesting
consequence that the orbit will inevitably decay and lead to a merger, since the energy loss
to GWs must be compensated by the orbital energy.  For some critical value of the separation
$R$, this can occur within the age of the Universe and lead to observable merger events. It is
notable that the GW maximum power, at the moment of the merger when $R = 2 GM$, has a
universal value $\sim c^5/(16^2 G) 10^{10}\,$W.

\subsection{Stochastic GW background}
Following the LIGO-Virgo-KAGRA merger discoveries, numbering in the 300's at this writing,
another kind of gravitational wave was discovered by pulsar timing arrays (PTAs).  These are
collaborations of radio telescopes at different locations on Earth, who compare their
respective arrival timings of collections of pulsar signals.  A given pulsar's radio bursts
may travel slightly different times to get to different telescopes, not just because of the 
distance differential provided by being at different locations on Earth, but also because they
have passed through very long wavelength gravitational waves that contribute to the distances.

For this detection, it does not suffice to compare timing residuals of individual pulsars;
instead the residuals are cross-corrlelated between pairs of pulsars, that subtend an angle
$\theta$ in the sky. In 1983, Hellings and Downs famously derived that the correlation strength should
vary roughly as $\cos\theta$ if it is caused by gravitational waves \cite{Hellings:1983fr}. 
This was the Holy Grail signal being sought by the pulsar timing arrays, and it was first
observed in 2023 at 4$\sigma$ significance by the NanoGRAV collaboration, including telescopes
at VLA, Green Bank, CHIME, and the now defunct Arecibo \cite{NANOGrav:2023gor}.  
Since then there has been corroborating evidence from the Parkes PTA, the European PTA, and the Chinese PTA, some of
which will combine their data to increase the significance of the detection.  The observed
signal is not necessarily dominated by single events, but probably arises from the
superposition of many, accumulating over the last 10 billion years; hence it is a stochastic
gravitational background.

The favored astrophysical source for producing such GWs is the merger of supermassive black
holes SMBHs at the centers of galaxies.  Gravitational N-body simulations that simulate the
formation of structure in the Universe show that mergers of galaxies are a typical means of
structure growth, and this leads to the mergers of central black holes that were already
present in the original galaxies.  One can model the expected rate of such mergers, which are
distributed over their total masses, the ratio $Q$ of the individual black hole masses, and 
the redshift $z$ when the merger occurred.   The PTA events have frequencies in the range of 
$0.04-0.6$\,y$^{-1}$, which are much smaller than the LIGO events, corresponding to the fact
that the SMBHs can have masses up $3\times 10^9\,M_\odot$, compared to LIGO's
50-100\,$M_\odot$ mergers.  The observed strains are of order $h\sim 10^{-14}$.

However, there is an outstanding problem: it has
been estimated that orbiting black hole binaries only efficiently lose orbital energy 
when they are separated by more than $\sim 1$\,pc \cite{Begelman:1980vb}.  At larger
distances, there is enough material (gas and stars) to allow for fast energy loss by 
gravitational friction \cite{Chandrasekhar:1943ys} or by three-body interactions.  The former
is caused by the back-reaction of material concentrated in the wake of a moving BH, which was
attracted to it and becomes more dense behind it, resulting in a backward pull.
But ultimately this material gets swept
away from the binary through the energy it must gain, creating a vacuum (``loss cone'') at
separations below $1$\,pc, which shuts off the energy-loss mechanisms.  Only at separations
$\lesssim 0.1$\,pc does gravitational wave emisssion become efficient for completing the
merger.  This bottleneck is known as the ``final parsec'' problem.  There are suggestions for ways to
overcome it, but so far no consensus.  It is possible that a more exotic source from the early
Universe is the origin of the PTA signals; GWs from inflation, first order phase transitions,
cosmic strings, and domain wall collapse have been suggested.

\subsection{Gravitational memory}
An interesting difference between gravitational and electromagnetic waves is that once a source of
electromagnetic waves has turned off, the system returns to its original state before the radiation
started.  But with a transient source of gravitational waves, there can be an offset between the metric at
early times and later after the event that produced the waves has finished.  This arises because of the
proportionality between the metric perturbation and the stress energy tensor of the source.  The simplest way to
see this is to consider two point masses orbiting each other.  Suppose (artificially) that the two bodies
started orbiting each other at some moment, and they stop at a later time $t_1$, when they are oriented
along the $x$ axis.  The gravitational wave signal does not entirely disappear at the moment of
the retarded time corresponding to $t_1$. By continuity, the metric remains constant at later times, retaining
the memory of the wave at the time when it stopped oscillating.  While the effect is
theoretically interesting, it is beyond the reach of present experiments, given the extremely
small magnitude of observed GW amplitudes.

\subsection{Problems}

6.1.  Compute the {\it static} perturbation for the system of two masses $M$ rotating around each other, each with linear velocity $v$, in the far-field limit $r\gg 
R$, including the energy density, and the pressure caused by the rotation of the source, which is the diagonal trace term
subtracted from (\ref{Tjkeq}). This amounts to taking the time-average of $T^{\mu\nu}$ over one period.
How is the force on a particle moving nonrelativistically in the gravitational field affected?
And the time dilation?
Can you interpret the difference with our previous weak-field results in terms of the total energy of the source?
Does the pressure matter?\\

\noindent 6.2. Fill in the missing steps to derive Eq.\ (\ref{Gret}).  First do the $\omega$ integral by adding the two residues; then do the integral over
$\cos\theta$, where $\theta$ is the angle between $\vec k$ and $\vec x - \vec y$.  The final integral over $k$ can be extended to negative $k$ by
the symmetry of the integrand.\\

\noindent 6.3.  Estimate the maximum GW amplitude before the merger of two orbiting black holes of mass
30\,$M_\odot$ each, at a distance of 400\,Mpc from Earth.  1\,pc (parsec) =
3.26 light-years. \\ 
%Answer: $|h^{jk}|\sim 10^{-21}$. 

\noindent 6.4.  Consider a source with two equal masses $m$ oscillating along the $x$ direction, with positions $x_\pm = \pm (R_0 +R\cos\omega t\,)$,
where $R_0$ is the rest length of the spring that connects them, and $R < R_0$ so they don't bump into each other.\\
(a) Compute the stress-energy tensor, integrated around a small volume containing the sources, including the kinetic energies and the
potential energy of the spring in $T^{00}$.  For convenience define $v=R\omega$. Notice that you don't need to know the
spring constant.\\
(b)  Find the corresponding solution for  $\bar h^{\mu\nu}$, for an observer who is located far away at a distance $z$ on the $z$
axis.
Do not subtract the trace from $T^{ij}$, like we did in Eq.\ (\ref{GWsol}), since we want to check in part (c) that this gets accomplished
by a coordinate transformation.\\
(c) For the remainder of the problem, we want to split $h^{\mu\nu}$ into the static part and
the GW part.  To do this, notice that $\sin^2(x)$ can be split into a constant plus an
oscillatory wave that is not squared.  We can ignore the constant parts, whose effect was
studied in problem 6.1, and just focus on the oscillatory part, which is the GW.
Find a gauge transformation that gets rid of the trace of the oscillatory part of
the above solution, and that does not produce any other 
unwanted components.  (Hint: $\xi^0$ and $\xi^z$ are sufficient.)  Does the resulting solution correspond to one of the standard GW polarizations?  Hints: split the oscillating part of
$h^{xx}$ into a constant plus a wave, since problem 6.1 shows there should be a constant 
contribution.  If your gauge transformation produces
small unwanted terms of $O(1/z^2)$, we can neglect them since we are already ignoring such corrections to the GW.\\
(d) Compute the quadrupole moment of the source, and its second time derivative.  
Check the proportionality between it and your GW solution, written in terms of 
$\bar h^{\mu\nu}$. Compare to Eq.\ (\ref{ddquad}).\\

\noindent 6.5. (a) Fill in the missing steps that give the estimate for the power loss to 
GW emission for two orbiting masses $M$ with 
orbital radius $R$ (separation $2R$), below Eq.\ (\ref{GWeden}). \\
(b) Find the total orbital energy of the system, neglecting any losses, as a function of $G$, $M$ and $R$.\\
(c) By equating the time derivative of the orbital energy to the power loss, compute the time-dependence of $R$.\\
(d) Suppose $M = 10^{10}\,M_\odot$, the size of a supermassive black hole that could be found in the center of a galaxy.
How small must $R$ be in order for the black holes to merge within 1 Gyr ($10^9$ years)?  Express your answer in pc (parsecs).\\

\noindent 6.6. Simulations of core-collapse supernovae show that they release about $10^{53}$\,ergs in neutrinos \cite{Abdikamalov:2020jzn}. 
Because of turbulent instabilities that quickly develop, the neutrinos can be emitted 
 anisotropically. Even though the neutrinos are relativistic, we can still model the GW source as two equal masses being ejected in opposite
directions along the $x$ axis, with velocities $\pm v$, such that $Mv^2$ equals the total kinetic energy.  In the end, it is only the total
energy that matters, not whether it is relativistic.
 The emitted wave is not sinusoidal,
but rather a burst whose duration is of order $\Delta t = 1\,$s.\\
(a) Estimate the order of magnitude of the maximum strain $|h^{ij}|$ of the GW burst, as a formula, depending upon the distance
from the supernova.\\
(b) Estimate the total energy emitted in GWs in ergs.  How many factors of $c$ must be restored in order to get the correct units from your formula?

\section{Curved space}
Weak gravity is a perturbation on Minkowski space, which like its Euclidean counterpart, is flat: it has no curvature.  As a first
step toward developing the nonlinear theory of gravity, we would like to understand
the geometry of spaces that have curvature, like the surface of a sphere.  For
simplicity we will start with purely spatial geometries and ignore the time direction
(hence the metric signature is $(+,+,+)$.)
The mathematics that describes curved space generalizes straightforwardly to curved
spacetime.

Our starting point is the line element for purely spatial dimensions
\be
	ds^2 = g_{ij} dx^i dx^j
\ee 
in any number of dimensions $d$.  Euclidean space has $g_{ij} =\delta_{ij}$.  
A familiar example of a curved space in $d=2$ is the unit sphere, 
\be
	ds^2 = d\theta^2 + \sin^2\theta\, d\phi^2
	\label{sphmet}
\ee
which therefore has metric $g_{ij} = {\rm diag}(1,\sin^2\theta)$ in spherical
coordinates.  One might think that the circle is a curved space in $d=1$ with 
the single metric component $g_{\theta\theta}=1$, but this is exactly the same as the
metric for a straight line.  In fact, for $d=1$, we could take the metric to be 
an arbitrary function $f(x)$ for line element $ds^2 = f(x) dx^2$, but this is just a
coordinate transformation of the trivial metric, since we can always locally define
a new coordinate $dy = f(x)^{1/2}dx$ that makes $g_{yy}=1$.  Hence there is no such
thing as curvature in $d=1$, although there can be topology.  The
circle is a closed space while the line is open.

This goes against our notion that there is an obvious difference between a
circle and a straight line.  The circle is curved, after all!  We must make a
distinction between {\it intrinsic} and {\it extrinsic} curvature.  The first is a
property of the space by itself, regardless of any higher dimensions it might be
embedded in.  The second one is a measure of how a lower-dimensional surface appears
to be curved with respect to a higher dimensional space in which it is embedded.
Consider the circle in the absence of any other dimensions, as though our world were
one-dimensional.  Any being traveling along that surface would see it as being just as
flat as if it were a line; indeed there is no distinction, other than topology (the
ant will notice that it keeps coming back to the same point when walking on a circle).  In this chapter we will be
concerned with intrinsic curvature, since our ultimate goal is to understand curvature
of the full 3+1 dimensional spacetime, which need not be embedded in any higher
dimensions---although in string theory, it is, in 10 dimensions as a matter of fact.  

For example, consider the curve $y = x^2$ in the $x$-$y$ plane.  We can define it as the
solution to $g(x,y) = y-x^2 = 0$.  The normal vector is $\nabla g = (-2x,y)$ along the
curve.  The fact that its direction is changing as one moves along the curve is a way of
distinguishing that it has extrinsic curvature in the $x$-$y$ plane, as opposed to a straight
line $g = y - ax = 0$, whose normal vector is a constant.  But if we forget about the 
normal direction and consider only the metric induced on the surface of the curve,
\be
	ds^2 = dx^2 + dy^2 \to dx^2\left(1 + \left({dy\over dx}\right)^2\right) 
	= dx^2(1 + 4x^2)
\ee
then there is no intrinsic curvature, which becomes obvious by defining a new coordinate $dz = dx\sqrt{1+4
x^2}$ which has a trivial metric.

Notice that we cannot reduce the sphere to the Euclidean plane by a similar trick,
{\it e.g.,} defining new coordinates such that $d\theta = dx$ and $dy = \sin\theta
d\phi$.  There is no change of coordinates $\theta = f(x,y)$, $\phi = g(x,y)$ which
would allow this.  Let's try it and see what goes wrong.  First compute the differentials
\be
	d\theta = {\partial f\over \partial x}dx + {\partial f\over \partial y}dy\,;
	\quad 	d\phi = {\partial g\over \partial x}dx + {\partial g\over \partial y}dy\,.
\ee
Then we plug them into the metric (\ref{sphmet}) and demand that it revert to Euclidean space:
\bea
	\left(\partial f\over\partial x\right)^2 + \sin^2 f\left(\partial
g\over\partial x\right)^2 &=&1,\nn\\
	\left(\partial f\over\partial y\right)^2 + \sin^2 f\left(\partial
g\over\partial y\right)^2 &=& 1\nn\\
	{\partial f\over\partial x}{\partial f\over\partial y} + \sin^2 f\,
	{\partial g\over\partial x}{\partial g\over\partial y} &=&0
\eea
This is three differential equations for two functions, which is generically
overdetermined.  It is not a rigorous proof that no solution exists; that might be 
rather difficult to prove.   

More generally, supposing we were presented with some
metric $g_{ij}(x,y)$ (continuing in $d=2$ dimensions for simplicity), we would
need to determine whether there exist functions $f,g$ such that 
\bea
	g_{xx} &=& \left(\partial f\over\partial x\right)^2 + 
\left(\partial g\over\partial x\right)^2\,,\\ \nn
		g_{yy} &=& \left(\partial f\over\partial y\right)^2 + 
\left(\partial g\over\partial y\right)^2\,,\\ \nn
	g_{xy} &=& {\partial f\over\partial x}{\partial f\over\partial y}
+{\partial g\over\partial x}{\partial g\over\partial y}
\eea
in order to know whether it was just a reparametrization of flat space.  But in fact there is
an easier way: we just need a measure of curvature that does not depend on a choice of
coordinates.  If the curvature of a given metric is nonzero, then no coordinate
transformation exists that would turn it into Euclidean space.

Since we are more familiar with extrinsic than intrinsic curvature from daily life,
we can borrow the intuition that an extrinsically curved surface, say parametrized by
a function $y=f(x)$ in the $x$-$y$ plane, can be characterized by the fact
that $f''(x)$ is nonvanishing.  This suggests that appropriate derivatives (in
particular, second derivatives) of the
metric might be able to tell us something about curvature.  For example, by taking two
derivatives and contracting indices in the right way, we could perhaps form a scalar quantity
that would be a measure of the curvature. It would be coordinate invariant by virtue of not having
any indices.

\subsection{Covariant derivatives} 

However, contracting indices is guaranteed to produce coordinate-independent results only if
the object transforms like a tensor
under coordinate transformations, and we immediately encounter a problem when we take 
derivatives of the metric tensor, or any other kind of tensor, as will be seen.  
Consider
\be
	g_{ij,k} = {\partial g_{ij}\over \partial x^k}
\ee
We would like to know how it behaves under a general coordinate transformation
$x^i\to x^i(y)$.    For clarity, let's denote the metric in the new coordinate system
as $\bar g_{ij}(y)$, and $\bar g_{ij,k} = \partial\bar g_{ij,k}/\partial y^k$.  We want to relate
it to $g_{ij,k}$ in the original $x$ coordinates.  The relation is
\bea
	\bar g_{ij,k} =  {\partial\over \partial y^k}
\left({\partial x^m\over \partial y^i}g_{mn} {\partial x^n\over \partial y^j}\, 
\right)\, &=& \partial_k (JgJ)_{ij} \equiv (JgJ)_{ij,k}\nn\\
	&=& (J_{,k}gJ)_{ij} + (Jg_{,k}J)_{ij} +  (JgJ_{,k})_{ij}\,.
\label{gtrans}
\eea
The problem is that $\partial/\partial y^k$ operates in general not only on $g_{mn}$, but also on
the Jacobian matrices $J$, so that extra terms get generated which do not have the form of
a tensor coordinate transformation, but instead involve the derivative of the $J$ matrix. 
If it weren't for these extra terms, $g_{ij,k}$ would transform
by the ordinary rule: each index gets contracted with a $J$ matrix.

This problem is not particular to the metric tensor; it also
occurs for any tensor.  It is completely analogous to the problem in electrodynamics that the 
derivative of a charged field does not transform in a simple way under a gauge
transformation.  For example, the electron wave function $\psi$ transforms as $e^{i\theta(x)}\psi$
under a gauge transformation, but $\partial_\mu \psi \to e^{i\theta}\left(\partial_\mu
+ i\partial_\mu\theta\right)\psi$.   To counteract this, we recall that the gauge
field transforms as $A_\mu \to A_\mu + \partial_\mu\theta$, and we therefore define the 
{\it covariant derivative}
\be
	D_\mu\psi = (\partial_\mu -iA_\mu)\psi
\label{covderEM}
\ee
which does transform simply, $D_\mu\psi \to e^{i\theta} D_\mu\psi$.  In the same way,
we would like to define a covariant derivative with respect to general coordinate
transformations (here conventionally denoted by $\nabla_\mu$), by finding the appropriate extra terms to add to $\partial_\mu$.
The answer is not so obvious in this case.  Unlike the electrodynamic example, the
form of $\nabla_\mu$ depends on what kind of tensor it is operating upon, since each
index of the tensor generates extra terms when the partial derivative acts on the 
Jacobian matrix ${\partial x^m/\partial y^i}$.  Let us first write
the general result for a one-index contravariant tensor $A^j$:
\be
	\nabla_i A^j = \partial_i A^j + \Gamma^{j}_{\, ik}A^k \equiv A^j_{\, ;i}\,.
\label{covder}
\ee

One can show (problem 7.1) that the extra terms involving
derivatives of ${\partial x^m/\partial y^i}$ cancel out, so that $A^j_{\, ;i}$ indeed
transforms as a 2-index tensor (with one contravariant and one covariant index).  We notice that this formula is easy to remember just by putting the indices in the
right places.  Since $\Gamma^j_{\, ij}$ is symmetric in the lower indices, it doesn't
matter in which order we write them.  Second, the generalization to any number of
contravariant indices is straightforward: just add the appropriate extra term for each
index while leaving the others unchanged,
\be
	\nabla_i A^{jk\dots} = \partial_i A^{jk\dots} + \Gamma^{j}_{\, il}A^{lk} 
	+ \Gamma^{k}_{\, il}A^{jl} + \dots
\label{covder2}
\ee
To differentiate a tensor with covariant indices, the formula is
\be
	\nabla_i A_j = \partial_i A_j - \Gamma^{k}_{\, ij}A_k \equiv A_{j;i}\,.
\label{covder3}
\ee
Again, the positions of the indices are dictated, and only the relative sign must be
remembered.  This sign comes from the fact that it is now the inverse of the Jacobian 
matrix that is being differentiated.  Again, the generalization to additional indices 
is obvious, as well now as the combination of different kinds of indices.  In this
context, the Christoffel symbol is often called the ``affine connection;'' you could
use this language if you would like to impress people with your mathematical
sophistication.	  Later we will give a geometrical interpretation of the connection
that explains the origin of this terminology.

For a 0-index tensor (scalar), the covariant and partial derivatives are the same,
since there is no index to make us invoke the Jacobian matrix in the first place.

\subsection{Geometric significance of covariant derivative}
To get a better feeling for why the covariant derivative should have more physical significance
than the ordinary partial derivative, consider the following simple example.  Let $f(x,y) =
\sfrac12(x^2 + y^2)$ in 2D Euclidean space.  Then $f_{,ij} = \delta_{ij}$.  As a geometrical 
object, we could write it as $\hat x\hat x + \hat y\hat y$.  In polar coordinates, on the other
hand, $f(\rho,\theta)=\sfrac12\rho^2$, and $f_{,ij} = \delta_{i\rho}\delta_{j\rho}$, which
geometrically is $\hat\rho\hat\rho$.  But the unit tensor is $\hat\rho\hat\rho
+\hat\theta\hat\theta$, which is the same as $\hat x\hat x + \hat y\hat y$, as one can readily
verify.  These are quite different objects.  We want a notion of derivatives
that keeps its geometrical character independently of the choice of coordinate system.
This is what the covariant derivative does.  In the present example,
\be
	f_{;ij} = f_{,ij} - \Gamma^k_{ij}f_{,k}
\ee
with $\Gamma^\rho_{\ \theta\theta}= -\rho$ and $\Gamma^\theta_{\ \rho\theta} = 
\Gamma^\theta_{\ \theta\rho} = 1/\rho$.  We find that $f_{;ij} = {\rm diag}(1,\rho^2)= g_{ij}$.
This corresponds  to the Cartesian coordinate result, since
$\partial_\theta$, considered as a tangent vector is not normalized; instead
$\rho^{-1}\partial_\theta$ is the unit vector.  Then the geometrical form of $f_{;ij}$ is
 $\hat\rho\hat\rho +\hat\theta\hat\theta$ in terms of unit basis vectors, as expected.

Another way to describe the covariant derivative is that it 
accounts for the 
variation of the vector field components from the changing directions (or lengths) of the basis vectors,
in a nonCartesian coordinate system.  What is left is the essential variation of the vector
field itself, which is independent of the choice of coordinates.  This explains why the metric
tensor has vanishing covariant derivatives.  Since it can be formulated as the inner products
of the basis vectors, its variation is completely owing to their nonuniformity, whose subtraction
is the job of the Christoffel symbols.  A trivial example is the
vector field $\vec A = \hat x$ in 2D Euclidean space.  It has no
variation, but in polar coordinates, its components depend upon
$\theta$:  $\vec A = \cos\theta\,\hat\rho -\sin\theta\,\hat\theta$.  The
covariant derivative is zero, as expected, even though the partial
derivatives $A^i_{\ ,j}$ are nonvanishing.  Conversely, a vector field
$\vec A = \hat\theta$, which has $A^i_{\ ,j} = 0$ is not uniform, and
the covariant derivative correctly describes its variation within the
tangent space where it lives.

Here is another simple example in 2D Euclidean space.  The function $f(x,y) =
(x-y)$ can be used to describe straight lines by taking $f = $ constant, and the normal
direction is given by $\vec n = \vec\partial f = (1,-1)$.  Obviously $n_{i,j}=0$; the normal does not
vary from one position to another.  Suppose we transform to coordinates $x = x' + x'^2$ and
$y = y' + y'^2$.  In the new coordinate system, $\vec\partial\,'f = (1+ 2x',-1-2y')$, and
$n_{i,j}$ is no longer zero: it has components $\left({2\atop 0}{\phantom{-}0\atop -2}\right)$.  It is the
same vector field, described in different coordinates.  The partial derivative is giving an
ambiguous idea about its variation.  

In general, the covariant derivative of a vector field considered as a geometric object 
$\vec A = A_j\hat x^j$ with covariant components is

\be
	\nabla_i \vec A = A_{j,i}\hat x^j + A_j \partial_i\hat x^j
\ee
and its $k$th component is
\bea
	A_{k;i} = \hat x_k\cdot\nabla_i \vec A &=& 
	A_{k,i} + A_j \hat x_k\cdot \partial_i\hat x^j\nn\\
	&=& A_{k,i} - \Gamma^{j}_{\ ik}A_j\,,
\eea
and similarly for the contravariant components (leading to Eq.\ (\ref{covder3})).  The Christoffel symbols merely describe the
variation in $\vec A$ from the nonuniform basis vectors.  Of course, in flat space
with Cartesian coordinates this is not an issue.  Coming back to the example above, it is
straightforward to show that $n_{i;j} = 0$ in the primed coordinate system, even though
$n_{i,j}$ is nonzero. 

\subsection{Curvature}
We now return to our earlier motivation for differentiating the metric tensor:
the search for a coordinate invariant (or at least covariant, in the case of a tensor)
quantity that can quantify the curvature of a spatial manifold.  The hope was that 
taking appropriate derivatives of the metric and contractions of indices might
accomplish this, and this led us to the definition of the covariant derivative.
However it turns out that the covariant derivative of the metric tensor vanishes
identically (see problem 8.1),
\be
	g^{ij}_{\ \ ;k} \equiv 0\,,
\ee
so we are back at square one.  Nevertheless the answer is related to $\Gamma^i_{\ jk}$;
the relevant object that transforms as a tensor can be constructed from 
derivatives and products of the Christoffel symbols.  It is called the Ricci curvature
tensor:
\be
	R_{jk} = \partial_i \Gamma^{i}_{\ jk} - \partial_j\Gamma^{i}_{\ ik}
	+\Gamma^i_{\ i m} \Gamma^m_{\ jk} - \Gamma^i_{\ jm}\Gamma^m_{\
\ ik}\,.
\label{ricci}
\ee
Although tedious, it is straightforward to show that all the unwanted terms cancel out
under a change of coordinates, 
so that $R_{jk}$ transforms as a tensor.   Despite appearances, it
is symmetric in its indices.  
 It fulfills our initial
expectation of involving second derivatives (plus products of first derivatives) of
the metric.  Moreover its trace $R = R^i_{\ i}$, known as the Ricci scalar, is
coordinate invariant, and is therefore a candidate for measuring a geometric
property of a space, namely its curvature, in a coordinate-independent way.  $R_{ij}$ and $R$
have dimensions of 1/(distance)$^2$, as we would expect for a measure of curvature.

Even though the Christoffel symbols are not tensors, one might wonder whether 
$R_{jk}$ could be obtained through differences of their covariant derivatives, pretending that they
were tensors.  Under this pretense, one would find that
\be
	\nabla_i \Gamma^{i}_{\ jk} - \nabla_j\Gamma^{i}_{\ ik} \ \stackrel{?}{=} \ 
\partial_i \Gamma^{i}_{\ jk} - \partial_j\Gamma^{i}_{\ ik} 
+2\Gamma^i_{\ i m} \Gamma^m_{\ jk} - 2\Gamma^i_{\ jm}\Gamma^m_{\
\ ik}\,.
\ee
Curiously, it ``almost'' gives the Ricci tensor, but the products of $\Gamma$'s come with a coefficient that is
twice too large.  In the next chapter, we will describe a derivation that relates the Ricci
tensor (more precisely, its predecessor the Riemann tensor) to commutators of 
covariant derivatives.

Ref.\ \cite{Tao-Ricci}  explains the geometrical meanings of $R_{ij}$ and
$R$ in the case of two dimensions. At a given point $x$ in the manifold, choose a direction $\hat
e^i$ specified by a unit tangent vector, and measure the area of a
small pie slice of radius $r$ along the $\hat e^i$ direction, with 
angular extent $\theta$.  In Euclidean space, the area would be
$A_E = \sfrac12\theta r^2$.  But in a curved space, it will differ.
One finds that the measured area $A$, compared to $A_E$, satisfies
\bea
\label{Ricci2}
	{A\over A_E} &=& 1 - \sfrac{1}{12} r^2
R_{\hat\imath\hat\jmath}\hat e^i\hat e^j + O(r^4)
%\\
%\label{Ricci-meaning}
%&=& 1 - \sfrac{1}{24} r^2R + O(r^4)\,,
\eea
for small $r$.\footnote{The hats on the indices are to emphasize that if one were to choose $\hat e$ as a
unit vector in the $i$ direction, then it would be an orthonormal basis vector and we would be
measuring $R_{\hat\imath\hat\imath}$ in that basis, rather than $R_{ii}$ in the coordinate basis.
The distinction between coordinate and orthonormal bases will be discussed in Chapter 8.}
Thus as $r\to 0$, the space always looks locally Euclidean,
but for any finite $r$, one can measure the difference and find that
positively curved surfaces have smaller area within a circle or
a section of a circle.  It must be noted that for this formula to work, the sides of the small
slice follow geodesics $\theta(r)$, which generically are not rays of fixed angle.  Hence the
angle $\theta$ appearing in $A_E$ is the limiting value close to the vertex, and the 
actual opening angle may
tend to increase or decrease away from $r=0$ depending upon the curvature or the choice of coordinates.

  In higher dimensions, one should replace areas
by $d$-volumes to generalize Eq.\ (\ref{Ricci2}), and the coefficient $1/12$ will be different.  
It is interesting that the result can depend upon the direction in which the centroid of the volume
is oriented, since $R_{ii}$ can have different magnitudes and signs depending on $i$.
One can also compute the total $d$-volume of a ball centered at some point and compare to its Euclidean
counterpart \cite{wiki-curvature},
\be
	{V\over V_E } = 1- {r^2 R(x)\over 6(d+2)} + O(r^4)\,. 
\label{Ricci3}
\ee

In problem 7.2, you will show that $R=2$ for a unit two-sphere, whose metric is
is $ds^2 = d\theta^2 + \sin\theta^2 d\phi^2$.  Physically, curvature should have units of
inverse distance squared.  For a two-sphere with radius $L$, the metric would be multiplied by
$L^2$.  Then following the powers of $g_{ij}$ in Eq.\ (\ref{ricci}), we find that $R = 2/L^2$.
This makes intuitive sense, since as $L\to\infty$, the sphere
becomes locally more and more flat.\\

\subsubsection{Example with negative curvature}

We saw that a sphere has constant positive curvature, which accords with our
intuition: the curvature is the same at every point.  This provides a proof that 
we cannot transform a sphere into Euclidean space by a coordinate transformation.
The Ricci curvature of Euclidean space vanishes since all its Christoffel symbols
are zero.  To confirm the picture, it would be nice to see that a negatively curved
surface has $R<0$.  A well known example is a hyperbolic surface, such as 
\be
	z = {1\over 2 L}(x^2-y^2)
\ee
embedded in three dimensions.  We can find its intrinsic metric in two dimensions by 
computing the line element in 3D, restricted to the surface (the
induced metric):
\be
	ds^2 = dx^2 + dy^2 + dz^2 = dx^2 + dy^2 + L^{-2}(x dx - y dy)^2\,.
	\label{hyperg}
\ee
%This is known as the {\it induced metric} on the surface.  However once we have derived it,
%we are free to forget about the original 3D space and consider (\ref{hyperg}) as the 
%intrinsic 2D metric of some peculiar spatial manifold.
The metric tensor and its inverse take the form
\be
	g_{ij} = \left({ 1+ (x/L)^2\atop -xy/L^2} {-xy/L^2\atop 1 + (y/L)^2}\right)\,;
\qquad
	g^{ij} = g^{-1}\left({ 1+ (y/L)^2\atop xy/L^2} {xy/L^2\atop 1 + (x/L)^2}\right)
\ee
where $g = 1 + (x/L)^2 +(y/L)^2$ is the determinant of $g_{ij}$.
To compute the Christoffel symbols, it is easiest to start with all covariant indices
and then raise the first one afterwards, using the inverse metric.  The nonzero ones are
\bea
	\Gamma_{xxx} &=& x/L^2\,,\nn\\
	\Gamma_{yyy} &=& y/L^2\,,\nn\\
	\Gamma_{xyy} &=& -x/L^2\,,\nn\\
	\Gamma_{yxx} &=& -y/L^2\,.
	\label{hypchrist}
\eea
To simplify calculations, let's work to leading order in $1/L^2$, which is relevant
for the region near the origin where $|x|, |y|\ll L$.  Then the inverse metric can be 
approximated as the unit matrix and we can raise indices trivially.  Moreover, we can
ignore products of Christoffel symbols in the Ricci tensor.  To leading order we find
that $R_{ij} \cong -(1/L^2)\,{\rm diag}(1,1)$, hence $R\cong -2/L^2$, confirming the
expectation that $R<0$ for a hyperbolic space.  This result is exact at the origin,
but we can expect that it 
will be corrected by powers of $x/L$
and $y/L$ at other positions, so that the curvature is a function of position for this metric. 
The exact answer is $R = -2 L^{-2}(1 + (x^2+y^2)/L^2)^{-2}$; the curvature vanishes at infinity.
  In problem 7.3 you will consider a case with constant negative curvature (which cannot be
embedded in 3D Euclidean space).

\subsubsection{The Riemann tensor and geodesic deviation}
\label{geo-dev-sect}

The Ricci tensor descends from a 4-index  tensor $R^i_{\ jlk}$ named after Riemann,
\be
	R^i_{\ klj} = \partial_l \Gamma^{i}_{\ jk} - \partial_j\Gamma^{i}_{\ lk}
	+\Gamma^i_{\ l m} \Gamma^m_{\ jk} - \Gamma^i_{\ jm}\Gamma^m_{\ lk}\,,
\label{riemann}
\ee
by contracting two indices: $R_{jk} = R^i_{\ jik}$. 
One could notice that by uncontracting appropriate pairs of indices in Eq.\ (\ref{ricci}),
the Riemann tensor results.  For the first two terms, there is only one contraction, so it is
obvious, while for the last two, there is a choice of contractions to undo.  But only one
of them is such that the remaining products of Christoffel symbols involve a contraction connecting them,
so the product doesn't ``fall apart,'' and this is the correct one to take.
Here is a mnemonic for constructing the Riemann tensor.  Pretend that $R^a_{\ bcd}$ is minus the covariant derivative
of $\Gamma^a_{\ bc}$ with respect to $x^d$, treating only the upper index $a$ as a tensor index, and then
antisymmetrize this expression with respect to $cd$. 

$R^i_{\ jkl}$ looks like it could have $d^4$ components in $d$ dimensions, but in fact they are not all 
independent, using the antisymmetry under $k\leftrightarrow l$, and also $i\leftrightarrow j$.  To make the 
latter precise, it is necessary to lower the first index and write $R_{ijkl}$ since it is only meaningful to
symmetrize or antisymmetrize under like indices. Furthermore, there is an identity $R_{ijkl}  = R_{klij}$
that further reduces the number of independent components.  In two dimensions, there is only one, hence 
the Riemann tensor carries
no further information than the Ricci tensor or scalar in that case.

The Riemann tensor contains the most detailed information about the curvature of a manifold, and perhaps the
best illustration is through the {\it geodesic deviation equation}, which describes how the distance between
two neighboring geodesics evolves as one follows them along equal distances.  To state the equation, we must
define the covariant derivative along a path, denoted by $D/Ds$.  Suppose we have a geodesic $x^a(s)$ with
tangent vector $\dot x^a = dx^a/ds$, where $s$ is the proper distance that parametrizes the geodesic.  Then
\be
	{D\over Ds} = \dot x^a\nabla_a
\ee
where $\nabla_a$ is the covariant derivative.  Suppose we have two neighboring geodesics, $x_1^a(s)$ and
$x_2^a(s)$, infinitesimally close to each other.  Their coordinate separation is $\Delta x^a = x_2^a-x_1^a$.
The geodesic deviation equation asserts that
\be
	{D^2\over Ds^2}\Delta x^a = -R^a_{\ ijk}\Delta x^j \dot x^i\dot x^k
\label{geo-dev}
\ee
where $\dot x^i$ can be either $\dot x_1^i$ or $\dot x_2^i$, since the two paths are supposed to be arbitrarily
close to each other.  In flat space, the geodesics would be straight lines and $\Delta x^a$ would be linear
in $s$, except in the special case where they are exactly parallel.  In curved space, geodesics will tend to
converge for positive curvature and diverge for negative curvature.   Eq.\ (\ref{geo-dev}) tells us that
this curvature effect could have different signs, depending upon which plane the neighboring geodesics lie in.

For example, suppose the geodesics are moving in the $z$ direction with $s=z$, and they lie in the 
$x$-$z$ plane.   Then 
\be
	{D^2\over Ds^2}\Delta x = -R^x_{\ zxz} \Delta x 
\ee
Hence the sign of $R^x_{\ zxz}$ determines whether these geodesics tend to move closer together or farther
apart.   If we considered two geodesics close to the same starting point but lying instead in the $y$-$z$ plane,
their behavior would be governed by $R^y_{\ zyz}$, which need not have the same sign as $R^x_{\ zxz}$.  Hence it
is possible to have opposite behaviors within the same manifold, even near the same point, depending on the
plane containing the two geodesics.  

The simplest example of the geodesic deviation equation is for two geodesics emanating from the north pole of a
unit 2-sphere with coordinates $(\theta,\phi)$:  $x_1^a = (\theta, \phi_1)$, $x_2^a = (\theta,\phi_2)$ for 
fixed $\phi_i$.  The separation vector is $\Delta x^a = (0,\phi_2-\phi_1)$, which is a constant.  Therefore 
its derivatives would vanish if we took ordinary derivatives, $d/ds = d/d\theta$ (notice that $s=\theta$ in this
example).  However, $R^\phi_{\ \theta\phi\theta} = 1$ for the unit 2-sphere.  Hence 
$D^2\Delta x^\phi/Ds^2$ does not vanish, 
but gets contributions from the Christoffel symbols appearing in $D/Ds$.

\subsubsection{Local geodesic (Riemann normal) coordinates}
The volumes and areas referred to in Eqs.\ (\ref{Ricci2},\ref{Ricci3}) are not arbitrary
coordinate volumes; they are bounded by geodesics.  Therefore one must know the geodesics not only for the 
centroid of these volumes, but also along their boundaries.  Typically, this is difficult to do in an arbitrary
coordinate system.  But there is a special choice of coordinates that one can make in the vicinity of a given
point $P$ with coordinates $x^a_P$, known as local geodesic coordinates, which makes it easy.  
In this system, the metric is made to be as close as possible
to flat Euclidean (or Minkowski) space, and the geodesics are straight lines, actually rays emanating from $P$.
Of course in a curved space, they only look straight because of the special choice of coordinates; their
separations correspond to distances that increase faster or slower than they would in flat space, if the
manifold is curved.  

The simplest demonstration of normal coordinates starts with the two-sphere or its negatively curved
counterpart, whose line elements are
\be
	ds^2 = dr^2 + \sin^2\!r\, d\phi^2 \hbox{\ or\ } \sinh^2\!r\, d\phi^2
\ee
I have used $r$ instead of $\theta$ here to emphasize the relation to polar coordinates in flat space near
$r=0$.  We could parametrize a more general metric with azimuthal symmetry, in the vicinity of $r=0$ by 
\be
	ds^2 = dr^2 + r^2(1 + a r^2 + \dots) d\phi^2\,,
\ee
where $a = \pm 1/3$ respectively for the $\sinh$ and the $\sin$ by Taylor expanding.  Clearly, there is a
continuum of geometries that interpolate between, or beyond, these two, by different choices of $a$.
Defining $F = (1 + a r^2)$ as the correction factor of $g_{\phi\phi}$ relative to flat space, the Christoffel
symbols are
\bea
	\Gamma^r_{\ \phi\phi} &=& -\sfrac12r(2F + r F') = -r(1 + 2 a r^2) \nn\\
	\Gamma^\phi_{\ r\phi} &=& \Gamma^\phi_{\ \phi r} = {F'\over 2F} + {1\over r} = {ar\over F} + {1\over r}\,.
\eea
Recall the geodesic equation along the radial direction: $\ddot x^i + \Gamma^i_{\ rr}\dot r\dot r = 0$.
Since $\Gamma^i_{\ rr}=0$,  rays moving in the $r$ direction are exact geodesics, that keep $\phi$ fixed.  In this coordinate system,
we know precisely how to bound the area referred to in Eq.\ (\ref{Ricci2}), and the volume in Eq.\
(\ref{Ricci3}).  Furthermore, it is straightforward to generalize this example to geometries that are not
azimuthally symmetric, or in higher dimensions.  For example, we can let $F = 1 + a r^2 f(\phi)$ to break the
azimuthal symmetry.  In higher dimensions, we can write
\bea
	ds^2 &=& dr^2 + r^2 F(r,\theta_i,\phi)d\Omega_n^2\,;\nn\\
	d\Omega^2_n &=& d\theta_1^2 + \sin^2 \theta_1( d\theta_2^2 + \sin^2 \theta_2(d\theta_3^2 + \dots +
\sin^2\theta_{n-2}(d\theta_{n-1}^2 + \sin^2\theta_{n-1}
	d\phi^2)\dots ) \label{n-sphere}
\eea
where $F = 1 +a r^2 f(\theta_i,\phi)$.  I do not know if this is the most general metric in normal coordinates,
but it at least illustrates a wide range of possibilities.

Riemann normal coordinates have many nice properties.  In the vicinity of $P$, the metric can be put into
Euclidean form, by going from spherical to Cartesian coordinates, plus quadratic corrections, that are related to 
the Riemann tensor evaluated at $P$ \cite{normal-coords}:
\be
	g_{ij} = \delta_{ij} + \sfrac13 R_{iklj}\,x^k x^l + O(x^3)\,.
\label{lgmet2}
\ee
 The Christoffel symbols vanish at
$P$ in  normal coordinates, up to linear in $x$ corrections.\footnote{One should ignore the $1/r$ term in 
$\Gamma^\phi_{\ r\phi}$ in polar/spherical coordinates which is related to the coordinate singularity at $r=0$
in these coordinates, and does not contribute to curvature.}   Hence their first derivative can be nonzero,
and is related to Riemann by \cite{Misner:1973prb}
\be
	\Gamma^i_{\ jk,l} = -\sfrac13\left(R^i_{\ jkl} + R^i_{\ kjl}\right)\,.
\label{nc1}
\ee
Similarly, by differentiating (\ref{lgmet2}),
\be
	g_{ij,kl} =  \sfrac13\left(R_{iklj} + R_{ilkj}\right)\,. 
\label{nc2}
\ee
The inverse relation is \cite{normal-coords}
\be
\label{nc3}
	R_{ijkl} = \sfrac12(g_{jk,il} + g_{il,jk} - g_{jl,ik} - g_{ik,jl})\,,
\ee
which apart from the prefactor one could guess from the symmetries of $R_{ijkl}$.  Another useful relation is 
for the volume element in normal coordinates,
\be
	\sqrt{{\rm det}\,g_{ij}} = 1 -\sfrac16 R_{ij}\, x^i x^j + O(x^3)\,.
\label{nc4}
\ee

\noindent\fbox{%
    \parbox{\textwidth}{%
{\bf A pathological geometry.}\ 
It would be interesting to see an example where the effects of curvature depend upon the direction
in which one is looking away from a point on a manifold.
To explore this in an interesting case where $R_{ij}$ is positive
along some directions and negative along others, consider the geometry
\be
	ds^2 = d\psi^2 + \sin^2(\psi)( d\theta^2 + \sinh^2(\theta)d\phi^2)
\label{posnegex}
\ee
with coordinates $(\psi,\theta,\phi)$ such that $\psi,\theta\in[0,\pi]$ and $\phi\in[0,2\pi]$.
The Ricci tensor is 
\be
	R_{ij}= 2\,{\rm diag}(1,\,-\cos^2\!\psi,\,-\cos^2\!\psi\sinh^2\!\theta)\,.
\ee
Hence if we measure small volumes oriented along the $\psi$ direction from some point, they should be smaller
than their Euclidean counterpart, and those along the $\theta$ or $\phi$ directions should be larger or equal,
depending on the values of $\psi$ and $\theta$.  The point $\psi=0$ might look like a convenient one to check, 
since from there, one can only move in the $\psi$ direction, and the metric seems to be in the form of normal 
coordinates there.  However, this is not the case: the internal $(\theta,\phi)$ manifold has negative curvature,
which goes as $-4/\psi^2$ as $\psi\to 0$.  This is a curvature singularity, which makes the manifold sick
at this point.  This demonstrates why the internal manifold should be such that the metric corresponds to flat space
in hyperspherical coordinates in the vicinity of $r=0$, in Eq.\ (\ref{n-sphere}).  

\quad Nevertheless, the metric (\ref{posnegex}) is sensible everywhere away from the singularities at $\psi=0$ and
$\psi=\pi$, and one could construct Riemann normal coordinates around such points to explore the question
mentioned above.
}}

\smallskip
Having seen the nice properties of normal coordinates, one would like to know how to construct them.  In
principle, it is straightforward \cite{Misner:1973prb}: for a given point $Q\neq P$ in the manifold, find the unit vector $\hat v$
pointing away from $P$ which is tangent to the geodesic passing through $Q$, and choose an orthonormal basis at
$P$ such that $\hat v = (x_Q,y_Q,z_Q)$, in the case of a 3-dimensional manifold, for illustration.  If $\lambda_Q$ is the distance from $P$ to $Q$, then assign the
coordinate $(x,y,z) = \lambda_Q  (\hat x_Q,\hat y_Q,\hat z_Q)$ to $Q$.  One sees that all points along 
this geodesic lie on
a straight line in these coordinates.  Therefore the geodesic equation says
$\Gamma^i_{\ j k}(x)\hat x_Q^j \hat x_Q^k = 0$,
since $\dot x^j = \hat x_Q^j$.   At $x^i=0$ (the point $P$), this implies $\Gamma^i_{\ j k}(0) = 0$, since 
$\Gamma^i_{\ j k}(0)\hat x_Q^j \hat x_Q^k = 0$ for all choices of $Q$.  Although it is easy to picture and to describe,
this method is not very practical to implement.

To actually construct the normal coordinates for a given metric and point $P$, 
the first step is to find a coordinate transformation that makes the Christoffel symbols vanish
at $P$.   The 
explicit transformation that does so is given by \cite{Ohanian:1995uu}
\be
	x'^a = \Delta x^a +\sfrac12\Gamma^a_{\ bc}(x_P)\Delta x^b \Delta x^c\,,
\label{lgct}
\ee
where $\Delta x^a = x^a - x^a_P$. In the new coordinate system, $\Gamma'^i_{\ jk}=0$ at $P$, and it
 deviates 
from zero linearly in $x'^a$ away from $P$.   However, it does not yet satisfy the condition that
$\Gamma'^i_{\ jk}x'^j x'^k = 0$ at points away from $P$, which is needed to get straight-line geodesics.
Assuming one has chosen $x'^a$ to be orthonormal at $P$ (so the metric is the Kronecker delta plus corrections of
order $(x'^a)^2$), then the remaining needed transformation is
\be
	x'^a = y^a - \sfrac16 \Gamma'^a_{\ jk,l}(0)\,y^j y^k y^l\,.
\label{nct}
\ee
where $\Gamma'^i_{\ jk}(x')$ is computed  in the $x'^a$ system.
This can be derived by keeping the $O(x'^2)$ corrections in $g'_{ij}$ and requiring 
$\Gamma^i_{\ jk}(y)y^j y^k = 0$ to vanish at $O(y^3)$. In fact, one finds that 
$\Gamma^i_{\ jk}(y)y^j y^k = 0$ to all orders in $y$ if $g'_{ij}$ is only quadratic in $x'$.
The process could be iterated to account for higher powers of $x'$ in $g'_{ij}$, that would generate
higher powers of $y^i$ in $\Gamma^i_{\ jk}(y)y^j y^k$.  In this way, one could successively
straighten the geodesics out to arbitrary distances.  However, significant computational power is needed to
go to higher powers, and some of the available Mathematica packages for GR computations are not coded efficiently enough to go beyond the
quadratic case.

\subsubsection{Other characterizations of curvature}

In two dimensions, one can say that a positively curved surface looks locally like a sphere, while a negatively
curved one looks like a saddle.  But these statements aren't obviously meaningful in higher dimensions.  
A more general characterization is that triangles in a positively curved space have interior angles that add up to
more than 180$^\circ$, while negatively curved spaces have the opposite behavior.  We must define what is meant by
triangle however. The geometrically sensible definition is to connect the three vertices by geodesics.  There is no problem to define the angles
close to the vertices since the space looks approximately Euclidean over infinitesimal distances; we can draw an infinitesimal
right triangle overlapping the large triangle at the vertex and determine its interior angle in the usual way, 
$\theta = \tan^{-1}\Delta y/\Delta x$ (being careful to use physical distances and not just coordinate intervals).  
In fact, we would not notice any excess in the angles for
an infinitesimal triangle, because the curvature effect involves the second derivative, hence requires going to second order in $\Delta x$
and $\Delta y$.  But for a large triangle the effect becomes obvious.  Take one vertex at the north pole and the other two on the equator,
at $\phi=0$ and $\phi=\pi/2$.  
The geodesics are great circles, so each angle is $90^\circ$, and they add up to $270^\circ$.

In problem 4.3 we saw that it is not always easy to determine the geodesics on a negatively curved surface.
Considering the metric (\ref{hyperg}), we could take one vertex at the origin, one at $(a,0)$ and the third at $(0,b)$.
The geodesics connecting the origin to the other two vertices simply lie along the coordinate axes, hence this vertex
has a right angle; but the geodesic
connecting $(a,0)$ to $(0,b)$ is not obvious.  Things simplify if $b=a$, as you will work out in problem 7.4.

A similar effect involves the relation between circumference and radius for circles in curved spaces.  A circle on drawn on a sphere
has a smaller circumference than in flat space, compared to the radius as measured along the sphere.  Again, the effect is opposite
if the curvature is negative.  The cone provides an interesting example of positive curvature, for a circle that contains the vertex.
The circumference is smaller than $2\pi R$ by the factor $\sin\alpha$, where $\alpha$ is the half-angle of the cone.  But locally, the
cone has Euclidean geometry.  All the curvature is concentrated at a single point, the vertex.  The Ricci scalar is proportional to
$\delta^{(2)}(\vec x)$, which has the right dimensions.
  This becomes clear if one replaces the
sharp tip with a smoothed-out geometry, whose curvature is obviously positive.  It is interesting that the circle knows about this 
enclosed curvature, even
though the space is completely flat at every point the circle traverses.  A circle on the cone that does not enclose the vertex has the 
usual flat-space properties.

\subsection{Symbolic manipulation}
Once you have done a few computations of Christoffel symbols and curvature tensors,
it starts to become tedious, especially in higher dimensions.  Packages have been written for 
symbolic algebra software such as Mathematica, which can save us the effort.  One package that
can be downloaded is called GR \cite{GR-math}.  Another is GRQUICK \cite{GRQ}, which has the
advantage of coming with some documentation!

To install Mathematica, go to
\href{https://mcgill.service-now.com/itportal?id=kb_article_view&sysparm_article=KB0010741}{this
site} and follow the instructions for Mathematica.  Download Mathematica onto your computer. Once
you have activated it, download the
\href{https://library.wolfram.com/infocenter/MathSource/8847/}{GR package} and unzip it to some
convenient directory.  Make a copy of the sample notebook 
``Schwarzschild.nb'' so that you can use it as a template for making modifications.

Open your sample notebook in Mathematica and run the first three commands.  You can delete the
next two groups of commands which are not needed for the example you can try to get started,
the unit two-sphere.  Redefine the signature to be $(1,1)$, and modify the assumptions as
appropriate for the coordinates $\phi$ and $\theta$.  In Mathematica, you could either spell out
the coordinate names, or use the escape key to make the Greek symbol, for example
{\esc f \esc}
for $\phi$ and {\esc q \esc} for $\theta$.   Skip the stress tensor computations and compute 
the Christoffel symbols, and then the Riemann tensor.  Verify that these give the expected 
Christoffel symbols for the two-sphere.

The commands for computing various other tensors depend upon where you want their indices to
appear.  For example, for the Ricci tensor, one has {\tt covRicciTensor}, {\tt conRicciTensor}
and {\tt mixRicciTensor}.  The Ricci scalar is just {\tt ricciScalar}.  The commands and variable
names in Mathematica are case-sensitive.
The GR package does not seem to come with covariant derivatives built in, but GRQUICK does.
On the other hand, coding your own covariant derivative with the provided Christoffel symbols is
a good (and not too difficult) exercise. 

\begin{figure}[t]
\leftline{\includegraphics[width=\textwidth]{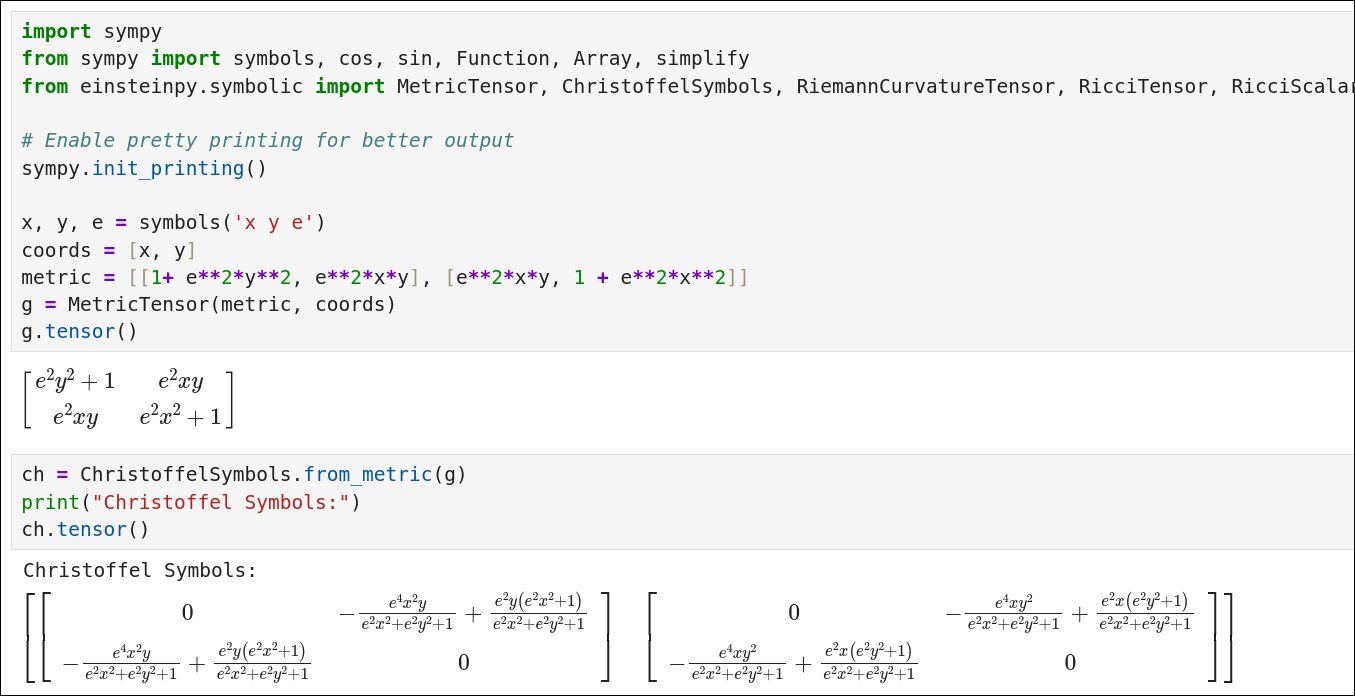}} 
\leftline{\includegraphics[width=\textwidth]{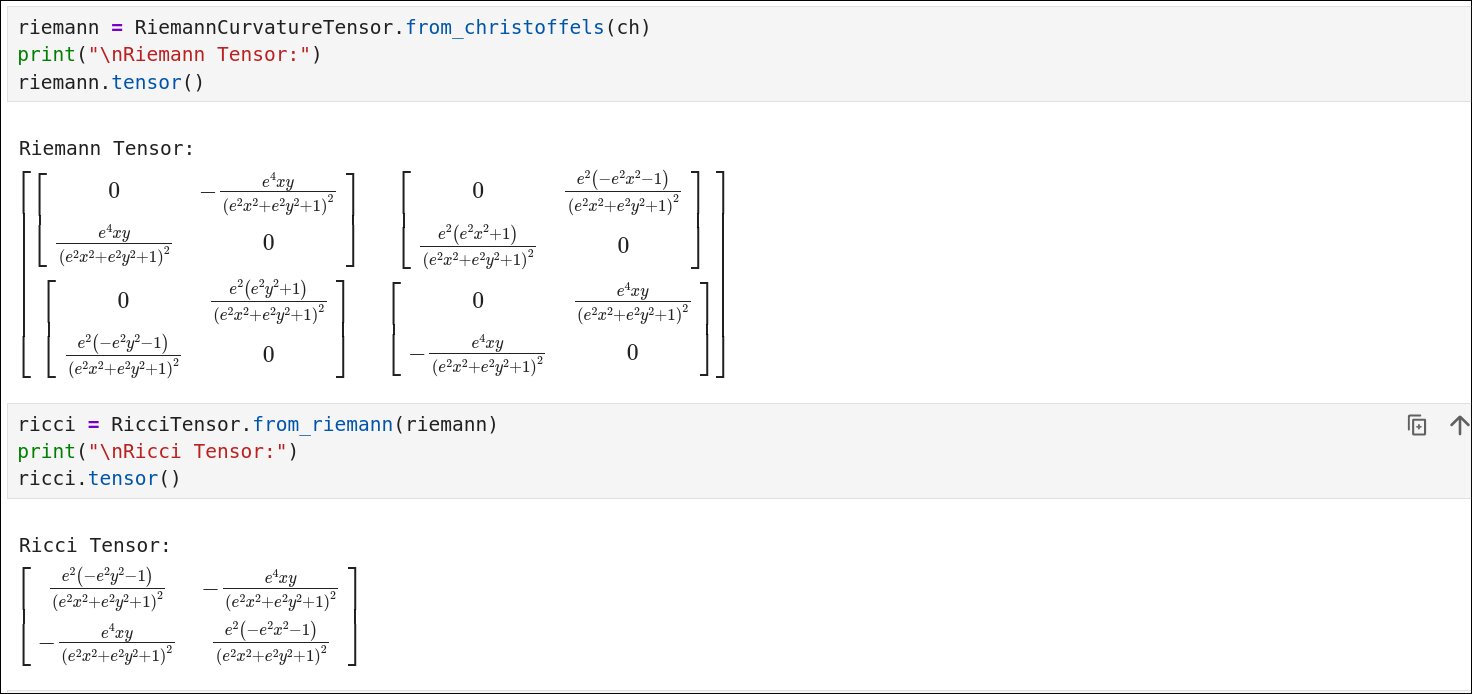}} 
%\leftline{\includegraphics[width=0.1\textwidth]{py3.jpg}}    
    \caption{Sample Python notebook for computing curvatures.  This
example has negative curvature everywhere.
}
    \label{fig:python_code}
\end{figure}

Since the above packages were developed, Mathematica has added its own native capabilities
for computing Christoffel symbols, the Riemann and Ricci tensors and the scalar $R$.
 For a nonproprietary alternative, Python offers symbolic manipulation tools, in particular 
the package EinsteinPy \cite{einsteinpy}.  
It is well documented and is likely to have steady development.  
An example is shown in Figure \ref{fig:python_code}.

\subsection{Problems}
7.1.  We would like to calculate the extra inhomogeneous term by which Christoffel symbols change under a coordinate transformation, 
in addition to the usual Jacobian factors that would be there for a tensor.  {\it I.e.}, the relation between $\bar\Gamma^i_{\, jk}$ in the $y$
coordinate system and $\Gamma^i_{\, jk}$ in the original $x$ coordinates is
\be
	\bar\Gamma^i_{\, jk}  = \hat J^i_a J^b_j J^c_k\, \Gamma^a_{\ bc} +\hat J^i_a J^a_{k,j}\,,
\ee
where $J^i_a = \partial x^i/\partial y_a$ is the Jacobian matrix, $\hat J^i_a = \partial y^i/\partial x_a$ is its inverse,
and $J^a_{k,j} = J^a_{j,k} = \partial^2 x^a/\partial y^j\partial y^k$.  It is not necessary to distinguish right from left indices
in $J$ since the upper and lower positions of the indices always tell us how they must be contracted.  \\
(a) First compute $\bar\Gamma_{ajk}$ using Eq.\ (\ref{gtrans}) for each of the three terms in $\bar\Gamma_{ajk}$.
It may be helpful to rewrite the bottom line of Eq.\ (\ref{gtrans}) in the form
\be
	J_{i,k}gJ_j + J_ig_{,k}J_j  +  J_igJ_{j,k}
\ee
to emphasize which factors of $J$ each lower index is associated with.  The unwritten indices are implied by matrix multiplication,
as a way of avoiding writing all the indices.  
Argue that we can ignore the terms where the derivative acts on $g$, since these will just go into the homogeneous transformation
of $\Gamma^i_{\ jk}$.  Further argue that the order of the matrices can be reversed in the above products.  Show that four of the
six terms cancel, leaving only two, which are equal to each other.\\
(b) Raise the first index to get the inhomogeneous contribution to the normal Christoffel symbol, using the coordinate-transformed 
inverse metric factor $(\hat J\hat g\hat J)^{ia}$.\\
(c) Now we want to prove that Eq.\ (\ref{covder}) does what it is claimed to do when we rewrite $\bar A^j = \hat J^j_k A^k$ to relate
the $y$ and $x$ coordinate systems.  Why does $A^j$ transform with $\hat J$ whereas the metric tensor transformed with $J$?
In order to deal with derivatives of $\hat J$, use the fact that $\hat J J$ is the unit matrix to relate the derivative of $\hat J$ to
the derivative of $J$.  Then show that the covariant derivative transforms covariantly, by the inhomogeneous terms canceling each other.
You are invited to look up explicit computations on the internet to see whether you agree that this way of organizing the calculation
is more transparent
than writing out all the implicitly summed indices.\\

\noindent 7.2. (a) Compute the nonvanishing Christoffel symbols for the unit sphere metric,
with $ds^2 = d\theta^2 + \sin^2\theta\, d\phi^2$.\\
(b) Compute the Ricci tensor.\\
(c) Compute the Ricci scalar.\\
%  Answer: (a)
%$\Gamma^\phi_{\ \phi\theta} = \Gamma^\phi_{\ \theta\phi} = \cot\theta$,
%$\Gamma^\theta_{\ \phi\phi} = -\sin\theta\cos\theta$.  (b) $R_{ij} = {\rm
%diag}(1,\sin^2\theta)$, which happens to be the same as the metric tensor.  (c) $R =
%g^{ij} R_{ij} = 2$.

\noindent 7.3. (a) Show that the metric of problem 4.3 can be obtained from the 2-sphere metric,
up to the overall sign and scaling, by letting $\theta \to i\theta$. (Notice that the overall sign of the metric does not affect the
Christoffel symbols.)  Do this analytic continuation 
on the Christoffel symbols you computed in problem 7.2 and compare to the results from 4.3.  How do you
explain the relative factors of $i$?\\
(b) Compute $R_{\theta\theta}$,  $R_{\phi\phi}$ and $R$, after removing any spurious factors of $-i$ from the Christoffel
symbols.\\

\noindent 7.4. (a) Derive the geodesic equations for the Christoffel symbols of Eq.\ (\ref{hypchrist}), keeping only
the leading order in $1/L^2$ contributions.\\
(b) We want to find the geodesic connecting the points $(a,0)$ and $(0,a)$.  
Show that the straight-line path solves the geodesic equations.  For definiteness, let it start at $(a,0)$ when $\lambda=0$ and go to
$(0,a)$ when $\lambda = 1$.\\
(c) If you draw the triangle on the $x$-$y$ plane, it looks like all the angles add up to $180^\circ$, but one must
remember that the distances are distorted by the metric.  By determining the interior angles from the physical distances
corresponding to the coordinate intervals $\Delta x$ and $\Delta y$, find the deficit angle by which they fail to add up 
to $\pi$ in radians.  Assume that $0<a\ll L$. \\

\noindent 7.5. (a) Rewrite the line element (\ref{hyperg}) of the hyperbolic space in polar coordinates,
$x= r\cos\theta$, $y = r\sin\theta$.\\
(b) We want to define a circle of physical radius $R$ centered at the origin.  This corresponds to a trajectory
$r(\theta) = R + \dots$ which is not constant.  Find the leading correction in $1/L^2$ to $r(\theta)$.\\
(c) By substituting this path into the line element, compute the circumference of the circle, to first order 
in $1/L^2$.  By what factor is it greater than $2\pi R$? \\

\noindent 7.6. (a)  Compute the components $A_{r;r}$, $A_{r:\theta}$, $A_{\theta;r}$,
$A_{\theta;\theta}$ of a vector in 2D polar coordinates.
Recall that $\Gamma^r_{\ \theta\theta} = -r$ and $\Gamma^\theta_{\ r\theta} = 1/r$.
\\
(b) Using the results of (a), find $A_{r;rr}$, $A_{r;\theta\theta}$, $A_{\theta;rr}$
and $A_{\theta;\theta\theta}$.\\
(c) Put these together to obtain the covariant Laplacian $g^{ij}\nabla_i\nabla_j A_k$ for 
the covariant
components of $\vec A$.  It is known as the vector
Laplacian.  Hint: you can consider $A_\theta$ to be dimensionless, and $A_r$ to have dimensions
of $1/r$, as a way of checking dimensions.  I get 
\bea 
\nabla^2 A_r &=& A_{r,rr} +
r^{-2}(A_{r,\theta\theta}- 2r^{-1}A_{\theta,\theta} + r A_{r,r} - A_r)\,,\nn\\
\nabla^2 A_\theta &=& A_{\theta,rr} +r^{-2}A_{\theta,\theta\theta} - r^{-1}A_{\theta,r} +
2r^{-1}A_{r,\theta}\,.
\eea
(This does not agree with Wolfram MathWorld, because the latter assumes a basis where the
$\theta$ unit vector is normalized to 1, whereas $\hat x^\theta\cdot\hat x^\theta =
g^{\theta\theta} = 1/r^2$ here.  In other words, $A_\theta$ is divided by $r$ in Wolfram to give it the
same dimensions as $A_r$.  See the box on coordinate versus orthonormal bases below.)\\

\noindent 7.7. (a) Using the results from Problem 7.2 (or looking up the needed quantities), compute $D\Delta^a/Ds$
for the geodesics discussed at the end of section \ref{geo-dev-sect}.  Show that only the $a=\phi$ component
is nonzero.\\
(b)  Compute $D^2\Delta x^a/Ds^2$, recalling that $\dot x^c \Delta x^b_{\ ;c}$ has three indices that must be
accounted for when taking the covariant derivative.  Verify that it matches $-R^\phi_{\ \theta\phi\theta}
\Delta x^\phi$ in accordance with the geodesic deviation equation.\\

\noindent 7.8. Consider the metric $ds^2 = dr^2 + r^2(1 + a r^2 \cos\phi)d\phi^2$ in the vicinity of
$r=0$.\\
(a) Using a symbolic manipulation package, compute the mixed Riemann tensor and
show that it is 
\be 
        R^\phi_{\ r r \phi}=3a\cos\phi,\ 
	R^\phi_{\ r \phi r} =-3a\cos\phi,\ 
\ee
at $r=0$.\\
(b) We would like to verify the geodesic deviation equation for two radial geodesics following nearby
trajectories $\vec x_i(r) = (r,\phi_0-\Delta\phi/2)$, $(r,\phi_0+\Delta\phi/2)$. Consider $\Delta\phi$ to be
infinitesimal.  Write the predicted result for
the right-hand-side of the equation.\\
(c) Compute the left-hand-side of the deviation equation to show that it agrees.  
Use symbolic manipulation
to compute the required Christoffel symbols.  Only keep the leading corrections to the Christoffel symbols
relative to their $r=0$ values since we are interested in the small-$r$ region.\\
(d) Compute the physical separation between the geodesics as a function of $r$, call it $\Delta s^\phi$.
Show that the geodesic deviation equation in the form $d^2\Delta s^\phi/dr^2 = - R^\phi_{\ r\phi r}\Delta
s^\phi$ is satisfied to leading order in $r$ where the ordinary derivative now appears on the left hand side.\\

\noindent 7.9. Consider the metric $ds^2 = (1 + ay^2) dx^2 + (1+ax^2) dy^2 -2a\, xy\, dx\, dy$.  It is in normal form,
so we can test the relations (\ref{nc1}-\ref{nc4}).\\
(a)  Use symbolic manipulation to compute the Riemann curvature and Christoffel symbols.  Show that Eq.\ (\ref{nc1}) is
satisfied for those Christoffel symbols whose first derivative is nonvanishing at the origin.\\
(b) Check Eq.\ (\ref{nc2}) for the three metric tensor elements.\\
(c) Verify the formula (\ref{nc4}) for the metric determinant.\\
(d) Compute the Ricci scalar, and use the result from (c) to verify Eq.\ (\ref{Ricci3}), changing to polar
coordinates.\\

\noindent 7.10.  The unit 3-sphere can be embedded into 4D Euclidean space with coordinates
 $(w,x,y,z)$ by the constraint $x^2 + y^2 + z^2 + w^2 = 1$.\\
(a) Show that the constraint is satisfied by parametrizing the Euclidean coordinates by
\be
	w = c_\psi\,,\quad x = s_\psi s_\theta c_\phi\,,\quad y = s_\psi s_\theta s_\phi\,,
	\quad z = s_\psi c_\theta
\ee
where $c$ and $s$ stand for cosine and sine of the subscripted angles $1\le \theta,\psi \le \pi$
and $0\le\phi \le 2\pi$. \\
(b) By computing the line element in Euclidean space, show that the induced metric is given by
${\rm diag}(1, s^2_\psi, s^2_\psi s^2_\theta)$ for the coordinates $(\psi,\theta,\phi)$. 
 Hint: start with $dx^2 + dy^2$ to isolate the $d\phi^2$ term, then add
$dz^2$ to get the $d\theta^2$ term.  Add $dw^2$ last.
  Find
the volume element and compute the volume of the 3-sphere.\\
(c) Plug the metric into a symbolic manipulation package and show that the Ricci curvature is $R=6$.
Also find $R_{ij}$.\\
(d) Compute the volume $V$ of a small region $\psi < \epsilon$ to $O(\epsilon^5)$
and compare it to the corresponding volume $V_E$ of a sphere of radius $\epsilon$ in flat (Euclidean)
space.  Thereby check Eq.\ (\ref{Ricci3}) for the 3-sphere.\\

\noindent 7.11. This problem is best done with symbolic manipulation.
 Consider the mixed-curvature metric $ds^2 = d\psi^2 + \sin^2\psi(d\theta^2 + \sinh^2\theta
d\phi^2)$.  At the point $P=(\psi,\theta,\phi) = (\pi/2, \theta_0, 0)$, it has Christoffel symbols
$\Gamma^\theta_{\ \phi\phi} = -\sfrac12\sinh2\theta_0$, $\Gamma^\phi_{\ \theta\phi} = 
\Gamma^\phi_{\ \phi\theta} =\coth\theta_0$.\\
(a) Compute $g_{ij}$ in the vicinity of $P$, to quadratic order in small deviations
$(\Delta\psi,\Delta\theta,\Delta\phi)$ away from $P$.  We will need
this later on.  Since the metric is block diagonal with only the last two components acting nontrivially,
use the notation $\bar g_{ij}$ to refer to the lower $2\times 2$ block, here and in the following.\\  
(b) Use Eq.\ (\ref{lgct}) to find the  coordinates $(\psi',\theta',\phi')$
based around $P$ where $\Gamma^i_{\ jk}(P)=0$, expressed in terms of $(\Delta\psi,\Delta\theta,\Delta\phi)$ and $\theta_0$.  What form does $g_{ij}$ in the original coordinates from part (a) 
take when the small deviations are rewritten in terms of the primed coordinates?\\
(c) Invert the relations between the coordinates to find $(\Delta\psi,\Delta\theta,\Delta\phi)$ in terms of 
$(\psi',\theta',\phi')$ to quadratic order, so that you can compute $\partial x^i/\partial x'^j$, and thereby
$g'_{ij}$, by using the tensor transformation of $g_{ij}$.  Show that the terms linear in $x'$ vanish so that
$g'_{ij}$ is a constant plus quadratic terms, and determine them.  Finally, rescale $\sinh\theta_0\phi' = f'$
to make the metric $g_{ij}=\delta_{ij}$ at $P$, and define $\cosh\theta_0 = K$.
Answer:
\be
	\bar g'_{ij} = \left(\begin{array}{cc} 1-\psi'^2& 0\\ 0 & 1+\theta'^2-\psi'^2 \end{array}\right) 
+ K^2 \left(\begin{array}{cc} f'^2 & -\theta'\phi'\\
	-\theta'\phi' & 2(f'^2  - \theta'^2)\end{array}\right)
\ee
(d)  In the new coordinates $(\psi',\theta',f')$, determine $\Gamma^i_{\ jk,l}(P)$ in order to carry out the 
coordinate transformation (\ref{nct}).   These are the normal coordinates around $P$.  Denote them by $y_i$ where
$i=(1,2,3)$ or $(\psi,\theta,f)$.  Show that the transformed metric is\\
\be
g_{ij} = \delta_{ij} + 
\frac13 \left(
\begin{array}{ccc}
   -y_2^2- y_3^2 &  y_1 y_2 & 
   y_1 y_3 \\
  y_1 y_2 & - y_1^2+ y_3^2 & -
   y_2 y_3 \\
  y_1 y_3 & - y_2 y_3 & -
   y_1^2+ y_2^2 \\
\end{array}
\right)
\ee
(d) Compute $\sqrt{g} =\sqrt{\det g_{ij}}$ to order $O(y_i)^2$ in order to find the volume $\int d^{\,3}y\sqrt{g}$
of 
a small pyramid along 
the $y_i$ direction, with extent $y_j, y_k \in [-\epsilon y_i/2, +\epsilon y_i/2]$ in the other
two (transverse) directions $j\neq k\neq i$.  Find the volume as a function of 
$r = y_i$ for small $r$, keeping corrections of order $r^2$ times the Euclidean result for the volume.
Show that the corrections vanish for $i=2,3$ (considering $\epsilon$ to be arbitrarily small), but not for $i=1$, and that they are in agreement with 
Eq.\ (\ref{Ricci2}), apart from a factor of 2 from being in 3 dimensions rather than 2.\\

\noindent\fbox{%
    \parbox{\textwidth}{%
{\bf Coordinate versus orthonormal bases.} You may be accustomed to seeing the gradient of
a scalar function in polar coordinates being expressed with components 
$(f_{,r}, r^{-1}f_{,\theta})$,
while in tensor calculus one would say that its components are simply 
$(f_{,r}, f_{,\theta})$.  The latter is the result we would get by transforming the Cartesian
components $(f_{,x},f_{,y})$ with the Jacobian matrix.  On the other hand, 
$(f_{,r}, r^{-1}f_{,\theta})$
is what one gets by transforming $(f_{,x},f_{,y})$ by a rotation.  The difference between the two
choices is the normalization of the basis vector in the $\theta$ direction.  If one does a
rotation, then $\hat\theta$ is a unit vector.  If one does a coordinate transformation, 
$\tilde\theta$ is not a unit vector, but instead has length $1/r$ (called a basis 1-form).  The rationale for doing this is
explained in the next chapter, but it has to do with the statement that
$\tilde\theta\cdot\tilde\theta = 1/r^2 = g^{\theta\theta}$.  Of course, the gradient as a geometric object, $\vec\nabla f$, is the
same regardless of the choice of basis.
}}

\section{Parallel Transport and Differential Geometry}
In the previous chapter we introduced the covariant derivative, but it turned out to
be unnecessary for our goal of quantifying curvature.  In fact we will see that the
covariant derivative {\it can } be used to derive the Ricci tensor, but first we
would like to get another perspective on the construction of $\nabla$.  Let's think about a vector field
living on a 2-sphere.  We will take basis vectors $\vec e_1 \propto \hat\theta$ and 
$\vec e_2
\propto \hat\phi$ at each
point, tangent to the sphere and pointing along the appropriate directions.  The
vector field is $\vec A = A^i\vec e_i$.  It is not right to say that it lives on
the sphere, since when viewed in an embedding space, only its tail can be on the
sphere; its head lies outside.  We say that it lives in the {\it tangent space} at each
point on the sphere, and the basis vectors span this tangent space.   It is a
Euclidean space, so if we want to take the dot product between two basis vectors, it
is not necessary to introduce any metric for contracting their tangent space indices;
this is just the normal dot product that we are used to.  However, it turns out to be
convenient to not insist that the basis vectors be orthonormal.  Rather,  we
can choose them such that
\be
	\vec e_i\cdot \vec e_j = g_{ij}\,;
\label{dot-prod}
\ee
hence in general they need not be unit vectors, nor even orthogonal.  We can view
$e_i^a$ (choosing some orthogonal directions in the tangent space for defining the
tangent space indices $a$---this choice is arbitrary) as being a kind of square root
of the metric tensor.  It is a matrix which when contracted with itself on the 
$a$ index gives $g_{ij}$.  Another name that is commonly given to it is an $n$-bein:
zweibein in two dimensions, fierbein in four, using the German names.  The reasons for
choosing this seemingly strange normalization of the basis vectors will become apparent
soon.

When taking derivatives of a vector field on a curved manifold, we have a complication not present in
flat space: not only are the components $A^i$ functions of position on the sphere, 
but so are the basis vectors.  In the usual definition of a derivative, we compute a
function at two nearby values and take its difference.  In curved space, that
operation becomes complicated by the fact that the basis vectors are changing
direction.  It is not a problem if we deal with the full vector $\vec A$ since then
the difference between its values at two different points is unambiguous, if we have a
way of expressing the rate of change of the basis vectors at a given point.  But it
now becomes apparent that this gives an additional contribution to the derivative of
the vector field, beyond the partial derivatives of its components $A^i$.  These extra
contributions are precisely what the Christoffel symbols encode.  They tell us about
the rate of change of unit vector directions as one moves on the manifold.

To be precise, when we take the difference between two nearby vectors and express it
as another vector, we have to decide which basis to use: the unit vectors defined at
one point or the other.  Alternatively, we can imagine transporting the vector at the
position $\vec x+ d\vec x$ back to the point $\vec x$ before subtracting, so that both
vectors are expressed in terms of the same basis vectors.  This is known as parallel
transport.  We could transport the vector in any direction, but let's consider doing
it along the $\hat x_i$ direction.  Then the difference of the two vectors after
transporting the second one back to the initial position $\vec x_0$ is 
\be
	\Delta \vec A = A^j(\vec x_0 + \epsilon \hat x_i)\,\vec e_j(\vec x_0 + \epsilon
\hat x_i) - A^j(\vec x_0)\,\vec e_j(\vec x_0)
\ee
with the understanding that $\vec e_j(\vec x_0 + \epsilon
\vec e_i)$ is to be expressed in terms of $\vec e_k(\vec x_0)$.  It must be some
linear combination, which in the previous chapter we saw is given by the Christoffel symbols:
\be	
   \vec e_j(\vec x_0 + \epsilon
\vec e_i) = \vec e_j(\vec x_0) + \epsilon \Gamma^k_{\ ij}\vec e_k(\vec x_0)
+ O(\epsilon^2)
\label{basis-change}
\ee
The result is the covariant derivative along the $\vec e_i$ direction,
\be
	\nabla_i\vec A = {\lim_{\epsilon\to 0}}{\Delta\vec A\over \epsilon}
	= \left(\partial_i A^k + \Gamma^k_{\ ij}A^j\right)\vec e_k
	\label{covDvec}
\ee
The Christoffel symbols show us what is the ``connection'' between two nearby basis
vectors, and thereby take into account this contribution to the differential 
change in the vector.  The simple result (\ref{basis-change}) is one reason for
normalizing the basis as in Eq.\ (\ref{dot-prod}). If we took an orthonormal basis, we would
have a different (and more complicated) expression for Eq.\ (\ref{basis-change}).  A special
case of Eq.\ (\ref{covDvec}) is when $\vec A = \vec e_j$, in other words $A^i = \delta^i_j$.
Then
\be
	\nabla_i \vec e_j = \Gamma^k_{\ ij}\vec e_k\,.
\label{basis-variation}
\ee

\noindent\fbox{%
    \parbox{\textwidth}{%
{\bf Conundrum.}  Consider the metric
\be
	g_{ij} = \left(\begin{array}{cc}1+\alpha y^2 & \alpha x y\\ \alpha x y & 1 + \alpha x^2
\end{array}\right)
\label{con-met}
\ee
with $\alpha$ taken to be small.  To first order in $\alpha$, we can construct the basis vectors
\be
	\vec e_x \cong (1 + \sfrac12\alpha y^2)\,\hat x + \sfrac12\alpha x y\, \hat y,\qquad
	\vec e_y \cong (1 + \sfrac12\alpha x^2)\,\hat y + \sfrac12\alpha x y\, \hat x\,,
\label{basis-vecs}
\ee
that give $\vec e_i\cdot\vec e_j = g_{ij} + O(\alpha^2)$.  The nonvanishing Christoffel symbols are
$\Gamma^x_{\ xy} \cong \alpha y$ and $\Gamma^y_{\ xy} \cong \alpha x$ at leading order.  Then in 
contrast to Eq.\ (\ref{basis-variation}), we find that
\be
	\partial_y \vec e_x \cong \alpha y\,\hat x + \sfrac12\alpha x\,\hat y \cong \Gamma^x_{\ xy}\vec
e_x + \sfrac12\Gamma^y_{\ xy}\vec e_y + O(\alpha^2)\,,
\ee 
and similarly for $\partial_x\vec e_y$ (just interchange $x\leftrightarrow y$).  Moreover,
\be
	\partial_x \vec e_x = \sfrac12\alpha y\, \hat y
\ee
while $\Gamma^i_{\ xx} = 0$ exactly.  The discrepancies cannot be attributed to using $\partial_i$ instead of 
$\nabla_i = \vec e_i \cdot \vec\partial$, since the difference is higher order in $\alpha$. 
What is going wrong?  The problems seem to stem from the off-diagonal metric components, and the consequent
nonorthogonality of $\vec e_x$ and $\vec e_y$.  Hint: show that $\nabla_a \hat x$ and $\nabla_a \hat y$ are
nonvanishing, and take the right values to remove the discrepancies.
}}

\subsection{General tensor fields}
In the previous derivation, it was convenient to make use of the vector field $\vec A
= A^i\vec e_i$ that includes the basis vectors.  Since all its indices are contracted,
it is a coordinate-independent object, unlike its components $A^i$.  We can make a
similar definition for higher index tensors: just contract all (contravariant) indices
with basis vectors.  It is a geometric object analogous to vectors, although harder to
picture.  What if the tensor has covariant indices?  It is always possible to switch
between one and the other by raising or lowering with $g^{ij}$ or $g_{ij}$,
respectively.  The same must be true for the basis vectors.  There is a dual space
spanned by objects 
\be
	\tilde\omega^i = g^{ij}\vec e_j
\ee
that we call ``basis 1-forms.''  They  have the properties
\be
	\langle\tilde\omega^i,\,\vec e_j\rangle = \delta^i_j\,;\quad \tilde\omega^i\cdot\tilde\omega^j
	= g^{ij}\,.
\label{dual-dot-prod}
\ee
(The notation $\langle \tilde a, \vec b\rangle = a_ib^i$ is taken from MTW \cite{Misner:1973prb}, which 
distinguishes the inner product of a 1-form and a vector from that of two one-forms or two vectors.)
This is another reason that it is convenient to define the basis vectors as having
inner products (\ref{dot-prod}) rather than being orthonormal.  With the latter
choice, $\vec\omega^i\cdot\vec\omega^j$ would be the product of two inverse metrics
(and how would we contract the indices correctly?) rather than the intuitive expression in (\ref{dual-dot-prod}).

We can have higher-dimensional $n$-forms by contracting a covariant tensor
$A_{ijk\dots}$ with $\tilde\omega^i\tilde\omega^j\tilde\omega^k\dots$.  There is no
essential difference between the $n$-form and the $n$-component tensor; they are
geometric objects that do not depend on how we express them in coordinates.  However
it is natural to think of forms and tensors as objects that can be combined to give
scalars.  For instance a vector $\vec A$ combines with a 1-form $\tilde B$ as
$\langle\tilde B\,,\vec A\rangle  = A^i B_j \,\langle\omega^j,\,\vec e_i\rangle = A^iB_i$.

From the formula (\ref{covder3}) for covariant differentiation of covariant index tensors, we can
deduce the covariant derivative of a basis one-form:
\be
	\nabla_i \tilde\omega^j = -\Gamma^j_{\ ik}\tilde\omega^k\,.
\ee

\subsection{Differential Geometry}
Further motivation for our choice of coordinate bases rather than orthonormal ones comes from the intuition that the covariant
derivative of a scalar function $f$ is a one-form, $\nabla f =
f_{,i}\,\tilde\omega^i$, whose components are just the partial derivatives.  If we could
arbitrarily choose how to normalize the basis 1-form, the definition of the covariant
derivative would be ambiguous.  To motivate why it makes sense to have basis vectors
or forms that are not just unit vectors, consider again the line element
\be
	ds^2 = g_{ij}\,dx^i dx^j\,.
\ee
Suppose we defined the basis 1-forms so that they pointed in the direction of $dx^i$,
and their length was 1 unit as measured in the manifold.  For example, on the
2-sphere, it would give $2 = g_{\theta\theta}\, 
\tilde\omega^{\theta}\cdot\tilde\omega^\theta +
g_{\phi\phi}\,\tilde\omega^\phi\cdot\tilde\omega^\phi$, since we are summing over two
basis directions.  This agrees with the normalization in (\ref{dual-dot-prod}).
Differential geometers use the following shorthand for this logic:
\be
	\tilde\omega^i = dx^i
\label{omegadef}
\ee
To the unitiated, it looks mysterious, but it is just notation.  Master it, and you
will be admitted to the exclusive club.  What about the basis vectors?  Obviously
their normalization has to do with rewriting the line element as $ds^2 = 
g^{ij}\,dx_i dx_j$.  But what do we mean by $dx_i$?  It is the inverse of $dx^i$,
in other words $\partial_i = \partial/\partial x^i$.  Therefore
\be
	\vec e_i = \partial_i
\ee
is the counterpart to Eq.\ (\ref{omegadef}).  The inverse relationship of the two
kinds of objects is written by differential geometers as $\partial_i dx^j =
\delta^j_i$, or equally well $dx^j\partial_i  = \delta^j_i$.  They are both equivalent
to the first relation in Eq.\ (\ref{dual-dot-prod}).  If you are bothered by this
notation, don't worry.  Just keep telling yourself ``it is true, it is true, what an
ingenious system'' and you will eventually get used to it.  Section 9.2 of MTW \cite{Misner:1973prb}
gives a more persuasive, but also entertaining explanation:\\

\centerline{\includegraphics[width=0.8\textwidth]{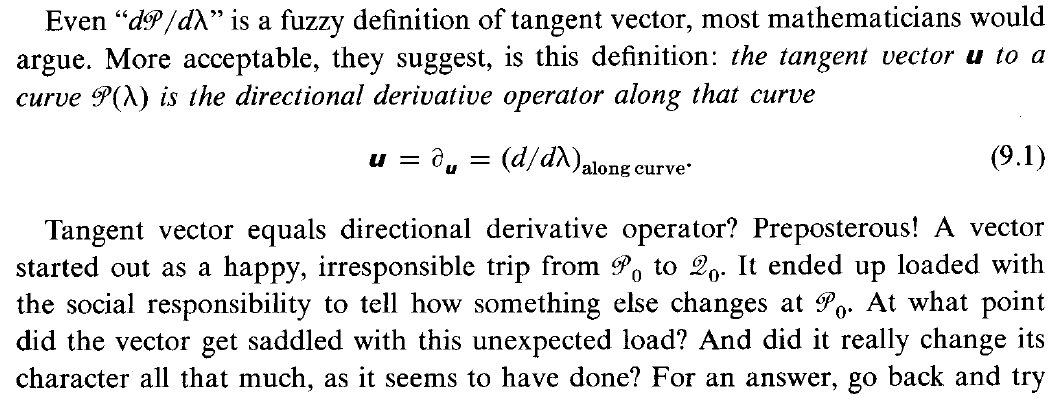}\raisebox{0.6em}{$\!\!\!\dots$}}

\smallskip
\noindent In short, 
$\partial_i$ has a more rigorous mathematical definition than the intuitive idea of a basis vector
inherited from flat space thinking.

An advantage of differential geometry is that it allows in many cases derivations
that are simpler and more efficient, alleviating the need to write out indices.
Although I am not giving an adequate introduction to differential geometry in this
course, it can be a powerful calculational tool, 
avoiding the cumbersome aspects of index notation.

It should be remarked that not all forms are {\it differential} forms.
For example, the metric 2-form cannot be directly integrated since
terms such as $\int h_{xx}(dx)^2$ are not meaningful.  On the other hand,
the electromagnetic field-strength 2-form $F_{\mu\nu}dx^\mu dx^\nu$ can
be integrated on a surface.  For example $\int dx\, dy\, F_{xy}$ is the
flux of a magnetic field $B_z\hat z$ through the $x$-$y$ plane.  In
differential geometry, the products $dx^\mu dx^\nu$ are specified as
wedge products, $dx^\mu\wedge dx^\nu$, which are explicitly
antisymmetric.  Another important example is the volume 4-form
$\sqrt{|g|}\epsilon_{\alpha\beta\gamma\delta}\,dx^\alpha dx^\beta
dx^\gamma dx^\delta/4!$.  It can be contracted with a vector-field
$n^\alpha$ to make a 3-form
$\sqrt{|g|}n^\mu\epsilon_{\mu\alpha\beta\gamma}$ which gives the
volume of a region orthogonal to $n^\mu$.

\subsection{Small closed paths for measuring curvature}
The notion of parallel transport can help us to derive the curvature of the manifold
from covariant derivatives.  Suppose we have a vector field $A^i$, which for 
simplicity
is taken to have constant components.  We will transport it from the origin to a
nearby point $x^i = (\epsilon_1,\epsilon_2,0,\dots,0)$ in two steps, first along the $x^1$
direction, then along $x^2$.  This will give rise to a deviation $\delta_{12}\vec A$
relative to its value at the origin.  Then we will repeat the operation along the path
starting in the $x^2$ direction, followed by $x^1$.  On a flat space, the result would
be the same but in general the second deviation $\delta_{21}\vec A$ is not
the same.  We will find out that the difference is given by
\be
	\delta_{12}\vec A - \delta_{21}\vec A =  \epsilon_1\epsilon_2 R^i_{\ j 2 1}A^j \vec
e_i
	\label{del_curv}
\ee
where $R^i_{\ klj}$ is the previously introduced Riemann curvature tensor (\ref{riemann}).
One can
derive Eq.\ (\ref{riemann}) from Eq.\ (\ref{del_curv}) for an arbitrary vector
field.  It can be written more generally as 
\be
	A_{j;kl} - A_{j;lk} =  R^i_{\ jkl}A_{i}\,,
\label{Riemann-definition}
\ee
underscoring the fact that, unlike partial derivatives, covariant derivatives do not commute,
at least not on a curved manifold.

An analogous situation exists in
electrodynamics.  It has to do with transporting an electron along different paths in a
background electromagnetic field.  The covariant change in the electron wave function
$\psi$ along the 1-2 path is $(\epsilon_2 D_y)(\epsilon_1 D_x)\psi$, where $D_i$ is the EM
covariant derivative, $D_\mu = \partial_\mu - i A_\mu$ (Eq.\ (\ref{covderEM})).  Then the
difference between the two paths is given by
\be
	(\delta_{12} - \delta_{21})\psi = \epsilon_1\epsilon_2 (D_y D_x - D_x D_y)\psi
	= -i\epsilon_1\epsilon_2(\partial_y A_x - \partial_x A_y)\psi
	= -i\epsilon_1\epsilon_2(F_{yx})\psi
\ee
where $F_{yx} = B_z$ is the field strength tensor.  This gives the famous result that when
transporting an electron around a closed loop, it gains a phase equal to the magnetic flux
enclosed by the loop.   In gravity, the commutator of the covariant derivatives is related 
to the curvature of the manifold.  A difference is that the EM connection $A_\mu$ is just a
number, which commutes with itself, whereas $\Gamma_{\mu\alpha\beta}$ is a matrix (in the
extra indices $\alpha,\beta$), which does not commute with itself; hence the commutator
contains $\Gamma$ quadratically, with no derivatives, in addition to the linear term
that is differentiated.
It is thus a straightforward calculation to arrive at (\ref{del_curv}), once we realize that
it comes from the commutator of the covariant derivative,
$	\delta_{12}A^\mu - \delta_{21}A^\mu =  \epsilon_1\epsilon_2 \left(A^\mu_{\ ;12} - 
	A^\mu_{\ ;21}\right)$.  I leave it as an exercise for you. 

\subsection{Lorentzian manifolds}
For simplicity, we have considered purely spatial manifolds for the above discussion.
Our ultimate interest is to apply this formalism to spacetime.  How do these results change
when we replace the spatial indices $i,j,k,\dots$ by Lorentz indices $\mu,\nu,\alpha,\dots$?
The answer is simple: {\it they do not change}.

\subsection{Problems}

\noindent 8.1. The metric tensor is a 2-form that can be written as 
$\tilde g = (\vec e_i\cdot \vec e_j)\, \tilde\omega^i\tilde\omega^j$.
From this prove that $\nabla_a \tilde g = 0$.  Use the covariant
derivatives of the basis vectors and 1-forms in terms of Christoffel
symbols and show that they cancel.\\

\noindent 8.2. Carry out the computation in Eq.\ (\ref{Riemann-definition}) to derive the 
explicit form of the Riemann tensor.  Suggestion: first write out $A_{j;kl}$.  Then show that the terms
involving one or two derivatives of $A$ are symmetric under $k\leftrightarrow l$ so they don't contribute to the
commutator.\\

\noindent 8.3. Consider the weak-field metric $ds^2 = (1+2\phi)dt^2 - (1-2\phi)d\vec x^{\,2}$ around a spherically
symmetric source where $\phi = -GM/r$. However we will not assume $\phi$ takes this particular form, only that it is
time-independent.  Put this metric into a symbolic manipulation program, writing $\phi$ as
$\phi[x,y,z]$.  Insert a power counting factor $e$ 
multiplying $\phi$ so that you can expand to linear order in
the perturbation.  (In Mathematica, this could be done using the 
\texttt{Series[argument,\{e,0,1\}]} operation.  Mathematica can accept
entire tensors as the argument, but sympy/Einsteinpy can only compute
Taylor series on individual components of tensors.  For example if
you have extracted the [1,1] component of the metric as
\texttt{g$\_$xx}, then \texttt{g$\_$xx.series(e,sympy.0,n=1)} will
expand it to first order in \texttt{e}.
)\\
(a) Compute some covariant components $R_{\mu\alpha\nu\beta}$ of the Riemann tensor to lowest order in $\phi$ and show that they 
happen to agree with Eq.\ (\ref{nc3}), even though we are not in a Riemann normal coordinate system.  You don't
need to check every component, just enough representative ones to convince yourself that it works.\\
(b) Now assume the solution  $\phi = -GM/r$, and a geodesic that is heading toward $r=0$ along the
$z$ axis, with a speed $|\dot z| \ll 1$.  A nearby geodesic is separated from the first one by a 
distance $\Delta x$ in the $x$ direction.  Consider $\Delta x$ to be small so that the two geodesics are
approximately parallel.  Find the dominant contributions in the geodesic deviation equation to determine
$\Delta \ddot x$, and show the relevant components of the mixed Riemann tensor $R^\mu_{\ \alpha\nu\beta}$.  Supposing the two geodesics represent worldlines of particles that are part of a semi-rigid
body, what is the physical interpretation of your result?\\
(c) Repeat the exercise for two geodesics that are lying on top of each other, but separated in the $z$
direction by $\Delta z$.  {\it I.e.,} one particle is ahead of the other.  Explain the qualitative difference
relative to the previous situation.\\

\noindent 8.4. Fill in the details to resolve the ``conundrum''
described below Eq.\ (\ref{basis-variation}).

\section{Nonlinear gravity}

Throughout much of this course, we have discussed parallels between gravitation and
electromagnetism.  First, we noticed that both entail $1/r^2$ forces.  They both lead to
radiation, which even though is different in detail, can be derived from the same basic
d'Alembertian operator $\partial_t^2 - \nabla^2$ and its associated retarded Green's
function.  Both theories have a connection: the gauge field $A^\mu$, or the Christoffel
symbol $\Gamma^\mu_{\ \alpha\beta}$, which are used to construct covariant derivatives.
Both theories are based on local symmetries.  For EM, it is gauge symmetry, the transformation
of a charged field by a phase $e^{i\theta(x)}$ that can be an arbitrary function of
coordinates.  For gravity, it is symmetry under coordinate transformations, $x^\mu\to
x^\mu(y^\mu)$.  We will now exploit another similarity to try to guess what is the correct
equation of motion for the full nonlinearized metric in GR, by following the example from EM.  
Our approach will be to use the principle of least action: equations of motion correspond to
stationary points of the action.  

\subsection{Variational principle: example of electromagnetism}

The first task is to figure out what is the action that corresponds to a given field.  We
want it to be invariant under the symmetry transformation, since it is a physically meaningful
quantity (kinetic minus potential energy).  For EM, $A^\mu$ is not gauge invariant, but the
field strength $F_{\mu\nu}$ is.  The action should also be Lorentz invariant, so we must
contract the indices.  For EM, there is no trace since $F_{\mu\nu}$ is antisymmetric;
the simplest possibility is to contract it with itself.   Therefore the Lagrangian 
{\it density}, since it must be integrated over space as well as time, is proportional to 
$F_{\mu\nu} F^{\mu\nu} = 2(\vec B^2-\vec E^2)$.  The right normalization turns out to be
\be 
	S_{EM} = \int d^{\,4}x\, {\cal L} = -\sfrac14 \int d^{\,4}x\, F_{\mu\nu} F^{\mu\nu}
	= \sfrac12 \int d^{\,4}x\, (\vec E^2-\vec B^2)
\label{EMaction}
\ee
If we regard $\sfrac12 \vec E^2$ as the kinetic energy density and $\sfrac12 \vec B^2$
as the potential energy density, then the total energy density is $\sfrac12(\vec E^2 + \vec
B^2)$, in agreement with what we learn in EM.
One way to understand why $\vec E^2$ is kinetic energy is to recall that $\vec E = -\vec\nabla
A^0 + \partial_t \vec A$.  We could work in the gauge $A^0=0$.  Then $\vec E$ looks like a
``velocity'' associated to the ``coordinate'' $\vec A$.  On the other hand, $\vec B$ contains
only spatial derivatives of $\vec A$ (it is the curl), so it cannot lead to a kinetic energy.

The action (\ref{EMaction}) so far applies only to EM fields in the vacuum, without the 
presence of any sources.  There is an additional term for coupling the EM field to charged
matter, represented by the current 4-vector $J^\mu = (\rho, \vec \jmath)$, where $\rho$ is the
charge density and $\vec\jmath$ is the current density.  This extra term is simply 
\be
	S_{\rm source} = - \int d^{\,4}x\, J^\mu A_\mu\,.
\ee
Under a gauge transformation, $A_\mu\to A_\mu + \partial_\mu\theta$, the extra term that is
generated can be integrated by parts to give $\int d^{\,4}x\,\partial_\mu J^\mu$, which 
vanishes because of current conservation,
\be
	\partial_\mu J^\mu = \dot\rho - \vec\nabla\cdot\vec\jmath = 0\,,
\ee
It says that charge density can only change if the charge is transported by an appropriate
current.  The integrated form is that the time derivative of the charge contained inside some
closed surface is equal to the flux of current flowing through the surface.

Once we have the action, the equations of motion can be derived by looking for its stationary
points, like in classical mechanics.  One takes the variational derivative of $S$ with respect
to the dynamical field.  In classical mechanics, this field would be the position $\vec x_a(t)$
of some particles with positions $
\vec x_a(t)$ (here $a$ labels which particle), and the variational derivative would be defined
as
\be
	{\delta x^i_a(t)\over \delta x^j_b(t')} =  \delta_{ab}\,\delta_{ij}\,\delta(t-t')\,.
\label{funderiv}
\ee   
We would like to generalize this to fields such as the gauge field $A^\mu(x)$ or the metric
$g_{\mu\nu}(x)$.  They represent an continuously infinite number of degrees of freedom,
 now labeled by position in space.  The generalization of (\ref{funderiv}) is
\be
	{\delta A^\mu(x)\over \delta A^\nu(y)} =  \delta^\mu_\nu\,\delta^{(4)}(x-x')\,.
\label{funderiv2}
\ee  

Now we can compute the variation of the action.  It is convenient to first rewrite
$F_{\mu\nu}F^{\mu\nu} = 2F_{\mu\nu}\partial^\mu A^\nu $, using the antisymmetry of the
field strength tensor.  When we vary this quantity, we can just vary the factor 
$\partial^\mu
A^\nu$ and multipy by 2, instead of separately considering the variation of $F_{\mu\nu}$, since 
we know that its variation is just the same as that of the first factor of $F^{\mu\nu}$, 
apart from trivial reshuffling of indices.  Then
\bea
	{\delta S\over \delta A^\alpha(x')} &=& -\int d^{\,4}x\, 
[\partial^\mu \delta^{(4)}(x-x')\delta^\nu_\alpha]F_{\mu\nu} + J_\mu
\delta^{(4)}(x-x')\delta^\mu_\alpha\,\nn\\
 &=& \partial^\mu F_{\mu\nu} - J_\nu = 0
\eea
where we used integration by parts and ignored boundary contributions for the first term.
Decomposing it into time and spatial components, we obtain Gauss's Law and Amp\`ere's Law,
$\vec\nabla\cdot\vec E = \rho$ and $\vec\nabla\times\vec B = \vec\jmath$.  The remaining two
Maxwell's equations, the magnetic Gauss's Law $\vec\nabla\cdot\vec B = 0$ and Faraday's Law
$\vec\nabla\times\vec E + \partial_t\vec B = $ are contained within the relativistic
identity
\be
	\partial^\alpha F^{\mu\nu} + \partial^\nu F^{\alpha\mu} +\partial^\mu F^{\nu\alpha} =
0\,.
\ee
It follows from the definition of $F^{\mu\nu} = \partial^\mu A^\nu - \partial^\nu A^\mu$ rather than the action; hence it is considered
to be an equation of constraint, rather than a dynamical equation.

\subsection{Einstein's equations}
\label{EEqs}
We would like to imitate the above logic to motivate the form of the action for general
relativity.  The first step is to identify quantities that are invariant under coordinate
transformations, and which contain the expected number of derivatives in order to yield a
second order partial differential equation for the metric, as we know from the linearized
theory must be the case.  We learned that the Ricci scalar $R$ is such a quantity, hence it
appears to be a good candidate for building the Lagrangian.  Recall that it is the trace of
the Ricci tensor $R_{\mu\nu}$.   Another invariant we could construct, which superficially 
resembles the EM Lagrangian, is $R_{\mu\nu}R^{\mu\nu}$.  But this contains four powers of
derivatives, so it would lead to a higher-order differential equation.  $R$ is the only
possibility that has just two.    So we expect that $R$ will play a similar role to
$F_{\mu\nu}F^{\mu\nu}$.

For the source term, analogous to $J_\mu A^\mu$, one might expect the stress-energy tensor
$T_{\mu\nu}$ to be present.  But there is an important difference between $T^{\mu\nu}$
and its EM analogy $J_\mu$.  In EM, we could consider $J_\mu$ to be independent of $A^\mu$
when taking the variational derivative.  But in gravity, $T^{\mu\nu}$ itself depends  
on the metric.  Therefore we might not necessarily expect the source term in the 
gravitational action to be expressed as some simple simple function of $T^{\mu\nu}$ such as
its trace (see however problem 9.1). Instead, we will assume that the part of the action that depends on the
matter-energy content is some expression $S_m$, that we can compute in specific examples, 
and as suggested by Hilbert, its variational derivative {\it defines} the stress-energy tensor
through
\be
	{\delta S_m\over \delta g^{\mu\nu}(x)} = {\sqrt{|g|}\over 2}\,T_{\mu\nu}(x)\,.
\label{Tdef}
\ee
To see that this makes sense, consider a nonrelativistic particle of mass $m$ moving on the trajectory $\vec
x(t)$ in a potential $V$.\footnote{The concept of an arbitrary form of potential energy is not strictly speaking
relativistic, but in the present context where we have restricted from full general covariance to separate space and
time diffeomorphisms, it is consistent.}   For simplicity, I will assume $g^{0i}=0$.  
Its action can be written as
\be
	S_m = \int dt\,\sqrt{g_{00}}\,\left(-\sfrac12 m\, g_{ij}\,g^{00}\,\partial_t
x^i\partial_t x^j - m -
V(x)\right)
\label{mat-ex}
\ee
if it is moving at nonrelativistic speeds (recall that $-g_{ij}$ is positive in our metric
convention).  $S_m$ is invariant under restricted coordinate
transformations $t\to t(t')$, $\vec x\to \vec x(\vec y)$ that do not mix space and time.
Using (\ref{Tdef}), we find that
\be
	T_{00}(\vec x') = g_{00}{\delta(\vec x-\vec x')\over \sqrt{|g_{ij}|}}\left(\sfrac12 m \dot
x^2 + V(x)+ m\right)
\label{T00def}
\ee
where $\dot x^2 = -g_{ij}\,g^{00}\,\partial_t
x^i\partial_t x^j$ represents the physical velocity squared.  The presence of $g^{00}$ in the
kinetic term causes the relative sign change between the kinetic and potential energies when
we vary with respect to $g^{00}$.  The factor of $\sqrt{|g_{ij}|}$ in the denominator is needed
to cancel the variation from the delta function when we do purely spatial diffeomorphisms
$\vec x\to \vec x(\vec y)$, under which $T^{00}$ should be invariant.  It may at first be surprising that
there is an overall factor of $g_{00}$ in front, but this makes sense when one remembers that $T_{00}$ is a
component of a tensor, whereas the factors following $g_{00}$ in Eq.\ (\ref{T00def}) are invariant under 
reparametrizations of time.  We did not see this factor previously since we were working in Minkowski space when we first
derived the stress-energy of a point particle.  On the other hand,
the spatial components of the stress tensor are
\be
	T_{ij}(\vec x') = {\delta(\vec x-\vec x')\over \sqrt{|g_{ij}|}}\, {p_i p_j\over m}
\label{Tijdef}
\ee
where $p_i = m\sqrt{g^{00}}\,g_{ij}\partial_t x^j$ is the momentum.  So the definition (\ref{Tdef}) gives
us the expected result for a point particle.  In Problem 9.1 you will fill in the missing steps in the above
derivations.

There is an additional complication that did not occur for EM.  Namely, the integration measure
$d^{\,4}x$ is not invariant under coordinate transformations.  Under $x^\mu = x^\mu(y)$, it
transforms by the Jacobian determinant,
\be
	d^{\, 4}x = \left|{\partial x^\mu\over \partial y^\nu}\right|\, d^{\, 4}y\,.
\ee
This is easily remedied by considering how the metric tensor transforms, and hence its
determinant.  Define the Jacobian matrix $J^\mu_{\nu} = \partial x^\mu/\partial y^\nu$.
Then
\bea
	g_{\mu\nu}(x) &\to& {J^{-1}}^{\alpha}_\mu\, {J^{-1}}^{\beta}_\nu\, 
	g'_{\alpha\beta}(y)\,,\nn\\
	g = {\rm det}\,g &\to& J^{-2}\,g'
\eea 
using the product rule for determinants.  Therefore the invariant metric is
$\sqrt{|g|}\,d^{\,4}x$, with the absolute value since the determinant is negative for
metric signature $(1,-1,-1,-1)$.   The analogous result for the 2-sphere is a well-known example,
since the area element is $\sqrt{|g|}\,d^{\,2}x = R^2\sin\theta\,d\phi\,d\theta$ in spherical
coordinates.  The physical volume element in 4D is $dV = \sqrt{|g|}\,d^{\,4}x$, so it is
natural that this is the combination appearing in the action.

By this reasoning, one can deduce that the action for gravity, known as the 
Einstein-Hilbert action, takes the form\footnote{Of course, $\kappa$ here turns out to match the value
$8\pi G$ we have assigned to it previously.  The minus is a consequence of our choice of $(+,-,-,-)$
metric signature.}
\be
	S_g = -{1\over 2\kappa}\int d^{\,4}x\sqrt{|g|}R
\ee
for some constant $\kappa$, that must be experimentally determined.  The Euler-Lagrange
equation of motion is found by taking the variational derivative $\delta (S_g+S_m)/\delta
g^{\mu\nu}(x)=0$.  Thus we have to find out how to compute the functional derivatives
$\delta \sqrt{|g|(x)}/\delta g^{\mu\nu}(x')$ and $\delta R(x)/\delta g^{\mu\nu}(x')$. 
The latter is given by 
\be
	{\delta R(x)\over \delta g^{\mu\nu}(x')} = R_{\mu\nu}\,\delta^{(4)}(x-x')
	 + g^{\alpha\beta}{\delta R_{\alpha\beta}(x)\over \delta g^{\mu\nu}(x')}
\ee
Although it is a tedious calculation, one can show that the second term is a total
covariant derivative, so it gives just a boundary term when integrating by parts,
that we will assume is zero.  For the determinant, we can use the general matrix identity
$\ln\det A = \tr\ln A$.  Taking the variation gives
\bea
	\delta\ln\det A = {\delta\det A\over \det A} &=& \tr\delta\ln A\nn\\
	&=& \tr\left[\ln(A + \delta A) - \ln A\right]\nn\\
	&=& \tr\left[\ln(A(1 + A^{-1}\delta A)) - \ln A\right]\nn\\
	&=& \tr A^{-1}\delta A
\eea
Therefore $\delta g = g\, g^{\alpha\beta}\,\delta g_{\alpha\beta}$ and 
\be
	{\delta \sqrt{|g(x)|}\over \delta g^{\alpha\beta}(x')} = 
	-\sfrac12 \sqrt{|g|}\, g_{\alpha\beta}\, \delta^{(4)}(x-x')\,.
\ee
The minus sign comes from the fact that $\delta g^{\alpha\beta} = - g^{\alpha\mu}g^{\beta\nu}
\delta g_{\mu\nu}$, since $\delta (g_{\alpha\beta}g^{\beta\mu}) = 0$.
We thus have all the pieces needed to compute the variation of the action, and the result is
\be
	R_{\mu\nu} -\sfrac12 g_{\mu\nu}R = -\kappa T_{\mu\nu}
\ee
where $\kappa = 8\pi G$.  These are the Einstein field equations.  By defining the Einstein
tensor $G_{\mu\nu} = R_{\mu\nu} -\sfrac12 g_{\mu\nu}R$, they take the simpler form
$G_{\mu\nu} = -\kappa T_{\mu\nu}$.  

\subsubsection{GR equations of constraint}
Given all the parallels between EM and GR, we expect that there will be equations of 
constraint in addition to the Einstein equations.   They are called that
because, rather than determining the evolution of the fields, they constrain the form of
initial data that we are allowed to specify when treating the system as an initial value problem.
In GR, it would not be general to specify initial conditions at some fixed time $t=t_0$ that is the same
everywhere in space.  Supposing we did so, one could always perform a coordinate transformation 
such that the flat hypersurface given by $t=t_0$ in the original coordinate system becomes a
wavy surface in the new system.  The more general procedure is to specify a spacelike
hypersurface $f(x^\mu)=0$ as the surface on which initial data will be specified.  The equations
of constraint put four restrictions on how the metric specified this surface is related to
its derivative in the direction normal to the surface (approximately the time direction).

These relations are expressed in terms of the 3D spatial metric $^{(3)}g_{ij}$
that describes the geometry of the hypersurface, and the extrinsic curvature $K_{ij}$ of this
surface embedded in the $3+1$ dimensional spacetime.  Thus far, we have alluded to the concept of
extrinsic curvature, but we did not yet need to define it.  We start by defining a unit normal
vector to the hypersurface, 
\be
	n_\mu = f_{,\mu}/\sqrt{f_{,\mu}f^{,\mu}}\,.
\ee
The extrinsic curvature tensor is defined to be
\be
	K_{ij} = -n_{i;j}\,,
\ee
where the indices $i,j$ are regarded as living in the tangent space of the hypersurface.
It turns out that $K_{ij}$ is symmetric \cite{Misner:1973prb}.  The reason it lives within the
lower-dimensional space is that $n^\mu$ can only vary perpendicular to itself, considering its
fixed normalization.  Hence its derivative has no component in the $n^\mu$ direction.
And it makes no sense to ask what its directional derivative is along the $n^\mu$ direction 
since we have only defined $n_\mu$ on the surface.  The constraint equations turn out to be
\cite{Wald:1984rg}
\be
	K^i_{\ j;i} = K^i_{\ i;j}\,,\qquad ^{(3)}R + (K^i_{\ i})^2 = K_{ij}K^{ij}\,,
\ee
known as the momentum and Hamiltonian constraints, respectively.  We will not use them in this
course, but they are essential for dynamical evolution in numerical relativity.

\subsection{Conservation of the stress-energy tensor}
We used the fact that EM current is conserved, $\partial_\mu J^\mu =0$, to prove the gauge
invariance of the interaction Lagrangian in the EM example.  There is an analogous statement
in GR,
\be
	\nabla_\mu T^{\mu\nu} = 0\,.
\label{Tcons}
\ee
This is four equations, which intuitively could correspond to the conservation of energy,
and the three spatial components of momentum.  To develop this intuition, let's consider the $\nu=0$ component,
in flat space where we can treat the covariant derivatives as partial derivatives:
\be
	\dot\rho - \partial_i T^{i0} = 0\,.
\ee
This looks similar to charge conservation in EM, but $\rho$ is now the energy density.
This tells us that $T^{i0}$ must be a current describing the flow of energy.  If it is a
uniform current, there is no net change in $\rho$ since the flow out equals the flow in,
but if it has a divergence then there is net flow through a closed surface.  Recall from 
Eq.\ (\ref{Teq}) that $T^{i0}\sim m v^i$ for a (slowly) moving particle.  This accords with our
description, that it represents the flow of mass-energy.

For the spatial components, it is the same argument, but now applied to the components of the
momentum.  We have
\be
	\dot T^{0j}- \partial_i T^{ij} = 0\,.
\ee
This means that $T^{0j}$ represents not only the flow of energy in the $j$ direction, but
also the density of momentum component $p^j$.  Although they sound like different quantities,
they are numerically equal.  We can interpret $T^{ij}$ as the flow of momentum density of
component $p^j$ along the $i$ direction.  And since $T^{ij}$ is symmetric, this must be equal
to the flow of momentum density of
component $p^i$ along the $j$ direction, even though {\it a priori} it sounds like these two
quantities need not be the same.

If we take the covariant divergence of the Einstein equation, and use (\ref{Tcons}), it
implies that
\be
	\nabla_\mu R^{\mu\nu} -\sfrac12\nabla_\mu R = 0\,,
\ee
which is known as the Bianchi identity.  It is a mathematical identity, which follows from
the definition of $R^{\mu\nu}$, and ultimately from invariance under coordinate
transformations.  Therefore one can regard conservation of $T^{\mu\nu}$ as being similar
to EM current conservation, namely a consequence of the gauge symmetry underlying the
theory.

\subsection{Linearization}
We can now go back to our initial approach where we considered gravity in the weak-field
limit, $g_{\mu\nu} = \eta_{\mu\nu} + h_{\mu\nu}$ and linearized in $h_{\mu\nu}$.  We know what
the full equations of motion are, so there will be no need for hand-waving to find the 
exact form of the linearized EOM.  We need to calculate $R_{\mu\nu}$ to first order in 
$h_{\mu\nu}$.  This contribution arises from the terms linear in $\Gamma^\mu_{\ \alpha\beta}$
in Eq.\ (\ref{ricci}), since $\Gamma^\mu_{\ \alpha\beta}=0$ for flat space.  We have
\bea
	\Gamma_{\mu\alpha\beta} = \sfrac12\left(-h_{\alpha\beta,\mu}  + h_{\mu\alpha,\beta}
	+ h_{\beta\mu,\alpha}\right)\,\nn\\
	\Gamma^\mu_{\ \alpha\beta} = \sfrac12\left(- h_{\alpha\beta}^{\ \ \ ,\mu}  
	+ h^\mu_{\ \alpha,\beta}
	+ h^{\mu}_{\ \beta,\alpha}\right)\, \label{linChrist}
\eea
hence from Eq.\ (\ref{ricci})
\bea
	R_{\alpha\beta} &=&  \sfrac12\left(h_{\alpha\beta\ \ \mu}^{\ \ \ ,\mu}  
	- h^\mu_{\ \alpha,\beta\mu}
	- h^{\mu}_{\ \beta,\alpha\mu}\right) - 
 \sfrac12\left(h_{\mu\beta\ \ \alpha}^{\ \ \ ,\mu}  
	- h^\mu_{\ \mu,\beta\alpha}
	- h^{\mu}_{\ \beta,\mu\alpha}\right)\nn\\
	&=& 
\sfrac12\left(h_{\alpha\beta\ \ \mu}^{\ \ \ ,\mu}  
	- h^\mu_{\ \alpha,\beta\mu}
	 - 
 h_{\mu\beta\ \ \alpha}^{\ \ \ ,\mu}  
	+ h^\mu_{\ \mu,\beta\alpha}
	\right)
\eea
which is symmetric in $\alpha,\beta$, even though the definition (\ref{ricci}) is not obviously so.
The curvature scalar, from taking the trace, is
\be
	R = \Box h -h^{\alpha\beta}_{\ \ ,\alpha\beta}\,.
\ee
We can thus read off all the six terms that appear on the left-hand-side of the linearized
Einstein equations by taking $R_{\alpha\beta} -\sfrac12\eta_{\alpha\beta}R$.  If we transform to
a coordinate system where $h_{\alpha\beta}$ is traceless ($h=0$) and transverse
($h_{\alpha\mu}^{\ \ \ ,\mu}=0$), then all of them vanish except the simple wave operator
$\Box h_{\alpha\beta}$ term.  In this coordinate system, $T_{\alpha\beta}$ must also be
traceless for consistency.  As we discussed before, this is a good choice for gravitational
waves, but it might not be convenient for describing the effects of a static source.

\subsection{Problems}

\noindent 9.1. (a) Fill in the missing steps to derive Eqs.\ (\ref{T00def}-\ref{Tijdef}).  Hint: you will need to show that
$\delta g_{il}/\delta g^{jk} = -g_{ij}g_{kl}$.\\
(b) Show that the trace of $T_{\mu\nu}$ has the correct form to be identified with the negative of the
matter Lagrangian in the case of the nonrelativistic point particle, as discussed in section \ref{EEqs}.
{\it I.e.,} show that $-\int dt\,d^{\,3}{x'}\sqrt{|g|}g^{\mu\nu}T_{\mu\nu}$ reduces to the world-line action for the
nonrelativistic point particle.\\  
%(b) Show that the variation of 
%$\sqrt{|g|}T$ reproduces the expected $T_{\mu\nu}$ term in the Einstein equations.  You can
%restrict to the $00$ and $ij$ elements as in the example (\ref{mat-ex}).\\

\noindent 9.2.  The atmosphere of Earth can be characterized by its mass density $\rho(z)$ and pressure $p(z)$ along
a vertical $z$ axis above some point on the surface.  Take the perfect fluid form for the stress tensor,
$T^{\mu\nu} = {\rm diag}(\rho,p,p,p)$.\\
(a) Using Eq.\ (\ref{linChrist}), compute the covariant divergence $\nabla_\mu T^{\mu z}$ and demand its
conservation, to find a differential equation relating $\rho'$ and $p'$.  It is convenient to let $z=0$ correspond to the
center of the Earth, even though the equation is valid only above the Earth's surface.  Show that
$p_{,z} +\rho \,\phi_{,z} = 0$.\\
(b) To solve the equation, one needs a constitutive equation relating $p$ and $\rho$,  usually called the
equation of state.  For air we can use $dp/d\rho = c_s^2$, the sound speed squared.  Solve for $p(z)$.
How far above the surface must one go for the pressure to fall by $1/e$?

\section{The Schwarzschild and TOV metrics}
\label{BHsect}
We are now ready to look for  exact solutions to the full Einstein equations.  The simplest 
situation is the field around a mass $M$ approximated as a point particle, whose stress tensor
is $T^{00} = M\delta(\vec x) = M\delta(r)/(4\pi r^2)$, and $T^{ij}\approx 0$.  The source is spherically symmetric,
so we would like a coordinate system that takes advantage of this: $x^\mu = (t,r,\theta,\phi)$.
An ansatz for the metric could take the form
\be
	g_{ij} = {\rm diag}\left(e^{2\Phi}, -e^{2\Lambda}, -e^{2\Sigma},
-e^{2\Sigma}\sin^2(\theta)\right)
\ee
where $\Phi$, $\Lambda$ and $\Sigma$ are functions of $r$.   We can make one immediate
simplification by redefining the $r$ coordinate such that $e^\Sigma = r$.  Then the lower part of
the metric takes the more familiar form
\be
	g_{ij} = {\rm diag}\left(e^{2\Phi}, -e^{2\Lambda}, -r^2,
-r^2\sin^2(\theta)\right)
\label{Sph-ansatz}
\ee
like for Minkowski space in spherical coordinates.  Then we have only two unknown functions
to solve for.  From the weak field limit, we know that they must reduce to $\Phi = -\Lambda =
\phi$ for small $\phi = -GM/r$.

Computing the Einstein tensor for the ansatz, for example using software tools, one finds that
for the $tt$ component, 
\be
	G_{tt} = r^{-2}e^{2(\Phi-\Lambda)}\left(e^{2\Lambda} + 2 r\Lambda' -1\right)
\ = \ \kappa T_{tt} =  2GM\, r^{-2}e^{2\Phi}\delta(r)\,.
\label{Gtteq}
\ee
For all $r>0$, this implies $\partial_r(r(1-e^{-2\Lambda})) = 0$, hence $e^{2\Lambda} =
(1-k/r)^{-1}$
for some constant of integration $k$, whose value can be determined by integrating the Einstein
equation over an
infinitesimal region around $r=0$.  Or more easily, we can infer that $k=2GM$ from matching
the solution onto the weak-field limit.  Hence we have determined $e^{2\Lambda} = (1 -
2GM/r)^{-1}$.

Similarly, the $rr$ component of Einstein's equations gives
\be
	G_{rr} = r^{-2}\left(-e^{2\Lambda} + 2 r\Phi' + 1\right) \ = \ 0\,.
\label{Grreq}
\ee
This is easily integrated to find that $e^{2\Phi} = e^{k'}(1-2GM/r)$, where $k'$ is another
constant of integration.  Again matching onto the weak-field solution gives $k'=0$.  
We see that $\Phi = -\Lambda$ as we had anticipated from that limit.  In summary, the
Schwarzschild line element is given in terms of the Newtonian potential by
\be
	ds^2 = (1+2\phi)dt^2 -{dr^2\over 1+2\phi} - r^2\,d\Omega_2^2\,,
\label{Schw-metric}
\ee
in the chosen coordinate system, which in this context is known as Schwarzschild coordinates.
Here $d\Omega_2^2 = d\theta^2 + \sin^2\theta d\phi^2$ like for the 2-sphere.

One should keep in mind that the Schwarzschild solution applies not only for black holes, but also
exterior to any spherically symmetric mass distribution for which $2\phi > -1$, such as stars.
In this case, the exterior solution must be matched onto an interior solution, for which $T^{00}$ is
no longer zero.  In the following, we will assume that a black hole has formed, to explore what happens
when the mass distribution is so dense that it lies within $r = 2GM$, and collapses to form a black hole.
The case of normal stars will be treated in Section \ref{TOV}.

\subsection{The Schwarzschild radius}
At small $\phi$, the metric does not look exactly like what we expected from the weak-field 
solution,
\be
	ds^2 \approx (1+2\phi)dt^2  - (1-2\phi)(dr^2 + r^2\,d\Omega_2^2)\,,
\ee
because of our different choice of $r$ coordinate.  However by letting $r = (1-\phi)r'\approx
(r' + GM)$, we see that they will match to leading order in $\phi$, so they are 
physically equivalent at this order.  The interesting qualitative difference is that the coefficient of
$dr^2$ diverges as $r\to 2GM\equiv r_s$ in the Schwarzschild solution, while remaining finite in the
weak-field approximation, which of course is no longer valid for such large values of $|\phi|$. 
Something strange seems to be going on at this radius, which as you recall is the 
Schwarzschild radius.  However in reality, there is no physical singularity at $r_s$.  For
example, the proper distance from some radius $L> r_s$ to $r_s$ is finite,
\be
	\int_{2GM}^L {dr\over \sqrt{1- 2GM/r} } = 2GM\left(
\sqrt{a(a-1)}+\ln\left(\sqrt{a}+\sqrt{a-1}\right)\right)
\ee 
where $a = L/(2GM)$.  And the Ricci curvature vanishes everywhere for $r>0$:
\be
	R = r^{-2}e^{-2 \Lambda (r)} \left[2 r \left(r \Phi ''(r)-\left(r \Phi '(r)+2\right) \left(\Lambda
   '(r)-\Phi '(r)\right)\right)-2 e^{2 \Lambda (r)}+2\right] = 0\,,
\ee
as can be verified by substituting the solutions for $\Lambda$ and $\Phi$, and letting the
computer do the tedious algebra.

But something peculiar occurs when we consider a light ray, obeying $ds^2=0$ falling along
a radial trajectory at fixed angle.  Its equation of motion is
\be
	dt = {dr\over 1-2GM/r}\ \implies\ t = r + 2GM\ln\left({r - 2GM\over r_0}\right)\,,
\label{Sch-null}
\ee 
which diverges as the light ray approaches $r_s$. ($r_0$ is a constant of integration.)  A distant observer would say that it 
never reaches $r_s$.  And such an observer, keeping track of the ticks of a clock that is
approaching $r_s$, would measure that it ticks slower and slower, coming to a standstill as it
approaches $r_s$.

However, freely falling observers do not notice anything out of the ordinary as they cross the horizon.  There is
nothing to warn them that they are passing a point of no return, that anything which crosses inside
the horizon can never get back out.  Only a bit later will the direness of their situation start to sink in.
Inevitably, they will reach the singularity at $r=0$, but before that, they become shredded by the diverging
tidal forces to be analyzed belowl.  Here, we would first like to understand the nature of the 
singularity.  Contrary to expectations, it is not at a point in space; instead it happens everywhere in space,
at a moment in time.  This bizarre turn of events is a consequence of the sign change experienced by the 
$g_{tt}$ and $g_{rr}$ metric components below the horizon.  As a consequence, when $r<r_s$, $r$ becomes the time
coordinate, and $t$ is spatial.  The freely falling observer, indeed all such observers, will reach the
singularity at time $r=0$, for all values of the spatial coordinate $t$.

\subsection{Near-horizon limit}
The Schwarzschild coordinates are singular at the horizon, which obscures the physical interpretation of that
surface, and what happens as one crosses it.  In particular, at $r=r_s$, the coefficient of $dt$ in the line
element vanishes, and time stops flowing in these coordinates.  One can choose different coordinates to avoid
this problem in the vicinity of the horizon \cite{Tong}.  Let $r= r_s + \eta$ for small $\eta>0$ near the
horizon, and let $\eta = \rho^2/(4 r_s)$.  One can then define a timelike coordinate $T = \rho\sinh(t/2r_s)$
and spacelike one $X = \rho\cosh(t/2r_s)$ such that the line element becomes 
\be
	ds^2 \approx dT^2 - dX^2 - r_s^2 d\Omega_2^2
\label{rindler}
\ee
in the vicinity of the horizon.  This looks like Minkowski space in the $T$-$X$ subspace, in direct product with
two-spheres of radius $r_s$, reinforcing the
fact that nothing dramatic is happening locally near the horizon.  The horizon itself seems to have degenerated into a single
point $X=T=0$, but if we approach it through a limiting process $\rho\to 0$ while allowing $t$ to vary from
$-\infty$ to $\infty$, one sees that the curves of constant $r$ are hyperbolae that approach the lines
$T=X$ in the first quadrant and $T=-X$ in the fourth quadrant.  Therefore the horizon actually consists of
the rays $X = \pm T$: it is a lightlike surface, not spacelike as one would have normally associated
with a surface of fixed $r$.  In fact a light ray at $r=r_s$ can have closed circular orbits by moving in the
angular directions of the two-spheres.

Comparing the definitions of $T,X$ to Eq.\ (\ref{rindler-def}), we see that $t,\rho$ are a version of Rindler
coordinates, with $\rho = e^{a\xi}/a$.  This is an interesting example of Einstein's equivalence principle, that says a uniformly
accelerated observer sees the same physics, locally, as an observer at rest in a uniform gravitational field.
Notice that an observer sitting close to the horizon of the black hole (at fixed $\rho$) 
feels a gravitational acceleration of $GM/r_s^2 = 1/(2 r_s)$.  This matches the $a$ inside the hyperbolic trig
functions in Eq.\ (\ref{rindler-def}).  It is the acceleration a rocket would have to exert against the gravitational
force at the horizon, in order to keep the observer from falling through.  Notice that this force is
increasingly weaker, the larger the mass of the black hole.  Hence for a very large $M$, the gravitational
effects near the horizon are quite weak.  What should prevent someone from going back and forth across this
surface?  It is the fact that it is light-like: although an observer moving toward the horizon can pass through
as slowly as he likes, one trying to come back out would have to travel faster than light.  To substantiate
these statements, we need coordinates that have smooth behavior going across the horizon.  Eddington-Finkelstein
coordinates are the best-known example.

\subsection{Eddington-Finkelstein coordinates}

They are constructed in two steps, starting from the Schwarzschild coordinates.  First, we define a new radial
coordinate $r_*$, known as the tortoise coordinate, such that $dr/(1+2\phi) = dr_*$.  This is the same
differential equation we already solved in Eq.\ (\ref{Sch-null}); thus $r_*$ equals the time elapsed for
light to travel along a radial geodesic.  We take the integration constant to be $r_0 = r_s$, so 
\be
	r_* = r + r_s\ln|r/r_s-1|\,.
\ee
Then $r_*\to -\infty$ as $r\to r_s$, and $r_*$ has domain $(-\infty,\infty)$.
Light rays in this coordinate system are simply described by $t = \pm r_*$ plus a constant.  The tortoise
coordinate can also be used inside the horizon, thanks to the absolute value in the logarithm.  But for $r<r_s$,
it takes values $r_* \in (-\infty,0]$.  Notice that if $r_*<0$, one could be either inside the horizon, or
outside but close to it; hence one must keep in mind that it has different meanings in the two patches.  The
metric has the simple form
\be
	ds^2 = (1+2\phi)\left(dt^2 - dr_*^2\right) - r^2 d\Omega^2\,
\ee
where $r$ is considered to be an implicitly defined function of $r_*$

Next we define
a coordinate $v = t+r_*$, which is constant for light rays falling into the black hole.  Defining $r_*$ is just
an intermediate step for defining $v$; we are free to go back to the $r$ coordinate and simply trade $t$ for
$v$.  The Schwarzschild line element then becomes
\be
	ds^2 = (1+2\phi)dv^2 - 2\, dv\, dr -r^2d\Omega_2^2\,.
\label{EF}
\ee
These are called ingoing Eddington-Finkelstein coordinates since lines of fixed $v$ describe 
light rays falling into the black hole.

Compare the $dv$-$dr$ part of the metric to the corresponding $dt$-$dr$ sector in Schwarzschild coordinates.
It is better behaved at the horizon, where $1+2\phi = 0$, since $\det({\phantom{-}0\atop -1}{-1\atop \phantom{-}0})$ is nonvanishing.
In the Schwarzschild coordinates, the corresponding determinant is $0/0$ at the horizon, which requires more
care.  We can let $r$ continue past $r_s$ all the way to the singularity at $r=0$ in the Eddington-Finkelstein coordinates,
to see what happens below the horizon.  In the geometry (\ref{EF}), the null radial geodesics are given by 
$(1+2\phi)dv = 2 dr$ and $dv = d(t + r_*) = 0$.  The latter are the ingoing geodesics we already identified in
the Schwarzschild solution.  Even though $r_*$ diverges at the horizon, in the original $r$ coordinate it is
clear that nothing strange happens as $r$ passes below $r_s$.  In the $r$-$v$ plane, the geodesic is just
a straight line of constant $v$ that reaches $r=0$.  The ingoing light rays all reach the singularity.

The interesting feature is the other class of null geodesics, with trajectories
\be
	v = 2\int {dr\over (1+2\phi)} = 2r_*\,.
\ee 
For $r>r_s$, we have $v = t+r_* = 2 r_*$, so $t=r_*$, which is our normal expectation for an outgoing geodesic.
But for $r < r_s$, the integrand is negative, and so $v$ decreases as a function of $r$ instead of increasing.  The result is that
inside the horizon, the other null geodesic also approaches the singularity, albeit more slowly than the
first solution.  This is further explored in problem 10.2, but one can immediately see what is happening by
drawing the light cones bounded by slopes $dv/dr = 0$ and $dv/dr = 2/(1+2\phi)$ in the $r$-$v$ plane.  Below the 
horizon, there are no light rays traveling in the $+r$ direction. 

One can also define the outgoing counterpart $u = t- r_*$  of $v$, and express the geometry as
\be
	ds^2 = (1+2\phi)du^2 + 2\, du\, dr -r^2d\Omega_2^2\,,
\label{EF2}
\ee
which are the outgoing Eddington-Finkelstein coordinates.  It turns out that these coordinates cover a different
region of the spacetime (overlapping with the incoming coordinates outside of the horizon).  In this new region,
the singularity at $r=0$ appears in the past, and the light rays are emerging from it.  This solution is known
as a white hole.  It seems to have only academic interest, since astrophysical black holes correspond to the
ones with future singularities, since they arose from the collapse of normal stars with no singularity in the
past.  A white hole would have to be past-eternal.   

The full Schwarzschild geometry  has a causal structure much like Minkowski space relative to Rindler coordinates.   The
Schwarzschild coordinates are similar to Eqs.\ (\ref{rindler-def},\ref{rindler-sup}) in that they cover only
half of the full space, corresponding to the $R$ (outside of horizon) and $F$ (inside horizon toward the
future) patches.  These are the only physical regions expected to exist for an astrophysical black hole that
came from gravitational collapse.  Mathematically, one can extend these with the $L$ and $P$ patches, that
contain the white hole solution, whose description requires an extra copy of the original coordinates.  This
complete manifold is known as an eternal black hole.

\subsection{Kruskal coordinates}
Although the Eddington coordinates are somewhat better behaved than Schwarzschild, having no metric elements
that diverge at the horizon, the fact that $g_{vv}$ or $g_{uu}$ vanishes there is often inconvenient.  A
stepping stone toward a truly well-behaved coordinate system is to first replace both $t$ and $r$ with
$u$ and $v$.  The line element simplifies to 
\be
	ds^2 = (1+2\phi) du dv - r^2 d\Omega^2
\ee
where $r$ is regarded as an implicit function of $v-u = 2r_*(r)$.  This is still singular at the horizon,
but by defining
\be
	U = -e^{u/(2 r_s)},\quad V = e^{v/(2 r_s)}\,,
\ee
the problem is cured, by virtue of the fact that $e^{r_*/(2 r_s)} = |1-r/r_s|^{1/2}e^{r/(2 r_s)}$.  The
prefactor aborbs the unwanted $(1+2\phi)$ factor in the metric in the new system, whose line element is
\be
	ds^2 = 4 {r_s^3\over r} e^{-r/r_s} dU dV - r^2d\Omega^2\,.
\ee
Here $r$ is defined in terms of the null Kruskal coordinates by $UV = (1-r/r_s)e^{r/r_s}$.  The horizon thus
appears at $U=0$ or $V=0$ in these coordinates.  One can alternatively define timelike and spacelike
combinations, valid in the region outside the horizon,
\bea
	T &=& \sfrac12(V+U) = e^{r/(2 r_s)}(r/r_s-1)^{1/2}\sinh t/2 r_s\,,\nn\\
	R &=& \sfrac12(V-U) = e^{r/(2 r_s)}(r/r_s-1)^{1/2}\cosh t/2 r_s\,,
\label{TRdef}
\eea
for which the metric becomes
\be
	ds^2 = 4 {r_s^3\over r} e^{-r/r_s} (dT^2-dR^2) - r^2d\Omega^2\,.
\label{Kruskal-ext}
\ee
For the region inside the horizon, the metric has the same form, but the relation to Schwarzschild 
coordinates is instead \cite{Misner:1973prb}
\bea
	T &=& e^{r/(2 r_s)}(1-r/r_s)^{1/2}\cosh t/2 r_s\,,\nn\\
	R &=&  e^{r/(2 r_s)}(1-r/r_s)^{1/2}\sinh t/2 r_s\,.
\label{TRdef2}
\eea
The reason for the interchange of $\sinh$ and $\cosh$ is that $t$ is a spacelike coordinate inside the 
horizon, so if we fix $t=0$, say, the march of time $T$ should not be halted in the Kruskal coordinates.

Notice the similarity between Eq.\ (\ref{TRdef}) and the transformation (\ref{rindler-def}) between Rindler and
Minkowski coordinates.  The Kruskal coordinates are analogous to Minkowski.  An observer at rest in Kruskal
coordinates is uniformly accelerated in Schwarzschild coordinates, with acceleration $1/(2r_s)$, which is the
surface gravity of the black hole at the horizon.  This will be useful in Section \ref{hawking-sect}.

\subsection{Tidal forces}
\label{tidal-sect}  In problem 8.3 we introduced the internal stresses, known as tidal forces, that
would be experienced by an extended body falling into a $1/r$ gravitational potential.  It gets compressed in
the transverse directions and stretched along the $r$ direction.  The same thing will happen in the black hole
geometry, but the effects start to diverge as the object approaches the singularity.
Let's examine the tidal forces on radial
geodesics in the original Schwarzschild coordinates (\ref{Schw-metric}).   According to the geodesic deviation
equation, the tidal acceleration of a separation vector $\xi^\mu$ is given by 
\be
	{D\xi^\mu\over D\tau^2} = -R^\mu_{\ \alpha\beta\gamma}\xi^\beta \dot x^\alpha\dot x^\gamma
\ee
where $\tau$ is the proper time and $x^\mu(\tau)$ is a geodesic.  So we must first solve for the geodesics.
Let's consider radial ones, where $\theta$ and $\phi$ are held fixed, and we will now focus on the BH interior,
below the horizon.   Keep in mind that the roles of $t$ and $r$ are interchanged in this region: $t$ is
spacelike and $r$ is timelike.  Computing the Christoffel symbols, the geodesic equations are
\bea
\label{ddotteq}	-\ddot t &=& {r_s\over r(r-r_s)} \dot r\dot t\nn\\
\label{schwgeo}
\label{ddotreq}	-\ddot r &=& {r_s\over 2 r(r_s-r)} \dot r^2 + {r_s(r-r_s)\over 2 r^3} \dot t^2
\eea
Writing Eq.\ (\ref{ddotteq}) as $\ddot t/\dot t = d\ln t/d\tau = -\dot r/[r(1-r/r_s)]$, one can integrate to find that
$\dot t = C_t r/(r-r_s)$ where $C_t$ is a constant of integration.  Then Eq.\ (\ref{ddotreq}) becomes
\be
	\ddot r = - {r_s\over 2 r(r_s-r)}\left(\dot r^2 - C_t^2\right)
\label{ddotreq2}
\ee
Moreover, since $d\tau^2 = (1-r_s/r)dt^2 - dr^2/(1-r_s/r) > 0$, it is straightforward to prove that $\dot r^2 >
C_t^2$ below the horizon, for particles moving along timelike trajectories.  This means the time coordinate $r$
has $\ddot r < 0$, so that time is inevitably flowing in the $-r$ direction, toward the singularity.  This
confirms the claim that there is no escape from reaching the singularity, as the flow of time cannot be
reversed.  Since $\dot r < 0$ for a particle traveling forward in time, Eq.\ (\ref{ddotreq2}) insures that
if $\dot r^2 > C_t^2$ at horizon crossing, this condition remains fulfilled until it reaches the singularity.
For the special case $C_t=0$, we can integrate Eq.\ (\ref{ddotreq2}) once to find that $\dot r =
-C_r\sqrt{r_s/r-1}$, where $C_r$ is another integration constant.  $C_t=0$ represents the case where the particle
remains at rest in the spatial $t$ dimension that takes the place of the radial direction when crossing below
the horizon.

Taking $\dot t=0$ and $\dot r = -C_r\sqrt{r_s/r-1}$ for this class of geodesics, we can easily find the tidal
stresses along the spatial directions.  For $\xi^\mu = \xi^\theta$ and $\xi^\mu = \xi^t$,
\bea
	{D^2 \xi^\theta\over D\tau^2} &=& - R^\theta_{\ r\theta r}\,\xi^\theta \,\dot r^2 = -{GM C_r^2\over
r^3}\,\xi^\theta\nn\\
	{D^2 \xi^t\over D\tau^2} &=& - R^t_{\ r t r}\,\xi^t\, \dot r^2 = 2{GM C_r^2\over r^3}\,\xi^t
\eea
Notice that $C_r$ is dimensionless, so a typical geodesic has $C_r\sim 1$.  Then the tidal forces have the same
form as in the weak-field case, Problem 8.3: an observer oriented along the radial (now $t$) direction is
stretched in length, and compressed in the transverse directions.  These forces match smoothly onto
those experienced at the horizon if $C_r = 1$, lending further motivation to this choice.
As a byproduct, we learn how long it takes to travel from the horizon to the singularity in the observer's
time: $\tau = \int dr/\dot r = \int_0^{r_s} dr/ \sqrt{r_s/r-1} = \pi GM$.  
To get a feeling for the numbers, see
Problem 10.3.\footnote{The above calculation is a bit of a cheat, as explained in MTW \cite{Misner:1973prb}, pp.\ 822-823.
To obtain the coordinate-independent force experienced by the observer, one should make a transformation to the
static orthonormal coordinates at the event of interest.  They are related by a boost in the $r$ diredtion.
It turns out that the above Riemann components are unchanged by this boost, thanks to special properties of the
Schwarzschild geometry.}

\begin{figure}[t]
\centerline{\includegraphics[width=0.65\textwidth]{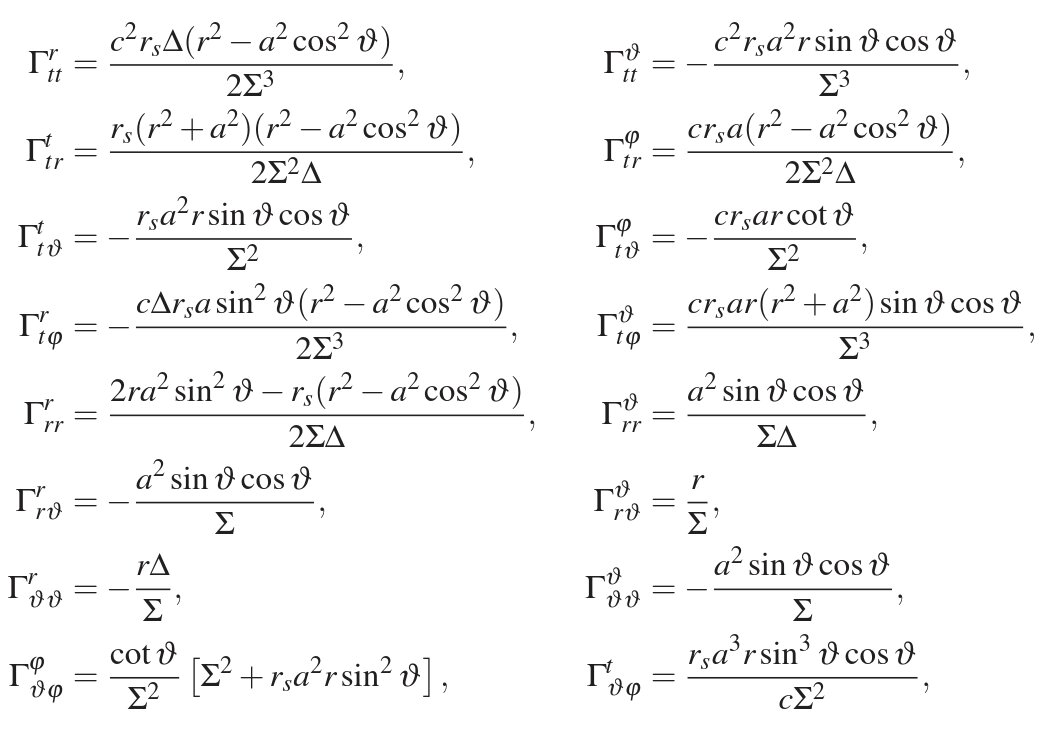}
\includegraphics[width=0.45\textwidth]{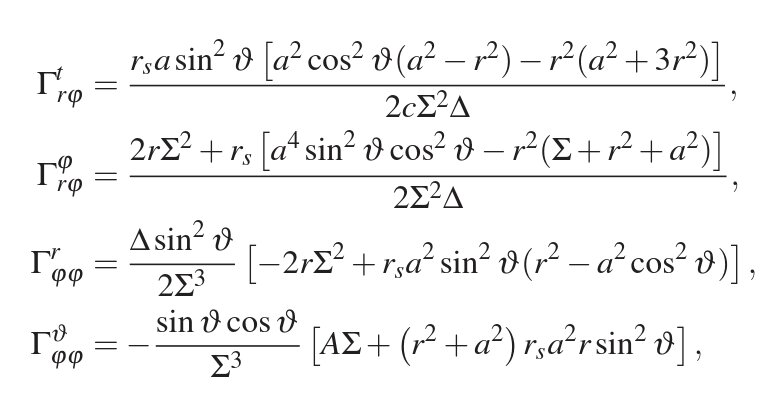}} 
%\leftline{\includegraphics[width=0.1\textwidth]{py3.jpg}}    
    \caption{Christoffel symbols for the Boyer-Lindquist metric, taken from Ref.\
\cite{Muller:2009bw},
and checked by me.
}
    \label{BLX}
\end{figure}

\subsection{Spinning black holes}
Most black holes in nature are not exactly Schwarzschild because they carry angular momentum $J$.  Even if they
initially formed with $J=0$, they commonly acquire companions or an accretion disk which are orbiting around
them.  As this rotating material gets absorbed the by the black hole, it gains not only the mass but also the
angular momentum, and gets spun up.  The information about the spin cannot be encoded in $T^{\mu\nu}$ since
a point mass has no moment of inertia.  In fact, even the construct of $T^0_{\ 0} = M\delta(r)/(4\pi r^2)$ is not
physically sensible for a black hole, since we now realize that $r$ is a timelike coordinate.  Instead, we
define what is meant by $M$ by matching the Schwarzschild solution at large distances onto the weak-field solution for a
nonsingular mass distribution.   The same can be done for a rotating star in order to identify the $J$ of a
rotating black hole solution, which still has $T^{\mu\nu}=0$.  This solution is known as the Kerr metric 
\cite{Kerr:1963ud}.  

It is not as simple as the Schwarzschild metric.  Kerr's original solution expressed it as a
generalization of outgoing Eddington-Finkelstein coordinates.  Boyer and Lindquist \cite{Boyer:1966qh} found
the coordinates which are closest to Schwarzschild, reducing to that metric in the limit $J\to 0$.
Introducing the parameter $a = J/M$ (which has dimensions of length in $c=1$ units), 
$\Delta = r^2-r r_s + a^2$ and $\Sigma = r^2 +
a^2\cos^2\theta$ (sometimes denoted by $\rho^2$), the line element is
\be
	ds^2 = 	\left(1-{r r_s\over\Sigma}\right)dt^2 - {\Sigma\over \Delta} dr^2 -\Sigma d\theta^2 -
\left(r^2 + a^2 + {r_s r a^2\over \Sigma}\sin^2\theta\right)\sin^2\theta\,d\phi^2 + {2 r_s r a\over
\Sigma}\sin^2\theta\, d\phi\, dt
\label{BLmetric}
\ee
The off-diagonal $d\phi\,dt$ term leads to the phenomenon of {\it frame dragging} in the azimuthal
direction, that is, the direction of rotation of the black hole.  It causes observers who are
initially at rest to start co-rotating with the black hole.  This effect becomes irresistible within
a radius 
\be
	r_0 = r_s/2 + \sqrt{r_s^2/4 - a^2\cos^2\theta}\,,
\ee
known as the ergosphere.  It surrounds the event horizon, whose 
radius gets reduced relative
to $r_s$ by the
rotation,
\be
	r_+ = r_s/2 + \sqrt{r_s^2/4 - a^2}\,,
\ee
revealing that there is a maximum angular momentum, $J= GM^2$, corresponding to $a = r_s/2$.
The horizon and ergosphere coincide at the poles $\theta = 0, \pi$.  It turns out that for equatorial
orbits, the frame-dragging effect becomes important even before reaching the ergosphere: at some
larger radius, it becomes impossible to orbit the black hole in the direction opposite to its spin---see problem 10.6.
The Christoffel symbols for this metric are shown in Fig.\ \ref{BLX}, to do this problem.

It is possible to generalize the metrics for rotating black holes to the case where the black hole
also carries an electric charge.   When the charge is maximal, the horizon disappears, and one is left
with a ``naked singularity,'' in violation of Penrose's cosmic censorship conjecture
\cite{Penrose:1969pc}.  Although many theorists get a thrill out of these naked sinularities, in the
physical world they are not expected to exist, since any black hole or other compact object with a
sizable net charge would discharge it quickly to surrounding gas \cite{Profumo:2025vqc}.
On the other hand, black holes with significant spins appear to be common in the
Universe, based on LIGO-Virgo-KAGRA determinations of properties of several hundred merging black holes
\cite{LIGOScientific:2025slb}.

\subsection{Interior of stars}
\label{TOV}
The Schwarzschild solution (up to coordinate transformations) is guaranteed to be the unique geometry outside of
a nonrotating, spherically symmetric mass distribution, a result known as Birkhoff's theorem.  But if the mass
giving rise to it has not collapsed into a black hole, the exterior solution must be matched onto an interior
solution, usually characterized by a perfect fluid stress tensor
\be
	T^\mu_{\ \nu} = {\rm diag}(\rho,p,p,p)
\ee
where $\rho$ and $p$ are functions of $r$.  We must reconsider Einstein's equations with these nonvacuum sources
on the right-hand side.  Because of the spherical symmetry, the metric ansatz (\ref{Sph-ansatz}) is still valid,
but the $00$ component of Einstein's equations becomes
\be
	G_{tt} = r^{-2}e^{2(\Phi-\Lambda)}\left(e^{2\Lambda} + 2 r\Lambda' -1\right)
\ = \ \kappa T_{tt} =  8\pi G\rho\, e^{2\Phi}.
\ee
It is straightforward to verify that this is satisfied by taking
\be
		e^{2\Lambda} = \left(1- 2G m(r)/r)\right)^{-1},
\ee
where $m(r) = 4\pi\int_0^r dr\, r^2 \rho$ is the mass enclosed at radius $r$.  This agrees with the case of
a point mass.

The $rr$ component becomes $G_{rr} = \kappa \,p \,e^{2\Lambda}$. Using Eq.\ (\ref{Grreq}), this implies 
\be
	\Phi' = {G m(r) + 4\pi r^3 p\over r(r-2 G m(r))}\,.
\ee
Futhermore, conservation of the stress-energy tensor gives a constraint on the pressure,
\be
	T^{r\mu}_{\  \  \ ;\mu} = e^{-2\Lambda}\left(p' + (\rho +p)\Phi'\right) = 0\,,
\ee
in other words, $p' = -(\rho +p)\Phi'$.  We can rewrite the previous results as three coupled differential
equations, known after Tolman, Oppenheimer and Volkov (TOV),
\bea
	{m'} &=& 4\pi\, r^2\,\rho\,, \nn\\
	p' &=& -(p+\rho){G m + 4\pi r^3 p\over r(r-2 G m)}\,,\nn\\
	\Phi' &=& -{p'\over p+\rho}\,.
\eea
They are not yet complete because we do not know the density profile $\rho(r)$ {\it a priori}.  Instead,
an equation of state (EOS) relating $p$ to $\rho$ is needed to close the system.  This depends upon the properties of
the material inside the star.  A common example is the polytropic EOS, $p = K\rho^\gamma$.  For neutron stars, 
typical values might be $\gamma=2.75$ and $K = 2\times 10^{-6}$ in cgs units.  Notice that the equations can be
solved independently of $\Phi$.  One can integrate for $\Phi(r)$ after $\rho(r)$ and $p(r)$ are determined, if
desired.

The TOV equations are easy to solve numerically.  One starts very close to the center, at some small radius
 $r=\epsilon$ where $m\approx 0$, and makes a 
guess for the central pressure $p_0$.  The pressure drops as one integrates outward, and when $p$ becomes
zero, we have reached the surface of the star.  Families of solutions can thus be found by varying the
central pressure.  One might think that in this way, one could generate solutions for arbitrarily massive stars
just by starting with a sufficiently large central pressure.  However, $p'$ is increasingly negative  as $p$
increases, so in practice the final mass of the star is not monotonic in $p_0$, and one finds that there is a
maximum mass for a given equation of state. Central pressures beyond this limit would correspond to unstable
solutions, since they could relax to equal-mass stars of lower pressure.   To have a very massive star, the EOS needs to be sufficiently
{\it stiff}, which means its pressure must grow fast enough with $\rho$.  Theoretical studies of the nuclear
EOS are constrained by the fact that in nature, no neutron stars with masses $\gtrsim 2M_\odot$ are observed.

%\begin{comment}
\subsection{Black Hole Evaporation}
\label{hawking-sect}
Although classically, nothing can ever escape from the black hole horizon, in 1975 S.\ Hawking discovered that
when treated quantum mechanically, black holes emit radiation, that would eventually cause their evaporation
\cite{Hawking:1975vcx}.  The spectrum of radiation is approximately thermal, with a temperature given by
$T = 1/(4\pi r_s)$ in natural units where $\hbar=c=k_B=1$.  There is a heuristic picture for understanding this
in terms of quantum fluctuations of positive and negative energy particles near the horizon, in which the 
negative energy state falls inward, reducing the mass of the black hole, while the positive energy fluctuation
radiates away.  For a more rigorous understanding, quantum field theory is needed.  Here I will give a
simplified account.  Readers who skipped section \ref{rindler-rad} should read it now.

Consider a massless scalar field $\phi(t,r)$ in the Schwarzschild metric background.  It is a model for more
realistic kinds of particles that would be emitted by the black hole.
We will restrict our attention to field configurations
that have no angular dependence for simplicity.  The action is
\be
	S = \sfrac12 4\pi\int dr\,r^2 \left(g^{tt}\dot \phi^2 - g^{rr}\phi'^2\right) = 
	2\pi\int dr\,r^2\left({\dot\phi^2\over (1-r_s/r)} - (1-r_s/r)\phi'^2\right)\,.
\label{sf-action}
\ee
By varying the action, the equation of motion is
\be
	\ddot\phi  - \phi''(1-r_s/r)^2 - {\phi'\over r}(1-r_s/r)(2-r_s/r)\,.
\ee 
We can do separation of variables to write $\phi_k = e^{\pm i\omega t} f_k(r)$, and find that $f_k(r)$ satisfies
\be
	(1-r_s/r)^2 f_k'' + (f_k'/r)(1-r_s/r)(2-r_s/r) + \omega^2 f = 0
\ee
where $\omega = |k|$.
This equation is difficult to solve in general, but there are two regions where it simplifies.  Far from the
horizon, $r\gg r_s$, it reduces to $f_k'' + f_k'/r +k^2 f_k = 0$, the usual equation for a spherical wave in flat
space.  We saw these solutions in chapter 3, Eq.\ (\ref{sph-wave}).
The general solution is then  
\be
	\phi  \cong  \int {dk\over 2\pi} \left( a_k {e^{-i(\omega t - kr)}\over r} 
	+ a^*_k {e^{i(\omega t -kr)}\over r}\right)
\label{far-field-phi}
\ee
Notice that the $f$ here is normalized differently from the $f$ in Eq.\ (\ref{sph-wave}).

The other simplifying regime is to work close to the horizon, $r = r_s + y$ with $y\ll r_s$.  Then the
equation of motion takes the form
\be
	y^2 f_k'' + y f_k' + (r_s\,k)^2 f = 0
\ee
This has solutions
\be
	\phi \cong  \int {dk\over 2\pi} \left( a'_k e^{-i(\omega t - r_s k\ln(r-r_s))} + 
	a'^*_k e^{i(\omega t - r_s k\ln(r-r_s))}\right)
\label{near-field-phi}
\ee
again with $|k| = \omega$.  Notice that the coefficients $a_k$ and $a'_k$ could be related by some linear
transformation that mixes different $k$ values.  But a crucial
observation is that the positive and negative
frequency solutions do not mix with each other in going between the two regions.  This is because the metric
is time-independent.
Thus there are no Bogoliubov
coefficients $\beta_{kk'}$ connecting creation and annihilation operators in the two regions, 
which would imply that particle production occurs
within this metric considered by itself.   Instead it is the comparison of vacuum states between different coordinate systems
that is important.

To see the particle production, we need a second coordinate system, where
an observer at rest in one coordinate system would be accelerating in the other one.
%, or by the equivalence principle, experiencing a gravitational force.  
This turns out to be the Kruskal coordinates.  Using the metric\ (\ref{Kruskal-ext}) for the region outside the
horizon, the scalar field equation of motion is
\be
	(\partial_\T^2-\partial_\R^2)\phi +g_{\T\T}\partial_\T(g^{\T\T}\partial_\T\phi) + 
g_{\R\R}\partial_\R(g^{\R\R}\partial_\R\phi) = 0
\label{kruskal-eom}
\ee
Once again, although exact solutions are not known, we can find approximate solutions in the regions of large
$r$ ($R^2\gg T^2$) and $r$ close to the horizon ($|R|\gtrsim |T|$).  In both regimes, the leading behavior of
the solutions are the simple plane waves $u_{k'} \cong (2\pi\omega_{k'})^{-1/2}e^{-i(\omega_{k'} T -k' R)}$, with small corrections coming from the
extra terms in Eq.\ (\ref{kruskal-eom}).   The Bogoliubov coefficients are given by
\be
	\beta_{k'k} = \int_0^\infty dR\,\,
v^*_{k}\!\!\stackrel{\leftrightarrow}{\partial}_{\!\T}\!u_{k'}\Big|_{T=0}\,,
\ee
where $v_{k}$ are the mode functions in Eqs.\ (\ref{far-field-phi},\,\ref{near-field-phi}).
Thus we need to transform the latter functions from Schwarzschild to Kruskal coordinates in order to do the
integral.  In the large-$r$ and near-horizon regions, the respective transformations are approximately
\bea
	r &\cong& r_s\ln(R^2-T^2)\quad \hbox{\ \ (large $r$)}\nn\\
	r &\cong& r_s(1 + R^2 - T^2)\quad \hbox{(near horizon)}
\eea
and (exactly)
\be
	t = r_s\ln\left(R+T\over R-T\right)\,.
\ee
Remarkably, both approximate solutions (\ref{far-field-phi},\,\ref{near-field-phi}) have the same form in 
Kruskal coordinates,\footnote{ignoring the weak $R$-dependence from $1/r\sim 1/(r_s\ln(R^2))$ in the 
large-$r$ region} giving
\bea
	\beta_{k'k} &\cong& {i\over 4\pi\sqrt{\omega_k\omega_{k'}}}\int_0^\infty dR\left(\omega_{k'} - {2r_s\over R}\omega_{k}\right) 
	\exp(2ikr_s\ln R + i k' R)\nn\\
	&\sim& \sqrt{\left| k\over k'\right|}e^{-\pi k r_s}({\rm sgn}\, k' +  {\rm sgn}\, k)\Gamma(2ik r_s)
\eea
up to an overall phase and constant factor.  Comparing with Eq.\ (\ref{rindler-beta}) for the case of Rindler
space, we see that the expressions are essentially the same, with the acceleration of the Rindler observer
replaced by $(2 r_s)^{-1}$, which is the surface gravity of the black hole at the horizon.  Therefore if the
true vacuum state of the system corresponds to the Kruskal coordinate system, then the Schwarzschild observer
will see thermal radiation with temperature $T = (4\pi r_s)^{-1}$, which is Hawking's famous result (performed
here less elegantly but perhaps more pedagogically than he did).  For
astrophysical black holes, the temperature is exceedingly small, with
$k_B T \cong 5\times 10^{-12}$\,eV for a solar mass black hole.

In the Rindler case, it was fairly obvious that the Minkowski vacuum
should be the correct choice, since it applies for any inertial
observer, while the Rindler one is for a special class of observers
with just the right acceleration.  For the black hole, one needs a
different argument, that is beyond the scope of the present
exposition.  One can compute the stress-energy tensor of the quantum
fluctuations in the two different vacuum states.  For the
Schwarzschild vacuum, these diverge near the horizon, which would
cause a large back-reaction on the geometry that is not accounted for
by the Schwarzschild solution \cite{Birrell:1982ix}.  This choice is
therefore not self-consistent, and seems unphysical.  On the other
hand, the Kruskal vacuum $|0\rangle_{\K}$ gives a well-behaved result
for $_{\K}\langle 0|T^{\mu\nu}|0\rangle_{\K}$, which corresponds to a
thermal radiation bath at large $r$.  This gives a negligible
back-reaction except when the black hole is about to disappear, hence
it is consistent to ignore its effect on the geometry.

The black hole behaves approximately\footnote{The emission of
long-wavelength radiation, with wavelengths $\lambda\gtrsim r_s$, is
modified by ``greybody factors'' \cite{Hawking:1975vcx} that suppress their emission.} like a black body radiator,
emitting energy at rate $P = \sigma T^4$, where $\sigma = \pi^2/60$ is
the Stefan-Boltzmann constant in natural units $c=\hbar=k_B=1$. 
Equating this to the rate of black hole mass loss gives 
\be
	{dM\over dt} = -{1\over 3840\,\pi\,r_s^2} = -{1\over 15360\,\pi
(GM)^2}\,,
\ee
which can be integrated to give the lifetime of the black hole whose
initial mass is $M$:
\be
	t = 5120\,\pi\, G^2 M^3\,.
\ee
Only black holes with mass $M\lesssim 10^{12}$\,kg would evaporate
within the age of the Universe, $\sim 14$\,Gy.

\begin{figure}[b]
\leftline{\includegraphics[width=\textwidth]{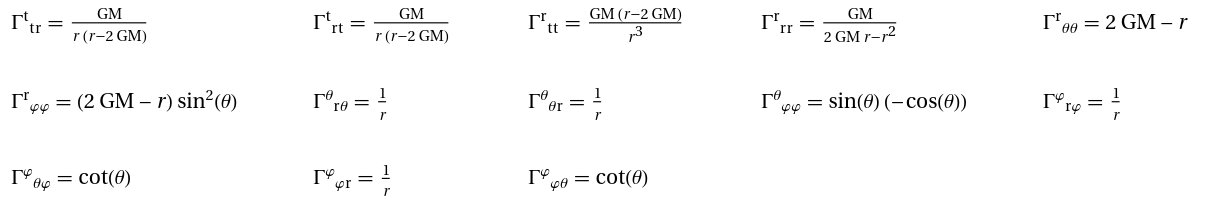}} 
%\leftline{\includegraphics[width=0.1\textwidth]{py3.jpg}}    
    \caption{Christoffel symbols for the Schwarzschild metric.
}
    \label{Xsfig}
\end{figure}

\subsection{Problems}

\noindent 10.1.  (a) Using symbolic manipulation, verify the
Christoffel symbols for the Schwarzschild metric shown in Fig.\
\ref{Xsfig}.\\
\indent (b) By comparing your results from (a) to Eq.\ (\ref{Xseq}),
corrected by footnote \ref{Xsfn}, find which of the latter differ from
the Taylor expansion of the exact result.\\
\indent (c) Show that making this correction accounts for the discrepancy
between problem 5.5 and the expected result for the perihelion
precession.\\

\noindent 10.2.  The clearest way to picture null geodesics in Eddington-Finkelstein coordinates is to define a new time
coordinate $v = t_* + r$.  Substitute this into the line element (\ref{EF}) and solve for the two slopes $dt_*/dr$ of
the light cones as plotted in the $r$-$t_*$ plane.  Show that at large $r$, the solutions approach 
$dt_*/dr = \pm
1$ as expected in Minkowski space, but below the horizon, both slopes are negative.  Plot the shapes of the
light cones as they evolve along the $r$ direction.\\

\noindent 10.3.  (a) Using symbolic manipulation tools, verify the Christoffel symbols and Riemann tensor
elements needed for section \ref{tidal-sect}.\\
\indent (b) It is also interesting to know what happens close to, but outside of the horizon.  Show that as $r\to r_s$,
$\dot r\to -1$ and $\dot t\to 1/(1-r_s/r)$ satisfies the geodesic equations.  (These are exact solutions; the
limit indicates that if you check a generic numerical solution, it tends toward these exact
solutions near the horizon.) 
Find the relevant Riemann tensor 
components to describe the longitudinal and transverse tidal forces and verify that they have the expected signs
(longitudinal stretching and transverse compression).  Show that the naive formula for the tidal accelerations
predicts that they diverge at the horizon.  This is not the physical stress experienced by the observer; it is a coordinate artifact. \\
\indent (c) To get the right answer, we first need to transform to an orthonormal coordinate system with basis
one-forms $d\hat t = \sqrt{1+2\phi}\,dt$ and $d\hat r =  dr/\sqrt{1+2\phi}$, whose metric looks flat at the
point that the observer is passing through.  In this basis, the covariant Riemann
components are given by $R_{\hat r\hat t\hat r\hat t}(d\hat r)^2(d\hat t)^2 = R_{rtrt}(dr)^2 (dt)^2$, and the
index can be raised using the Minkowski metric.  Show that $R^{\hat r}_{\ \hat t\hat r\hat t} = r_s/r^3$.
However $\dot{\hat t} = \dot t\sqrt{1-r_s/r}$, so the tidal acceleration still seems to diverge at the horizon,
though less badly than before.\\
\indent (d) One more step is needed to find the tidal force experienced by the observer (MTW p.\ 821 
\cite{Misner:1973prb}): we must do the Lorentz
boost from the static orthornormal system (which is at rest) to the traveler's frame, that is moving toward
the horizon.  In this frame, $\dot{\hat t}=1$ and $\dot{\hat r}=0$ instantaneously.  Using the symmetries of the
Riemann tensor, one can easily show that
$R^{\hat r}_{\ \hat t\hat r\hat t}$ does not change under the boost.  Prove this by Lorentz transforming just
the first two indices to start with.  Show that in the boosted coordinate system (denoted with a prime),
$R_{\hat r'\hat t'\alpha\beta} = (\Lambda_{\hat r'}^{\ \hat r}\Lambda_{\hat t'}^{\ \hat t}-
\Lambda_{\hat r'}^{\ \hat t}\Lambda_{\hat t'}^{\ \hat r})R_{\hat r\hat t\alpha\beta} = 
R_{\hat r\hat t\alpha\beta}$.  The other two indices transform in the same way.  Therefore 
the tidal acceleration remains finite at the horizon.\\
\indent (e) Take a representative human body to be 2\,m in height and 50\,cm in breadth.  Suppose that they can
withstand tidal accelerations of 100\,$g$ before being crushed or torn apart.  What is the
mass of the smallest black for which they could survive entry (feet first) into the horizon?\\
\indent (f) Inside the horizon, all the calculations are similar (don't redo them), but $t$ and $r$ exchange their roles.
 For what mass of black hole (expressed in solar masses) would the traveler be able to survive for one hour 
after crossing the horizon?  Assume the geodesic given in Section \ref{tidal-sect} with $C_r=1$.
Hint: for the given parameters, the observer reaches very close to the singularity in units of $r_s$. This saves
you from having to solve a transcendental equation.\\ 

\noindent 10.4.  (a) Verify the geodesic equations (\ref{schwgeo}), which we would like to solve for particles
falling radially toward the horizon starting from large distances.  Working in units where $r_s = 1$, suppose the
particle starts with proper radial velocity $\dot r_0$ at $t=\tau=0$.  Find the initial value $\dot t_0$
at $r=r_0$ such that $g_{\mu\nu}u^\mu u^\nu = 1$, where $u^\mu=(\dot t, \dot r)$.  \\
\indent (b) In Mathematica or Python, set up the system of geodesic equations as explained in class to numerically
solve for the geodesics, starting from $r_0 = 1000$, and with $\dot r_0 = 0$. 
How long does it take to reach the horizon in the proper time of the falling
massive particle? Call this value $\tau_{\rm max}$.  Determine it to at least three digits past the decimal
point (but don't round up since we don't want to pass the horizon).  
 Plot the solutions for $r(\tau)$, both for $\tau\in[0,\tau_{\rm max}]$ and for 
$\tau\in[\tau_{\rm max}-5,\tau_{\rm max}]$, to better see what is happening near the horizon. \\ 
\indent (c)  Plot the physical velocity $dr_*/dt_* = (\sqrt{|g_{rr}|}\dot r)/(\sqrt{g_{tt}}\dot t)$ 
and the corresponding Lorentz $\gamma$ factor (called $\gamma_{\rm mp}$ for ``massive particle''), and show
that $\gamma_{\rm mp}$ blows up as the object approaches the horizon.  Note: the Mathematica \texttt{Plot}
function will not show the full extent of the blowing up of $\gamma$ unless you add the directive
\texttt{,PlotRange->All} inside the command.  Otherwise it will zoom into what it considers to be the interesting
part of the plot.  Both ways can be useful.\\
A certain professor (who happens to be a condensed matter experimentalist) says, ``won't the observer get burned
up from behind by all the photons from starlight that are catching up with it, since they are getting infinitely blueshifted
as they approach the horizon?"  To answer this question:\\
\indent(d)  The limit that $\dot r_0\to -\infty$, $\dot t\to |\dot r_0|/(1 + 2\phi)$ describes a light ray,
so we can approximate a null (light-like) geodesic by taking $\dot r_0$ sufficiently large and negative.  Suppose
$\dot r_0 = -100$, and repeat parts (b) and (c), except you will notice you have to define $\tau_{\rm max}$ to 
5 or 6 significant figures to see the divergence in $t(\tau)$ near the horizon, and the interesting region
of the plots is $\tau\in [0.99,1]\tau_{\rm max}.$ \\
\indent Define $\tilde \tau = \tau_{\rm max} -
\tau$ as the time until reaching the horizon for this solution and for the one in part (b) so that both
solutions reach the horizon at the same $\tilde\tau=0$, and plot their ratio as a function of $\tilde\tau$.
(\texttt{LogLogPlot} can be useful to better visualize the behavior.)  
What is the maximum value of $\gamma_{\rm light}/\gamma_{\rm mp}$?
Does this help you to answer the professor's question?\\

\noindent 10.5.  Here you will modify your code from Problem 10.4 to use ingoing Eddington-Finkelstein coordinates.
In Mathematica, you will want to rename your results from \texttt{NDSolve} so that they don't overwrite
the solutions from Problem 10.4.  Then you will be able to compare the solutions for $r(\tau)$ in
 the two different coordinate systems without having to recalculate the Schwarzschild ones.\\
(a) Calculate the relevant Christoffel symbols to find the geodesic equations for $v$ and $r$.\\
(b) What will you use for $\dot v_0$ given initial values for $r_0$ and $\dot r_0$?\\
(c) Find the geodesic for a particle starting from $r_0=1000$ and $\dot r_0=0$ as in Problem 10.4, and show that
the solution for $r(\tau)$ is indistinguishable from the Schwarzschild case, except that it now extends all the
way to the singularity. By comparing $\tau_{\rm max}$ for this solution to that of Problem 10.4(b), how long
does it take the particle to go from the horizon to the singularity?  
Show that $v(\tau)$ is smooth all the way to $r=0$.\\
(d) Compute the ``physical velocity'' at the point where the traveler reaches the horizon,
and its associated gamma factor, and show that it is finite.
To do this, diagonalize the metric at $r=r_s$ to identify the timelike and
spacelike coordinate intervals (call them $dt'$ and $dr'$) in terms of $dv$ and $dr$.  \\

%\noindent 10.6.  Following Problems 10.4 and 10.5, to understand how the particle's $\gamma$ factor can diverge in one coordinate 
%system but stay
%finite in another, we need to show that the two systems are related by an infinite boost for points on the 
%horizon.  The local orthonormal frame in Schwarzschild coordinates (Problem 10.3(d)) is just Minkowski space.
%Define $(1+2\phi)=\epsilon$ in the Eddington-Finkelstein metric and diagonalize it for $\epsilon\ll 1$.  Show
%that the rotation matrix which diagonalizes it corresponds to a Lorentz transformation for an infinite boost when $\epsilon\to
%0$.  

\noindent 10.6. Consider a black hole whose spin is maximal, $a = r_s/2$, in the Boyer-Lindquist
coordinate system.  We want to solve for the equatorial orbits, $\theta = \pi/2$.\\
(a) Prove that inside the horizon ($r<r_s/2$), a massive particle must be moving in the 
$\phi$ direction in order for its 4-velocity to be timelike.\\
(b) Find the constraint equation which relates $\dot t$ to $\dot\phi$ for a massive particle on such
an orbit, using the line element (\ref{BLmetric}).  Write it in terms of the dimensionless variables
$v_\phi = r\dot\phi$ and $\rho = r/r_s$.  Then solve it for $\dot t$, approximately, in the limit of 
large $\rho$.  Using the Newtonian result for $v_\phi$, which of the small corrections should be most
important?  (It will turn out that none of them are important for part (d).)\\ 
(c) Using Fig.\ \ref{BLX} to find the appropriate Christoffel symbols, write the geodesic equation for
$r$ that will determine $\dot\phi$, using the same variables as in part (b).\\ 
(d) By combining the previous results, show that the allowed values of $v_\phi$ correspond to the
Newtonian result plus a small correction, whose sign is in the opposite direction from that expected
by rotational frame dragging!\\
(e) In order for the suprising result of (d) to be consistent with (a), it must be the case that 
the bias in $v_\phi$ against the black hole rotation disappears at some intermediate radius.  Using
a computer, evaluate the quadratic form for $\ddot r$ that is supposed to vanish, as a function of 
$\rho$ and $v_\phi$, once $\dot t$ is eliminated without making any approximations.  Plot it as a
function of $v_\phi$ at large $\rho$ and at $\rho\gtrsim 1$, and show that it crosses zero twice in the
former case but only once in the latter.  Show that the solution in which the particle orbits with 
opposite
angular momentum to the black hole spin disappears, requiring infinite energy in that limit.  At what critical value of $\rho$
 (to some number of
decimal places, by trial and error) does this happen?\\
(f) Can you find a way to analytically determine the critical radius?

\section{Cosmology}
One of Einstein's first applications of GR was to the Universe as a whole.  In 1917 it was not yet known that
the Universe was expanding, so he was looking for static solutions to the field equations, not in vacuum, but
with an average energy density $T^0_{\ 0} = \rho$.  The pressure from starlight today is negligible compared to
$\rho$.  We believe the Universe is spatially uniform on the large scales over which we averaged to approximate
$\rho$ as a constant.  But the metric can't be Minkowski space since this is a vacuum solution to Einstein's
equations.  The best guess is that the spatial part of the metric has constant curvature, a 3-sphere or a
hyperbolic space,
\be
	ds^2 = dt^2 - a^2(d\psi^2 + \sin^2\!\psi\, d\Omega_2^2)\hbox{\quad or\quad } 
dt^2 - a^2(d\psi^2 + \sinh^2\!\psi\, d\Omega_2^2)
\label{static-g}
\ee
where $a$ controls the size of the universe in the 3-sphere case, or the relative size for the hyperbolic
space, which is noncompact.

Consider the positive curvature ansatz.  The Einstein tensor for this metric is
\be
	G_{\mu\nu} = {1\over a^2}\,{\rm diag}(3,\,-a^2,\,-a^2\sin^2\psi,\,-a^2\sin^2\psi\sin^2\theta) \,
\ee
which does not resemble the desired $T_{\mu\nu}$.  However we notice that if we added a term proportional to the
metric, then it would have the desired form:
\be
	G_{\mu\nu} - {1\over a^2}\, g_{\mu\nu} = {2\over a^2}\,{\rm diag}(1,0,0,0)\,.
\ee
Then if we identified $2/a^2 = \kappa\rho$, our metric ansatz would be a solution to this modified version of
the Einstein equations.  How large would the Universe be?  The average density today is determined to be
$\sim 10^{-29}$\,g/cm$^3$.  Then one finds that $a = \sqrt{2/(\kappa\rho)} = 3350$\,Mpc  $\sim 10^{10} c$-y.
Therefore the new term in the field equation is exceedingly small with respect to the kinds of solutions we have
been concerned with previously; we would never have noticed its tiny effect on our solutions.  The spatial
curvature $3/a^2$ is so small that the metric is indistiguishable from Minkowski space to observers who can 
only probe a limited region of the Universe.  (They can rescale their local coordinates to absorb the factor of
$a^2$ in the metric.)   Einstein
therefore proposed to modify his original field equations to the more general form
\be
	G_{\mu\nu} - \Lambda g_{\mu\nu} = -\kappa T_{\mu\nu}\,,
\ee
where $\Lambda$ is known as the cosmological constant.   One can write it in a more intuitive way,
$G_{\mu\nu} = -\kappa(T_{\mu\nu} + (\Lambda/\kappa)g_{\mu\nu})$, which suggests that the new term represents an
extra contribution to the stress tensor, not associated with the matter in the Universe.  Instead, it is a
property of the vacuum.  Although this sounds strange at first, particle physicists later realized that quantum
fluctuations of the vacuum, consisting of momentary creation and annihilation of particle/antiparticle
pairs, should generically produce a nonzero value for $\Lambda$.  From this point of view, it is a contribution
to the total $T_{\mu\nu}$ that should in principle be present, and not just an {\it ad hoc} addition created to
agree with an outdated conception of cosmology.  The above solution is known as Einstein's static universe.
We can carry out the analogous procedure with the negatively curved metric.  One finds that $\Lambda = -1/a^2$
and the energy density of matter is $-\kappa/a^2$, which is unphysical.  Therefore only positive spatial
curvature gives rise to a sensible solution.

There are two problems with the static solution.  One is that it is unstable to small perturbations.
If one lets $a\to a+\delta a(t)$, then $\delta a(t)\to \pm \infty$ as a function of time.  The other is that E.\
Hubble and G.\ Lema\^\i tre discovered the expansion of the Universe in 1927.  Einstein was reported by G.\ Gamow
to have called $\Lambda$ the biggest blunder of his life, since if he had insisted on his original form, he
would have correctly predicted that the Universe cannot be static.  (Gamow was known for embellishing stories, so one
should perhaps take this claim with a grain of salt.)  In light of our modern understanding, the introduction
of $\Lambda$ was prescient, since we now have evidence that it is nonzero.  However, let's proceed in historical
order and construct time-dependent solutions with $\Lambda = 0$, which was the common belief once the Hubble
expansion was established.  We can use the same metric {\it ans\"atze} as in (\ref{static-g}) but now with
$a = a(t)$.  In addition there is a possible flat-space solution,
\be
	ds^2 = dt^2 - a^2(t)\, d\vec x^{\,2}
\ee
which is no longer equivalent to Minkowski space.  We can classify these three potential solutions by the
sign of their spatial curvature, $k = \pm 1, 0$.  It is then possible to find solutions to the original Einstein
equations with $\Lambda=0$.  Starting with the simplest case $k=0$, one finds
\be
	G_{00} = -3\left(\dot a\over a\right)^2;\qquad G_{ij} = \delta_{ij}(\dot a^2+ 2a\ddot a)\,.
\ee
Try a power-law solution, $a = a_0 t^p$.  Then $G_{ij}=0$ implies $p^2 + 2p(p-1)=0$, which has the 
nontrivial solution $p = -(2/3)$.   Hence $3(\dot a/a)^2 = 4/(3t^2) = \kappa\rho$.  We find a solution if
$\rho$ is decreasing as $1/t^2 \propto 1/a^3$.  This is exactly how $\rho$ {\it should} behave if it consists of
nonrelativistic matter, whose density is getting diluted by the expansion of the Universe.  According to this
solution, the present age of the Universe is related to its matter density by 
\be
	t_0 = \sqrt{1\over 8\pi G\rho_0}\sim 8\times 10^9\,{\rm y}
\ee
This is the same order of magnitude as the modern value 13.8\,Gy.  (One needs some nonzero contribution from
 $\Lambda$ to get the currently accepted value.)

For the curved metrics with $k=\pm 1$, the solutions are not so simple.  They must satisfy
\bea
\label{Friedeq}
	\left(\dot a\over a\right)^2 &\equiv& H^2 = -{k\over a^2} + {\kappa\rho\over 3}\,,\\
\label{Feq2}
	0 &=& k + \dot a^2 + 2a\ddot a\,.
\eea
These equations also correctly describe the $k=0$ case.  Eq.\ (\ref{Friedeq}) is known as the Friedmann equation
(or sometimes the first Friedmann equation, as there is a second one that can be written as a linear combination
of (\ref{Friedeq}-\ref{Feq2}) by eliminating $k$).  $H$ represents the relative rate of expansion (or
contraction) of the 3-volume, known as the Hubble parameter.  Its value today, $H_0$, is the Hubble constant.
This class of metrics is known as FRW for Friedmann, Roberton and Walker, or sometimes FLRW to include
Lema\^\i tre.

\subsection{The FLRW solutions}

We do not have closed-form solutions for Eqs.\ (\ref{Friedeq}-\ref{Feq2}) when $k=\pm 1$, but we can find the approximate
solutions near $t=0$, where they have the common feature that $a=0$.  This has a curvature singularity, known as
the big bang, which we will try to circumvent later on.  For now let's accept it as a necessary evil of the
solutions and look for power-law approximations $a= a_0 t^p$ near $t=0$.  This gives
\be
	{k\over a_0^2} + (p^2 + 2p(p-1))t^{2p-2} = 0\,.
\ee
For $k=-1$, this implies $p=1$ and $a_0 = 1$.  For $k=0$ it gives $p = 2/3$ and $a_0$ is undetermined, which
makes sense since for a flat infinite universe, we can always change $a_0$ by rescaling the coordinates.  (When
$k=-1$, even though space is infinite, $a_0$ controls its curvature, so its value is not arbitrary.)  For
$k=1$, we can satisfy the equation by taking $a = a_0 t^{2/3} + a_1 t^{4/3} +\dots$.  The terms going like
$a_0^2$ in (\ref{Feq2}) cancel as in the $k=0$ case, while those going like $a_0 a_1$ are constant, and cancel $k$ if
$a_0\,a_1 = -9/16$.  Once the behavior near $t=0$ is known, one can numerically integrate Eq.\ (\ref{Friedeq})
from some small $t>0$ to find the subsequent evolution.  It may depend on the initial condition for the matter
density, which we saw should get diluted as $\rho(t)= \rho_0/a(t)^3$.  There is a special value of $\rho$ today
called the critical density, $\rho_c$, defined in terms of the measured Hubble rate, such that the Universe has
$k=0$ if $\rho=\rho_c$.   It is just given by Eq.\ (\ref{Friedeq}) evaluated today:
\be
	\rho_c = {3 H_0^2\over 8\pi G} = 8.53\times 10^{-29}\,{\rm g/cm}^3 \approx 5\, m_H/{\rm m}^3\,
\ee
close to five hydrogen atom masses per cubic meter.  The present average density is determined to be close to
this number: the Universe seems to be spatially flat.  

By this definition, if $\rho < \rho_c$, $k=-1$, and if $\rho > \rho_c$, $k=+1$.  Without solving Eq.\
(\ref{Friedeq}) in detail, it is easy to deduce the qualitative behavior of the solutions.  If $k=-1$, 
the $k/a^2$ term will come to dominate as the Universe expands,  since $\rho$ is decreasing like $1/a^3$.
Then as $\rho$ becomes negligible, it will approach the solution $a = t$ as $t\to\infty$ and expand forever.  If $k=0$ and the Universe
consists only of nonrelativistic matter, the behavior $a\sim t^{2/3}$ persists forever.  If $k=1$, the curvature
term eventually cancels the $\rho$ term, at which point the expansion halts and reverses, resulting in a ``big
crunch'' at some future time.

In the early Universe, matter was relativistic, and could be regarded as a form of radiation, like photons,
since its energy far exceeded its mass energy, and one could approximate it as being massless.  This kind of 
energy density redshifts not as $1/a^3$, but rather $1/a^4$, because the wavelengths are getting stretched by
the expansion in addition to the particles' number density being diluted.  Even today, photons are
 present in the Universe, and make some small contribution to Eq.\ (\ref{Friedeq}).  Therefore one should 
write the total energy density as a sum of different components
\be
	\rho = \sum_i \rho_i = \sum_i {\rho_{i,0}\over a^{p_i}} 
\ee
where $p_i = 3$ for nonrelativistic matter and $p_i=4$ for radiation.   One could also pretend that the
curvature term is a form of energy density with $p_i= 2$ and thereby simplify the Friedmann equation 
to absorb $k/a^2$ into $\rho$.  There is one more important kind of energy density that must be reintroduced:
Einstein's cosmological constant.  It is indeed constant: it does not get diluted by the expansion.  It will
come to dominate over any other contribution to the energy density at late times, leading to exponentially
fast expansion.  In summary, the Friedmann equation can be expressed as
\be
	H^2 = {8\pi G\over 3}\left({\rho_{k,0}\over a^2} + {\rho_{r,0}\over a^4} + {\rho_{m,0}\over a^3} +
	\rho_\Lambda\right)\,.
\ee

\subsection{History of the Universe}

A convenient way of normalizing the scale factor $a$ is to define $a=1$ today so that the constants represent
their present values.  This requires rescaling $k$ away from $\pm 1$ if it is nonzero, but we can still refer to
$k=0,\pm 1$ as a shorthand for the sign of the curvature, and use $\rho_{k,0}$ to quantify its effect in the
present.  Another way of expressing the Friedmann equation is to introduce the fractions $\Omega_i = \rho_i/\rho_c$
and to write
\be
	H^2 = H_0^2\left({\Omega_k\over a^2} + {\Omega_r\over a^4} + {\Omega_m\over a^3} + \Omega_\Lambda\right)
\ee
By definition, $\sum_i\Omega_i = 1$.  The current measured values are 
(see section 2, ``Astrophysical Constants and Parameters'' of Ref.\ \cite{ParticleDataGroup:2024cfk})
\be
	\Omega_k = (7\pm 19)\times 10^{-4},\quad \Omega_r = 5.4\times 10^{-5},\quad
 \Omega_m = 0.315,\quad  \Omega_\Lambda = 0.685\,.
\ee
I will ignore $\Omega_k$ in the following.  
From these numbers, one might estimate that at early times when 
$1/a = 1/a_{\rm eq} = \Omega_m/\Omega_\gamma = 5400$, the
matter and radiation densities were equal.  This is an overestimate because it ignores the contribution of
neutrinos; a more careful determination gives $1/a_{\rm eq} = 3400$.  The temperature of the radiation at this time
was $T_0/a_{\rm eq} = 9200\,$K, using the
current temperature $T_0 = 2.7\,$K of the cosmic microwave background, corresponding to an energy of
$E_{\rm eq} =  kT_{\rm eq} = 0.3\,$eV, well below the ionization energy of hydrogen.  Thus the recombination of
electrons and protons into atoms happened somewhat earlier, at $1/a_{\rm rec} = 1100$.

Cosmologists prefer to talk about the redshift $z = 1/a -1$ instead of the scale factor when referring to
earlier epochs of the Universe. For $a\ll 1$, obviously $z\cong 1/a$.  It is so called since the wavelength
of light that has been traveling toward us since the time corresponding to that value of $a$ is stretched by
the factor $z$.

At temperatures $T\sim 0.1$\,MeV (in units where Boltzmann's constant $k_B=1$), the light elements first formed from binding of neutrons and protons,
by well-known nuclear physics properties.  This is known as big-bang nucleosynthesis (BBN), and it occurred at
$z_{\rm bbn} = 4.3\times 10^8$.  Such large redshifts are not very intuitive, and early-universe researchers
tend to refer to the temperature instead to characterize these earlier epochs.  We do not have direct
experimental evidence that the Universe was hotter than this, but theoretically it seems very likely, since it
is difficult to explain how the observed asymmetry of matter over antimatter could have developed at such low
temperatures, and one popular theory of baryogenesis (called leptogenesis) typically requires temperatures of 
at least $10^9$\,GeV.   In that case the next significant event before BBN was the QCD phase transition, in
which quarks and gluons combined into nucleons.  Before that, the electroweak phase transition occurred near
100\,GeV, when the Higgs boson got a nonzero vacuum expectation value, giving masses to all the Standard Model
particles.  Above that temperature, they were massless.  And somewhere along the way (we do not know when), the dark matter that
comprises 84\,\% of the total matter density emerged from the primordial plasma.

Characterizing the curvature of the Universe by a continuous parameter $\Omega_k$ is more physically reasonable
than the discrete choice $k=0,\pm 1$, since the 3D Ricci curvature at a given moment could take any real value.
There is something special about the choice $k=0$, of an exactly flat Universe: only in
this case is $\Omega_k/a^2$ well behaved into the past.  Put another way, suppose that $\Omega_k$ had a more
generic value of order unity.  If $\Omega_k\gtrsim 1$, the Universe could not survive to the present time, since
it would have already turned around and reached the big crunch.  On the other hand if $\Omega_k\lesssim -1$, it
would have become curvature dominated at an early time, and we would see galaxies accelerating away from each
other much faster now than is observed.  There seems to be a conspiracy leading to small curvature that is not
natural in the big bang.  This is known as the flatness problem.

\subsection{Conservation of stress-energy}
An important equation for FLRW cosmology can be derived from the
conservation of $T^{0\nu}$, that is $T^{0\nu}_{\ \ ;\nu}=0$:
\be
	\dot\rho = - 3H(\rho + p)\,.
\label{dotrhoeq}
\ee
Alternatively, it can be derived by combining the $00$ and $ii$
Einstein equations.  If there are no significant exchanges between the various 
contributions to the stress tensor (matter, radiation, vacuum energy),
then Eq.\ (\ref{dotrhoeq}) holds separately for each contribution
$\rho_i$ using its accompanying pressure $p_i$.  

Eq.\ (\ref{dotrhoeq}) has an intuitive
meaning:  the term $-3H\rho$ describes the decrease in energy density
caused by the expansion of the universe.  In the absence of pressure, it
would predict that $\rho\sim 1/a^3$ as expected for cold matter
particles.  The $-3Hp$ term describes the additional loss of energy
from the work that is done by the fluid as the Universe expands. 
For photons and other kinds of radiation, $p = \rho/3$ and one obtains
the $1/a^4$ redshifting of the energy density.  For the vacuum energy, $p =
-\rho$ and it remains constant.  In this case, the fluid is not doing
any work; rather work is being done on it to increase the total
energy, which exactly compensates the volume dilution effect.  This is
qualitatively similar to the stretching of a rubber band or membrane,
which also have negative pressure.

These results can be summarized by defining an equation of state
parameter
\be
	w = p/\rho
\ee
for each kind of energy density, with $w=0,\,1/3,\,-1$ respectively for
matter, radiation and vacuum energy.  If $w$ is constant in time, then
Eq.\ (\ref{dotrhoeq}) can be integrated to give $\rho \sim
a^{-3(1+w)}$.  More generally, one finds $\rho \sim a^{-3}\exp(-3\int
w\, d\ln a)$.  Theories of dark energy beyond Einstein's cosmological
constant are often characterized by their time-dependent equation of
state.

\subsection{Geodesics; the horizon problem}
The simplest and most informative geodesics in the FLRW geometry are those of photons, simply described by
\be
	ds^2 = dt^2 - a^2(t)\,d\vec x^{\,2}\quad\implies\quad {dx\over dt} = \pm {1\over a(t)}
\ee
independently of the direction of $\vec x$.  The solution is
\be
	x(t) - x_1 = \int_{t_1}^t{dt\over a(t)}
\ee
for a photon starting at $(t_1,x_1)$ and traveling to $(t,x)$.  Since the Universe has been approximately
matter-dominated for most of its existence, let's approximate $a(t) = (t_0/t)^{2/3}$, recalling that
$t_0 \sim 13\,$Gy is the present age.  For simplicity, set $x_1=0$; then $x(t) = 3(t - t_1^{2/3} t_0^{1/3})$.
For example, a photon that has been traveling since times $t_1\ll t_0$ has traversed a coordinate distance
$3t_0$, which equals the physical distance $a(t)x$ today since $a(t_0)=1$.  Hence the photon seems to have
traveled with speed $3c$.  This is an illusion, since at any given moment, it was traveling at $c$;
the space was expanding at the same time, leading to a larger distance traversed.  

Even two objects that are at
rest are moving away from each other as a result of this expansion.  This is how Hubble discovered the expansion:
distant galaxies are redshifted by an amount proportional to their distance, making them appear to be receding from
us with a speed that is also proportional to the distance, as indeed they are.  The redshift can be understood
as a Doppler shift, equivalent to the stretching of the wavelength by the expansion of the Universe.

We can learn an interesting feature of the FRW solutions from the null geodesics.  Consider the spherical surface
surrounding Earth at a distance corresponding to
$z=1100$, where recombination occurred and photons started to travel freely. This is called the surface of last
scattering (LSS), and constitutes the visible boundary of the Universe; even if there is no physical boundary, we 
cannot see beyond this radius $\sim 14\,$Gpc, from which the relic radiation of the big bang, the cosmic
microwave background photons, are reaching us.  
It corresponds to the time $t_{\rm
rec} = t_0/z_{\rm rec}^{3/2} \sim 360,000\,$y.  The farthest that a photon could have
traveled between $t=0$ and $t_{\rm rec}$ is $\Delta x = 3 t_{\rm rec}\sim 10^6\,c$y.  The physical area of such a region
today, projected onto the surface of last scattering, would be $\sim \pi(z_{\rm rec}\times 10^6\,c$y)$^2$, while the total area of
the LSS is $4\pi(14\,$Gpc)$^2$.  This means that the LSS is tiled by $\sim 7000$ causally connected regions,
that could not have communicated with each other at the time the CMB photons started traveling freely.

This fact becomes significant when we consider that the CMB photons coming from all different directions on the
sky are very nearly uniform, with fluctuations in their almost-perfect black body spectra that are only one part
in $10^{-5}$.  Why should CMB photons coming from opposite directions be so nearly the same, since they emerge
from regions that were never in contact with each other?  Generically one might expect the big bang to be a
chaotic state with order one fluctuations in particles' energies, rather than $O(10^{-5})$.  This seeming
conspiracy is known as the horizon problem, since the edge of each causally connected region represents a
boundary beyond which prior information should not have had time to propagate and homogenize the fluctuations.

\subsection{Cosmic Inflation}
We observed several unsatisfying features of the big bang model, beyond the fact that it has an initial
curvature singularity.  In 1979, physicists were worried about an additional problem: grand unified theories, that
achieved unification of the three forces of nature at a very high temperature, predicted that extremely heavy
particles, magnetic monopoles, should be produced in such great numbers that the energy density of the Universe
would far exceed $\rho_c$, leading to a closed Universe that collapsed after a brief existence 
\cite{Preskill:1979zi}.  This was known
as the monopole problem.  A.\ Guth, in searching for solutions to this problem, found one that he realized would
also solve the other problems of the big bang, which is the mechanism he dubbed as inflation \cite{Guth:1980zm}.  The idea is that
the Universe was temporarily dominated by vacuum energy at early times, leading to approximately exponential expansion
$a(t) \sim \exp(\sqrt{\Lambda/3}\,t)$, the solution known as de Sitter space for constant $\Lambda$.  
This vacuum energy would eventually decay into ordinary particles and radiation,
giving rise to a Universe resembling the big bang but without the singularity.  The period of exponential
expansion would mean that the entire observable Universe (and more) could fit within one of the causally
connected regions that were tiling the LSS in the big bang model, thereby solving the horizon problem.  And any
initial curvature would get quickly inflated away during inflation, solving the flatness problem.  Magnetic
monopoles would also get inflated away, if they existed before inflation began.

The problem is how to get vacuum energy to evolve with time instead of being a constant, like the cosmological
constant.  This requires invoking a new kind of field, analogous to electric or magnetic fields, but being a
scalar rather than a vector quantity.  It is called the inflaton, and in general could depend on time and space,
$\phi(x^\mu) = \phi(t,\vec x)$, but we will be interested in configurations that depend only on time, similarly
to our assumption about the spacetime metric.  (This behavior for $\phi$ will be at least partially
justified {\it a posteriori}.)   The Lagrangian for such a field is similar to the ones we have written
previously for weak gravitational fields and for the vector potential, but simpler since $\phi$ carries no
Lorentz index:
\be
	S = \int d^{\,4}x\sqrt{|g|}\left(\sfrac12 g^{\mu\nu}\partial_\mu\phi\partial_\nu\phi - V(\phi)\right)\,.
\label{Sinf}
\ee 
One important difference is that we can write a potential energy density $V(\phi)$ whose analog wasn't present for
$h^{\mu\nu}$ or $A^\mu$, apart from the linear source terms such as $A^\mu J_\mu$ or $h^{\mu\nu}T_{\mu\nu}$.  
This 
potential, which need not be linear, will play the essential role of providing a new source of vacuum energy, 
that is not just a
constant.  To see this, we can take advantage of Hilbert's definition (\ref{Tdef}) of the stress-energy tensor and compute
its contribution from the $\phi$ field,
\be
	T_{\mu\nu} = {2\over{\sqrt{|g|}}}{\delta S\over \delta g^{\mu\nu}} = \partial_\mu\phi\partial_\nu\phi 
	- g_{\mu\nu}\left(\sfrac12 g^{\alpha\beta}\partial_\alpha\phi\partial_\beta\phi  - V(\phi)\right)\,.
\ee
Suppose that $\phi$ is nearly uniform so that we can ignore the derivatives.  Then $T_{\mu\nu} = g_{\mu\nu}V$
has the same form as the cosmological constant, but $V$ can now vary in time as $\phi$ moves in its potential.

Consider the inflaton action in the background of the FLRW geometry, in the limit where its spatial gradients
are negligible and it only varies with time:
\be
	S = \int d^{\,4}x\, a^3(t)\left(\sfrac12\dot\phi^2 - V(\phi)\right)\,.
\ee
The Euler-Lagrange equation for $\phi$ is straightforward to derive:
\be
	\ddot\phi + 3H\dot\phi = -{dV\over d\phi}\,,
\label{infEOM}
\ee
where $H=\dot a/a$ is the Hubble parameter.  This looks like a particle in a potential, with a damping term.
In fact if $V$ is quadratic in $\phi$ and $H$ is constant, it is just a damped harmonic oscillator.  The above
equation of motion is coupled to the expansion rate through the Friedmann equation.  If we imagine that there is
initially no matter in the Universe and only $\phi$ provides the energy density, then it reads
\be
	H^2 = {8\pi G\over 3}\rho = {\kappa\over 3}\left(\sfrac12\dot\phi^2 + V(\phi)\right)\,.
\ee

In general, one cannot solve the coupled equations for $\phi(t)$ and $a(t)$ analytically, but there is a
situation where approximate analytic solutions exist.  This is known as the {\it slow-roll approximation}, which works
when $V$ and hence $H$ are so large that $|\ddot\phi|\ll H|\dot\phi|$, and $\dot\phi^2\ll V$.  Then $H\approx
\sqrt{\kappa V/3}$, and the inflaton EOM becomes
\be
	\sqrt{3\kappa V}\,\dot\phi = -V'
\label{SReq}
\ee
which can be reduced to quadrature.  Then 
Eq.\ (\ref{SReq}) can be written as
\be
	{\sqrt{V}\over V'}d\phi = -\sqrt{1\over 3\kappa} dt
\label{SReq2}
\ee
 Let's illustrate using the potential $V = \sfrac12 m^2\phi^2$, which is known as the {\it chaotic inflation}
model \cite{Linde:1981mu}\cite{Linde:1983gd}. Eq.\ (\ref{SReq2}) has the solution
\be
	\phi = \phi_0 - \sqrt{2\over 3\kappa} m(t-t_0)\,.
\ee
Notice that $\ddot\phi = 0$ in this case, so that the slow-roll approximation is exact, except that we ignored the
contribution to $H$ from the kinetic energy of the field.  When is this neglect justified?  We need
\be
	\sfrac12\dot\phi^2 = {1\over 3\kappa} m^2 \ll \sfrac12 m^2\phi^2\,,
\ee
which implies $\phi \gg \sqrt{2/3\kappa} = 1/\sqrt{12\pi G}$.  Recalling the definition of the Planck mass
(\ref{PlanckMass}), the condition becomes $\phi\gg 6M_p\sim 10^{20}\,$GeV.  This is an enormous energy scale,
where the classical description of gravity might be breaking down in favor of an as-yet unknown theory of 
quantum gravity.  Other models of inflation can circumvent this difficulty, but since the dynamics of 
chaotic inflation are so simple, let us ignore this potential problem for now and bravely forge ahead.

Knowing the time-dependence of $\phi(t)$, at least in the region where $\phi\gg M_p$, we can solve the Friedmann
equation for the scale factor.  For simplicity take $t_0 = 0$ to mark the beginning of inflation:
\be
	H = {\dot a\over a} \cong \sqrt{\kappa\over 6}m \left(\phi_0- \sqrt{2\over 3\kappa} m t\right)\,,
\ee
whose solution is
\be
	a(t) = a_0\exp\left(\sqrt{4\pi\over 3}\,{m\phi_0\over M_p}\,t - \sfrac16 m^2 t^2\right)\,.
\ee
At early times, we can ignore the $m^2 t^2$ term, and the solution behaves like that of a cosmological
constant, $a = a_0\exp(\sqrt{\Lambda/3}t)$.  Effectively, $\Lambda$ is decreasing with time as the field
rolls down its potential.  When it reaches values $\phi\lesssim M_p$, the slow roll approximation is no longer
valid.  In fact, the $3H\dot\phi$ term in the EOM becomes subdominant to $\ddot\phi$, and we can treat the
system like an undamped harmonic oscillator.  By this time, $a(t)$ will no longer grow exponentially with time;
one can show that it grows as $t^{2/3}$ like in a matter-dominated universe, because the oscillating field
energy damps as $1/a^3$, just like ordinary matter.  Inflation has ended and the universe resembles that of the 
big bang, but without having started from a singularity. 

How can the oscillating scalar field be said to resemble the big bang?  In fact, we need to assume that the
inflaton is coupled to photons and ordinary matter, such that these oscillations decay into the particles
we are familiar with.  This can be easily arranged, and the process of transferring the inflaton's oscillatory
energy into normal particles is called {\it reheating}.  The rationale for this name is the idea that the
Universe might have started in some hot thermal state before inflation, but because of the exponential expansion
during inflation, these original particles got so redshifted and diluted by the expansion that they were
inflated away to nothing.  For example, suppose that $\phi_0 = 30\, M_p$ and inflation continued until $\phi =
6 M_p$.  One finds that $a \cong e^{370}a_0 $, so the volume of the Universe increased by a factor of $e^{1110}$
during inflation, and the temperature decreased by a factor of $\sim 10^{160}$.  Any initial curvature was reduced
by the factor $10^{320}$, explaining why the Universe is so flat today.  If the reheating process was efficient,
then the oscillatory energy density of  the inflaton becomes the radiation energy density $\rho\sim T^4$ of the
plasma at its maximum, or reheating, temperature $T_{rh}$.  In the chaotic inflation model, $T_{rh}\sim
\sqrt{mM_p} \ll M_p$, depending on the mass $m$ of the inflaton particle.

If one repeats the exercise of light-like geodesics during the inflationary epoch, obviously they travel 
exponentially farther than in the matter-dominated universe, because of the rapid expansion of space.
This implies that the regions of causal contact at the surface of last scattering are no longer small, like in
the big bang model.  In fact they become so large during inflation, they completely dwarf the LSS, and it is now
clear that fluctuations in the temperature can be made very small.  The initial conditions during reheating were
extremely homogeneous because any spatial fluctuations in the inflaton that would have given rise to temperature
fluctuations during reheating have been stretched to enormous wavelengths, far beyond the size of the LSS at the
time when the cosmic microwave background formed.  This (in words) is how inflation solves the horizon problem.

But the above description makes it sound like perhaps there should have been no fluctuations at all, practically
speaking.  We don't
want the Universe to be that homogeneous, since after all, stars and galaxies need to eventually form, and they
need some small initial fluctuations to get started.  Remarkably, inflation also gives an answer to this puzzle.

\subsection{Inflationary origin of density perturbations}
\label{sect-inf-pert}

The previous description of inflation treated $\phi$ as a purely classical degree of freedom.  But the classical
description is only supposed to be an approximation to quantum mechanics.  When we consider the quantum behavior
of $\phi$, we find that it cannot be exactly homogeneous in space.  Suppose for the moment that space was
compact, say a three-torus of length $L$ in each direction.  Then $\phi$ could be decomposed in a Fourier series in each spatial direction and we
could write the general form at any given time as
\be
	\phi(t,\vec x) = \sum_{\vec k} b_{\vec k}(t)\,e^{i\vec k\cdot\vec x}\,,
\label{Fdecomp}
\ee
where $\vec k = (i,j,k)\pi/L$ for integers $i,j,k$.  
The perfectly homogeneous mode is represented by $b_{0}$ and the spatial fluctuations by all the other
terms.  Each of these terms can be considered as an independent quantum mechanical degree of freedom.
In the previous discussion, we set all the other $b_{\vec k}$'s to zero, but quantum mechanically, they cannot be
exactly zero.  There is the uncertainty principle, which tells us, roughly speaking
\be
	\Delta\dot b_{\vec k}\, \Delta b_{\vec k} \gtrsim {\hbar\over 2}
\label{uncert1}
\ee
This can't be exactly right, since the dimensions aren't correct.  To fix it, we need to look at the action for
the $b_{\vec k}$ degrees of freedom, and compare to a more familiar system.  Substitute (\ref{Fdecomp}) into the
action (\ref{Sinf}), for simplicity in Minkowski space and for the potential $V = \sfrac12 m^2\phi^2$.  Doing
the integral over space, it becomes
\be
	S = \sfrac12 L^3\sum_{\vec k}\int dt\left(|\dot b_{\vec k}|^2 - (m^2 + \vec k^{\,2})|b_{\vec k}|^2\right)\,.
\ee
We see that this is an infinite sum of harmonic oscillators of increasing frequencies.  For each mode, if we
identify $M=1/L$, $X_{\vec k} = L^2 b_{\vec k}$ and $\omega_k^2 = (m^2 + \vec k^{\,2})$, the action becomes
\be
	S = \sfrac12 M \sum_{\vec k}\int dt\left( |\dot X_{\vec k}|^2 - \omega_k^2 |X_{\vec k}|^2\right)\,.
\ee 
Therefore Eq.\ (\ref{uncert1}) should correctly read
\be
	L^3 \Delta|\dot b_{\vec k}|\, \Delta |b_{\vec k}| \gtrsim {\hbar\over 2}\,,
\label{uncert2}
\ee
which has the right dimensions, and takes into account that the Fourier coefficients are complex.

This was in flat space, but we are interested in the situation during inflation.  The crucial observation is
that the universe is expanding as $a\sim e^{Ht}$ with $H$ approximately constant.  The lightlike geodesics
obey $dx/dt = 1/a \sim e^{-Ht}$, which has the solution $x = H^{-1}(1-e^{-Ht})$.  Therefore $1/H$ acts as an
effective cutoff on the size of spatial regions; light cannot travel further.  It suggests that the uncertainty
relation (\ref{uncert2}) should be replaced by
\be
	\Delta|\dot b_{\vec k}|\, \Delta |b_{\vec k}| \gtrsim H^3{\hbar\over 2}\,
\ee
in the inflationary universe.  In natural units $\hbar = 1$, this can be satisfied if each Fourier
mode has quantum uncertainty
\be
	\Delta|b_{\vec k}|\sim H\,,\quad \Delta|\dot b_{\vec k}| \sim H^2
\ee
Therefore each Fourier mode in the inflaton field gets a quantum fluctuation of order $H$ (its value during
inflation).  These are the fluctuations that give rise to the small departures from homogeneity in the cosmic
microwave background, and provide the seeds for formation of stars and galaxies, through gravitational
attraction to the slightly more dense regions.

To the extent that $H$ is approximately constant, this predicts that the amplitude of fluctuations should be the
same on all distance scales.  To state this more quantitatively, one constructs a correlation function for
temperature fluctuations of the CMB as a function of their separation on two points on the sky, separated by
some distance $\vec y$,
\be
	\xi(\vec y) = \left\langle{\delta T(\vec x)\over T_0}\, {\delta T(\vec x + \vec y)\over
T_0}\right\rangle
\ee
where we average over positions $\vec x$.  Then Fourier transform it to obtain what is known as the power
spectrum of the fluctuations,
\be
	\int d^{\,3}y\, \xi(\vec y)\, e^{i\vec k\cdot\vec y} = P(k) = A_0 (k/k_0)^{n_s-1}\,.
\ee
The power-law parametrization is a conventional way to try to fit the observed spectrum, and inflation predicts
that $n_s$ is close to 1, in agreement with the approximate scale-invariance alluded to above:
\be
	n_s \cong 1 -6\epsilon + 2\eta\,,
\ee
where $\epsilon = \sfrac12 M_p^2 (V'/V)^2$ and $\eta= M_p^2 V''/V$ are known as the slow-roll parameters, and
they must be small in order for inflation to last sufficiently long.  The Planck satellite has measured
$n_s = 0.965$ \cite{Planck:2018vyg}.  Remarkably, this has ruled out the chaotic inflation models, except for
those with with small fractional powers like $V\sim \phi^{2/3}$ or flatter \cite{Planck:2018jri}.  Inflation
also predicts the magnitude of the power spectrum, $A_0 = H^4/(2\pi \dot\phi)^2 \sim 10^{-10}$, which gives a
constraint on the energy scale of inflation versus the flatness of the potential.

\subsection{Dark Energy}
Recall that the Universe is currently dominated by vacuum energy, $\Omega_{\Lambda} = 0.65$, which we associated
with a cosmological constant, for simplicity.  Inflation taught us that $\Lambda$ can be generalized by the slow
rolling of a scalar field; why not have such a field acting in the late Universe?  In this context, it is
known as quintessence \cite{Wang:1998gt,Carroll:1998zi}.  To distinguish it from a cosmological constant, it is
useful to compare the stress-energy tensors for the two cases:
\be
	T_{\mu\nu} = {\Lambda\over\kappa}\, {\rm diag}(1,-1,-1,-1) = {\rm diag}(\rho,p,p,p)
\ee
which tells us that $p=-\rho = \Lambda/\kappa$ for pure vacuum energy.  For a homogeneous scalar field,
one instead finds
\be
	\rho = \frac12\dot\phi^2 + V(\phi)\,,\qquad p  = \frac12\dot\phi^2 - V(\phi)\,.
\ee
The equation of state parameter $w = p/\rho$ summarizes the difference,
\be
	w = \left\{\begin{array}{ll} -1, &\hbox{cosmological constant}\\
	                        { \frac12\dot\phi^2 - V\over  \frac12\dot\phi^2 + V},& \hbox{quintessence}
	\end{array}\right.\,.
\ee
A simple parametrization to search for time-dependent departures of $w$ from the prediction of $\Lambda$CDM,
the standard
cosmological model with constant vacuum energy), is to suppose that $w(a) = w_0 + (1-a) w_a$ as a function
of redshift, for $a$ not too much smaller than $1$.

Until recently, cosmological data has been consistent with the $\Lambda$CDM prediction $w_0=-1$, $w_a = 0$.
The Dark Energy Spectroscopic Instrument (DESI) has measured redshifts of $\gtrsim 10^7$ galaxies out to $z
=4.2$ in order to constrain the expansion history $H(z)$ of the Universe.  They have found a $3\,\sigma$
preference for dynamical dark energy over cosmological constant, the strongest evidence so far the current
expansion may be incompatible with $\Lambda$CDM \cite{Gu:2025xie}.  

If this turns out to be correct, it exacerbates one of the most difficult theoretical discrepancies in physics.
Naive estimates in quantum field theory predict that $\Lambda$ should be 60 to 120 orders of magnitude larger
than the observed value.  For decades, theorists have searched for a mechanism whereby it could be naturally
vanishing, so far without any convincing success.  The urgency of this problem diminished when supernova
observations of $H(z)$ indicated that $\Omega_\Lambda \sim 0.7$ 
\cite{SupernovaSearchTeam:1998fmf,SupernovaCosmologyProject:1998vns}.  Then $\Lambda$ should not be zero, but
merely small.  There was already an anthropic reason for it to be small \cite{Weinberg:1988cp}: life as we know
it would not have been able to develop if $|\Lambda|$ was much larger than observed: the Universe would expand
or collapse too fast for structures to form.  Hence if one had a
theoretical framework that predicted the existence of many universes, with a range of values of $\Lambda$, 
then it would be natural to find ourselves in one where life was possible.   String theory appeared to do just
that, by providing a vast landscape of possible ground states, with varying constants of nature.  
But if $\Lambda$ is supplemented by a dynamical field, then we have to
explain not only why $\Lambda$ is so small, but also the smallness of
the additional
contributions to the energy from the field's kinetic and potential
energies.  {\it A priori,} there is no reason for them to have such small
values.

On the other hand, positive vacuum energy appears to be very hard to achieve in string theory, which prefers
$\Lambda < 0$. Perhaps the most plausible explanation is that the true minimum of the quintessence potential
corresponds to a large negative value of $\Lambda$, and we happen to be passing through the part of the
potential where $V + \Lambda$ is small but positive today. In that case, the Universe is on its way to the big
crunch originally predicted for $\rho>\rho_c$, even though we observe $\rho$ very close to $\rho_c$ today.

\subsubsection{Numerical solutions}
To test scalar field models of dark energy, one needs to solve the equation of motion, which has exactly the same
form as Eq.\ (\ref{infEOM}) for inflation.  Numerical solutions are generally required. For numerics, it is usually convenient to
rewrite the second-order o.d.e.\ as two coupled first-order equations by defining the canonical momentum
$\pi = d{\cal L}/d\dot\phi = \dot\phi$ and treating it as an independent variable.  Then
\bea
	\dot\phi &=& \pi\,;\nn\\
	\dot\pi &=& -3H\pi - {dV\over d\phi}
\eea
with $H^2 = (8\pi G/3)(\rho_m + \sfrac12\pi^2 + V(\phi))$.  Here $\rho_m$ is the density of matter (baryons plus
dark matter), assuming the
dark energy is only coming to dominate at late times when radiation makes a negligible contribution. 

In the above form, one must also solve the Friedmann equation $\dot a = a H$ as a third equation, since $a$
enters into $\rho_m = \rho_c\Omega_m/a^3$.  Furthermore, cosmic time $t$ is an awkward independent variable for numerical cosmology, since
it involves large numbers $\sim 13\,$Gyr and can vary by orders of magnitude.  Both of these annoyances can be circumvented by choosing a
better time variable, which I will call $u = \ln a$.  Then $du/dt = a^{-1}da/dt = H$, and one can easily show
that  the transformed equations are
\bea
	\phi' &=& {\pi\over H}\,;\nn\\
	\pi' &=& -3\pi -{1\over H}{dV\over d\phi}\,,
\label{EOMSquint}
\eea 
where primes denote $d/du$.  In this variable, cosmological evolution will take place over intervals of order
unity in $u$, rather than the orders of magnitude that could be entailed by using $t$, and it is no longer
necessary to solve explicitly for $a(t)$, although one can of course do so, if desired.

Nevertheless, Eqs.\ (\ref{EOMSquint}) are not yet fully optimized for numerical solutions since all the
quantities are still dimensionful, and $H$ involves numerically cumbersome constants like $G$ and $\rho_c$.
A useful
technique is to adopt {\it geometrized} units, for example $G=1$ and $c=1$.  Here I will instead choose
$8\pi G/3=1$ to maximally simplify the form of $H$.  We are already accustomed to $c=1$ units, which teach us
that lengths and times can be used interchangeably for denoting distances.  What does it mean to set $G=1$ or 
$G = 3/8\pi$?  Notice that $G/c^2$ has units of m/kg.  Therefore we see that mass or energy can equivalently be
expressed in units of length (or time) with this choice.  Just like with any such system of units (such as the
natural units introduced in chapter \ref{sect:static}), we merely
need to restore the necessary powers of $G$ (or $8\pi G/3$) and $c$ to get quantities back into the familiar units.

Thus, setting $8\pi G/3=1$, the Hubble parameter becomes
\bea
	H &=& \sqrt{\rho_c\Omega_m e^{-3u} + \sfrac12\pi^2 + V}\nn\\
	  &=& H_0\sqrt{\Omega_m e^{-3u} + (\sfrac12\pi^2 + V)/\rho_c}\nn\\
	  &\equiv& H_0\, \hat H(u)\,,
\eea
where we have used the fact that $H_0 = \sqrt{\rho_c}$ (since $8\pi G/3=1$) and defined the dimensionless Hubble
parameter $\hat H$.  The prefactor $H_0 = \sqrt{\rho_c}$ is an annoying number that we
would like to eliminate from the EOMs in our code.  This can be done by defining the dimensionless canonical
momentum $\hat\pi = \pi/\sqrt{\rho_c}$, and the dimensionless potential $\hat V = V/\rho_c$.  Notice that
$\phi$ has dimensions of $\pi/H_0$ = $\pi/\sqrt{\rho_c}=\hat\pi$ which is dimensionless, hence there is no need
to rescale $\phi$, but to be pedantic I will write $\phi = \hat\phi$.\footnote{This is consistent with Eq.\
(\ref{phi-phihat}) since we have already set $8\pi G/3=1$}\  
One readily confirms that the final form of the EOMs is
\bea
	\hat\phi' &=& {\hat\pi\over \hat H}\,;\nn\\
	\hat\pi' &=& -3\hat\pi -{1\over \hat H}{d\hat V\over d\hat\phi}\,
\label{EOMSquint2}
\eea
which of course is identical to Eqs.\ (\ref{EOMSquint}) except for the hats.  The advantage is that we have now
set $\rho_c=1$ in addition to the original choice of $G$.  Logically this is possible because we still had the
freedom to choose a unit of length or mass even after fixing $G$.  In this way, the EOMs become simple to code,
and all variables are expected to vary on scales of order unity. 

After solving the equations, one may wish to interpret them in terms of the usual dimensionful variables.  If one is interested
in the energy density, for example, it is simple: $V = \rho_c\hat V$, and no factors of $8\pi G/3$ are needed.
But in general the latter will be required too.  Consider $\phi$: since $\dot\phi^2$ is an energy density,
$\phi$ has dimensions  of (kg/m s$^2)^{1/2}$s = (kg/m)$^{1/2}$, which is the same as that of $c/\sqrt{G}$.
Therefore the relation between the original $\phi$ and its
dimensionless version is
\be
	\phi = 1\times\hat\phi = \left(8\pi G\over 3c^2\right)^{-1/2} \hat\phi\,.
\label{phi-phihat}
\ee 
If we further wish to convert to the natural units of particle physics
where $\hbar=c=1$, the conversion factor becomes
\be
	\left(8\pi G\over 3\hbar\, c^5\right)^{-1/2} = 4.2\times
        10^{18}\,{\rm GeV}\,.
\ee
Hence the natural scales for the values of the field are close to the
Planck scale, that characterizes quantum gravity.  In these units,
$\rho_c \cong 3.65\times 10^{-47}\,$GeV$^4$ is extremely small.
The kinetic energy of the scalar field $\sfrac12\dot\phi^2$ can be
comparable to $\rho_c$ despite $\phi$'s huge magnitude by virtue of
its extreme slowness: $\dot\phi \sim H\phi$.  
Restoring units to the Hubble parameter is straightforward: 
$H = \sqrt{(8\pi G/3)\rho_c }\,\hat H = H_0\, \hat H$, as we already knew.
In natural units, $H_0\cong 1.6\times 10^{-42}\,$GeV.  Hence
$\dot\phi^2 \sim \rho_c$, as must be the case if the quintessence
field kinetic energy is comparable to its potential energy, and is
dominating the total energy of the Universe today.

\bibliographystyle{utphys}
\bibliography{sample}

\begin{appendix}
\section{In-class problems}
\subsection{Chapter 1}

1.1. The Earth has mass $6\times 10^{24}\,$kg and radius 6400\,km.
Recalling that $g \approx 10\,$m/s$^2$, compute Newton's constant in
MKS units.  (To one significant figure if you want.)
\medskip

\noindent 1.2. Suppose a civilization at a distance $R$ from Earth wanted to send
gravitational signals like in Eq.\ (\ref{fret}), by shaking a large
mass $M$ with a displacement $x = A\sin(\omega t)$.  The observer at
Earth tries to measure the induced displacement on a test mass that is
free to respond to the force. (a) Estimate in order of magnitude the 
amplitude of the measured displacements.  (b) Estimate the number
using
the distance to Jupiter ($R = 10^9$\,km), $A = 10\,$m,  $M=1000\,$kg and $\omega =
2\pi$/s.  % Answer: GMA/(R^3 omega^2) = 1e-44 m
\medskip

\noindent 1.3. (a) Write down a solution to the wave equation
(\ref{eom1}) in the case where the source term on the right hand side
vanishes, and the wave is moving in the $x$ direction.  (b) Write the
corresponding solution for a spherical wave moving outwards from the
origin.
\medskip

\noindent 1.4. A physicist who likes to use $c=1$ units makes the
following peculiar statements.  Translate them into normal units.
(a) ``My house is a $2\times 10^{-5}\,$s bike ride from here.'' % 3km
(b) ``Last time I checked, I weighed $5\times 10^{18}$\,J.'' % 55.6 kg
(c) ``My baby girl is $2\times 10^{15}\,$m old today.''  % 3 months
(d) ``My car had a momentum of 56 milligrams when I crashed and totaled
it.'' % 17 kg m/s

\medskip
\noindent 1.5. Light passes by a star of mass $M$ with impact parameter $b$.
(a) Estimate the deflection angle in order of magnitude, using Newtonian
gravity.  (b) Put in the numbers for the Sun, 
$M=2\times 10^{30}\,$kg, and the orbit of Mercury, $b=47\times
10^6$\,km.  % GM/(b c^2) = 3e-8 rad

\medskip
\noindent 1.6. (a) A particle moves along the $x$ axis with energy $E$ and
momentum $p$.  Find its new energy and momentum in a reference frame
whose boost parameter is $\beta = v/c$, moving in the same direction
as the particle.   (b)  Consider $T^{\mu\nu} = {\rm
diag}(\rho,-\rho,-\rho,-\rho)$.  How does it transform under the above boost?

\medskip
\noindent 1.7.  Show that $\eta^{\mu\nu} - h^{\mu\nu}$ is inverse to
$\eta_{\mu\nu} + h_{\mu\nu}$ to linear order in the perturbation.

\subsection{Chapter 2}

\noindent 2.1. Consider the line element in 2D flat space, $ds^2 = dx^2 + dy^2$,
with metric tensor $\delta^{ij}$. (a) Transform to polar coordinates $x = \rho\cos\theta$, $y=\rho\sin\theta$ to
find the new form of the line element and metric tensor, by directly computing $dx$ and $dy$ in terms of
$d\rho$ and $d\theta$.  (b) Find the new metric tensor by formally transforming the metric tensor using the
matrices of partial derivatives.\\
(c) How could you change the metric of (b) to describe a cone with opening angle $45^\circ$ relative to
the symmetry axis?

\medskip
\noindent 2.2
(a) Compute the following quantities.
\be
	x^\mu_{\ ,\nu},\quad x^{\mu,\nu},\quad \partial_\mu(x_\alpha x^\alpha x^\beta)\, \quad \partial^\mu\partial_\mu(x_\alpha x^\alpha x^\beta)
\ee
(b) Invert the coordinate transformation $x^\mu = x'^\mu(1 + f(x'^2))$ to linear order in $f$.  Do the same for 
$x^\mu = x'^\mu(1 + f(x'^2))^{1/2}$\\
(c) Compute the Jacobian matrix $\partial x'^\alpha/\partial x^\beta$ for the transformation $x^\mu = x'^\mu(1 + f(x'^2))^{1/2}$, again
at linear order in $f$.

\subsection{Chapter 3}

\noindent 3.1. (a) Express the Minkowski metric tensor as a sum of outer products of the basis 4-vectors
$\hat t$, $\hat x$, $\hat y$, $\hat z$.\\  (b) If you convert each outer product into an inner product,
what is the result?\\
(c) Consider the 4-vector $\boldsymbol{ \delta x} = {\boldsymbol x_1}-{\boldsymbol x_2}$, the difference of
two 4-vectors that are close to each other.  What is the interpretation of ${\boldsymbol \eta}[\boldsymbol
{\delta x}, \boldsymbol{\delta x}]$?

\subsection{Chapter 4}

\noindent 4.1. (a) Eliminate $\lambda$ from the trajectory (\ref{lintraj}) to reexpress it in the form 
$\vec x(t)$. \\ (b) Compute $\dot x^\mu = dx^\mu/d\lambda$ and $\sqrt{\eta_{\mu\nu}\dot
x^\mu\dot x^\nu}$. \\ (c) Find the reparametrization $\lambda(\tau)$ needed to make
$\sqrt{\eta_{\mu\nu}\dot x^\mu\dot x^\nu}=1$ in the new parametrization (which is the proper
time). \\ (d)  What relation must exist between the two spacetime points in order for this
reparametrization to be physically well-defined?

\medskip
\noindent 4.2. Carry out the steps to go from Eq.\
(\ref{variation}) to (\ref{geodesic}). 

\medskip
\noindent 4.3. Supply the missing steps in Eq.\ (\ref{wfgeo}).

\medskip
\noindent 4.4. Compute the Christoffel symbols for the two-sphere, first from the 
general formula, and then using Eqs.\ (\ref{two-sphere}).

\medskip
\noindent 4.5.  Do the analogous calculation to Eqs.\ (\ref{two-sphere}) to find the
geodesic equations for a particle moving in a weak gravitational perturbation $h_{\mu\nu}$.

\subsection{Chapter 5}
\noindent 5.1. Estimate the value of $\theta$ for starlight that passed the edge of the sun at a radius of 1.2 
times the solar radius $R_\odot = 696,340\,$km.  The mass of the sun is $M_\odot = 2.0\times 10^{30}\,$kg $= 1.1\times 10^{57}\,$GeV.

\medskip
\noindent 5.2. Estimate the dilation factor for a clock at $r=\infty$ compared to a clock at
the Earth's surface.

\subsection{Chapter 6}
\noindent 6.1. Check the dimensions of Eq.\ (\ref{GWhest}).\\

\medskip
\noindent 6.2.  (a) Use the residue theorem to evaluate the integral $\int_{-\infty}^\infty dz {e^{iz}\over (z+5i)(z-3i)}$.\\
(b) Similarly, compute $\int_{-\infty}^\infty dz {e^{iz}\over (z^2+a^2)}$ where $a$ is real.\\
(c) Compute $\int_0^\infty dz {\cos z\over (z^2+a^2)}$.\\

\subsection{Chapter 7}

\noindent 7.1: Compute the Christoffel symbol for the 1D space with $ds^2 = f(x)dx^2$.
Show that the Ricci tensor vanishes.\\

\noindent 7.2: Compute the solid angle of the region around the north
pole of the unit sphere subtending small angle $\theta$, to
$O(\theta^4)$.  Check that Eq.\ (\ref{Ricci2}) gives the
expected result for the Ricci curvature $R$.

\noindent 7.3: Consider a vector field $\vec A$ on the Euclidean plane in polar
coordinates $(\rho,\theta)$, with constant
contravariant component $A^\theta$, {\it i.e.,} $\partial_i
A^\theta=0$. The nonvanishing Christoffel symbols are $\Gamma^\theta_{\ \rho\theta}
= \Gamma^\theta_{\ \theta\rho} = 1/\rho$, $\Gamma^\rho_{\
\theta\theta} = -\rho$.\\
(a) Compute the length $|\vec A|$ of the vector.\\ 
(b) Compute the covariant derivatives $\nabla_\rho\vec A$ and
$\nabla_\theta\vec A$.\\
(c) Give an intuitive interpretation to the previous results.  In
particular, why does $\nabla_\theta\vec A$ have a component in the
$\hat\rho$ direction, and explain its sign and dependence on $\rho$.\\

\noindent 7.4: Refer to the metric in Eq.\ (\ref{con-met}), taking $\alpha$ to be small.\\
(a) Show that $\Gamma^x_{\ xy} = \alpha y + O(\alpha)^2$ and $\Gamma^y_{\ xy} = \alpha x +
O(\alpha)^2$.  Further $\Gamma^i_{\ xx} = \Gamma^i_{\ yy} = 0$.\\
(b) Compute the Ricci tensor to linear order in $\alpha$; show it is $-\alpha$ times the unit
matrix.\\
(c) We would like to verify the formula (\ref{Ricci2}) for two nearby geodesics emanating from the
origin.  To calculate the area of the
small triangle, we need to solve for the geodesics that bound its sides.  Show that to leading order
in $\alpha$, they are
\[
	\ddot x = -2\alpha\, y\, \dot x\, \dot y; \quad \ddot y = -2\alpha\, x \dot x\,\dot y\,.
\]
Without doing any calculations, find four straight lines in the $x$-$y$ plane that are geodesics, and
sketch them.  Then sketch the qualitative shape of the geodesics in between these exact solutions,
based on the geodesic equations.\\
(d) To find the geodesics close bounding the triangle, parametrize them as
\[
	x = A\lambda +\delta x(\lambda),\quad y = B\lambda + \delta y(\lambda)\,,
\]
where $\delta x$ and $\delta y$ are of order $\alpha$.  Substitute this ansatz into the geodesic
equations and isolate the terms of $O(\alpha)$ to show that
\[
	\delta x = -\alpha AB^2\lambda^3/3, \quad \delta y = -\alpha BA^2\lambda^3/3\,.
\]
(e)  Invert the equation for $x(\lambda)$ to show that 
\[
	\lambda = (x/A)(1 + \alpha B^2\lambda^2/3) + O(\alpha^2)
\]
and therefore
\[
	y = B(x/A)(1 + (\alpha/3)(B^2-A^2)(x/A)^2 + O(\alpha^2)\,.
\]
Verify that these have the curvature that you previously predicted in your plot.\\
(f) Part (e) has the surprising result that the triangle appears to be getting smaller in area as one
moves away from the origin, whereas Eq.\ (\ref{Ricci2}) predicts it should get larger.  However, we 
have not computed the actual area, which depends upon $\int dx\,dy\,\sqrt{g}$ and not just $\int
dx\,dy$.  Show that $\sqrt{g} = 1 + \sfrac12\alpha(x^2+y^2) + O(\alpha^2)$.\\
(g) Consider the triangle whose central geodesic oriented along the $\hat x$ axis, and bounded by
geodesics with slopes $\pm B/A$ whose magnitude is small, hence $B\ll A$.  In this approximation, the
triangle's area differs negligibly from that of the corresponding pie slice, and its area is
\[
	A(x) = 2\int_0^x \int_0^{y(x)} \sqrt{g}
\]
as a function of $x$.  Show that Eq.\ (\ref{Ricci2}) is satisfied.

\subsection{Chapter 8}

\noindent 8.1:
(a) find a set of basis vectors for the tangent space of the 2-sphere that give rise to the
standard metric in spherical coordinates.\\
(b) Show that if one rotates them in the tangent space by an orthogonal $2\times 2$ matrix $O$ using
$e_j^a \to O_{ab}e_j^b$ to a new basis, it is also a valid choice.\\

\noindent 8.2:  Consider a vector field $\vec A$ whose covariant
components $(A_\theta, A_\phi)$ are constant on the unit 2-sphere.
The nonvanshing Christoffel symbols are $\Gamma^{\phi}_{\ \theta\phi}
= \Gamma^\phi_{\ \phi\theta} = \cot\theta$, $\Gamma^\theta_{\
\phi\phi} = \sin\theta\cos\theta$.  Hence 
\[	A_{\phi;\theta} = -\cot\theta A_\phi,\quad
	A_{\phi;\phi} = \sin\theta\cos\theta A_\theta,\quad
	A_{\theta;\theta} = 0
\]
(a) Compute $A_{\phi;\theta\phi}$.\\
(b) Compute $A_{\phi;\phi\theta}$.\\
(c) From these results, infer $R^\theta_{\ \phi\theta\phi}$.\\

\noindent 8.3: In the previous problem, someone argues that
$A_{\phi;\phi\theta} = \nabla_\theta(\sin\theta\cos\theta A_\theta)
= \cos2\theta A_\theta +
\sin\theta\cos\theta(-\Gamma^i_{\ \theta\theta}A_i)$, which is just 
$\cos2\theta A_\theta$ since the Christoffel symbols with those
indices vanish.  What is wrong with their argument?

\subsection{Chapter 9}

\noindent 9.1: From the action $S_m = -m\int d\tau\sqrt{g_{\mu\nu}\dot x^\mu \dot x^\nu}$,
compute the $T_{00}(\vec x)$ stress-energy tensor element for a particle at rest, at position $\vec
x'$.\\

\noindent 9.2: Using symbolic manipulation, for example, one can show that the Ricci tensor for the
weak-field metric around a point mass source is given by $\nabla^2\phi$ times the unit matrix,
where $\phi$ is the Newtonian potential, to leading order in $\phi$.  Compute the Einstein tensor to
this order.  It should have only $G_{00}$ nonvanishing.\\

\noindent 9.3: (a) Compute $\Gamma_{\mu\alpha\beta}$ near the surface of the Earth in terms of the
Newtonian potential $\phi(z)$.\\
(b) Compute $\nabla_\mu T^{\mu z} =0$ near the surface of the Earth, to find the differential equation
governing the atmospheric pressure variation with altitude.\\ 

\subsection{Chapter 10}

\noindent 10.1: (a) Sketch the light cones in the $r$-$t$ plane of the Schwarzschild metric.\\
(b) Do similarly in the $r$-$v$ plane for ingoing Eddington-Finkelstein coordinates.\\

\noindent 10.2: Approximately how long will an observer have before reaching the singularity, after
crossing the horizon of Sgr A$^*$ at the center of our galaxy?\\

\noindent 10.3: Estimate the tidal accelerations experienced by an observer of size $\sim 1\,$m near
the horizon of (a) Sgr A$^*$ and (b) a black hole of mass $1\,M_\odot$.\\

\noindent 10.4: Compute the radial and transverse tidal accelerations of an observer falling into a
black hole, using Newtonian gravity.\\

\noindent 10.5: Determine the Hawking temperature of a black hole with initial mass $M=10^{12}$\,kg
(whose lifetime is of order the age of the Universe).\\

\noindent 10.6: Estimate the lifetime of a black hole whose (a) mass or (b) radius is the same as
that of a baseball.

\end{appendix}

\centerline{\includegraphics[width=\textwidth]{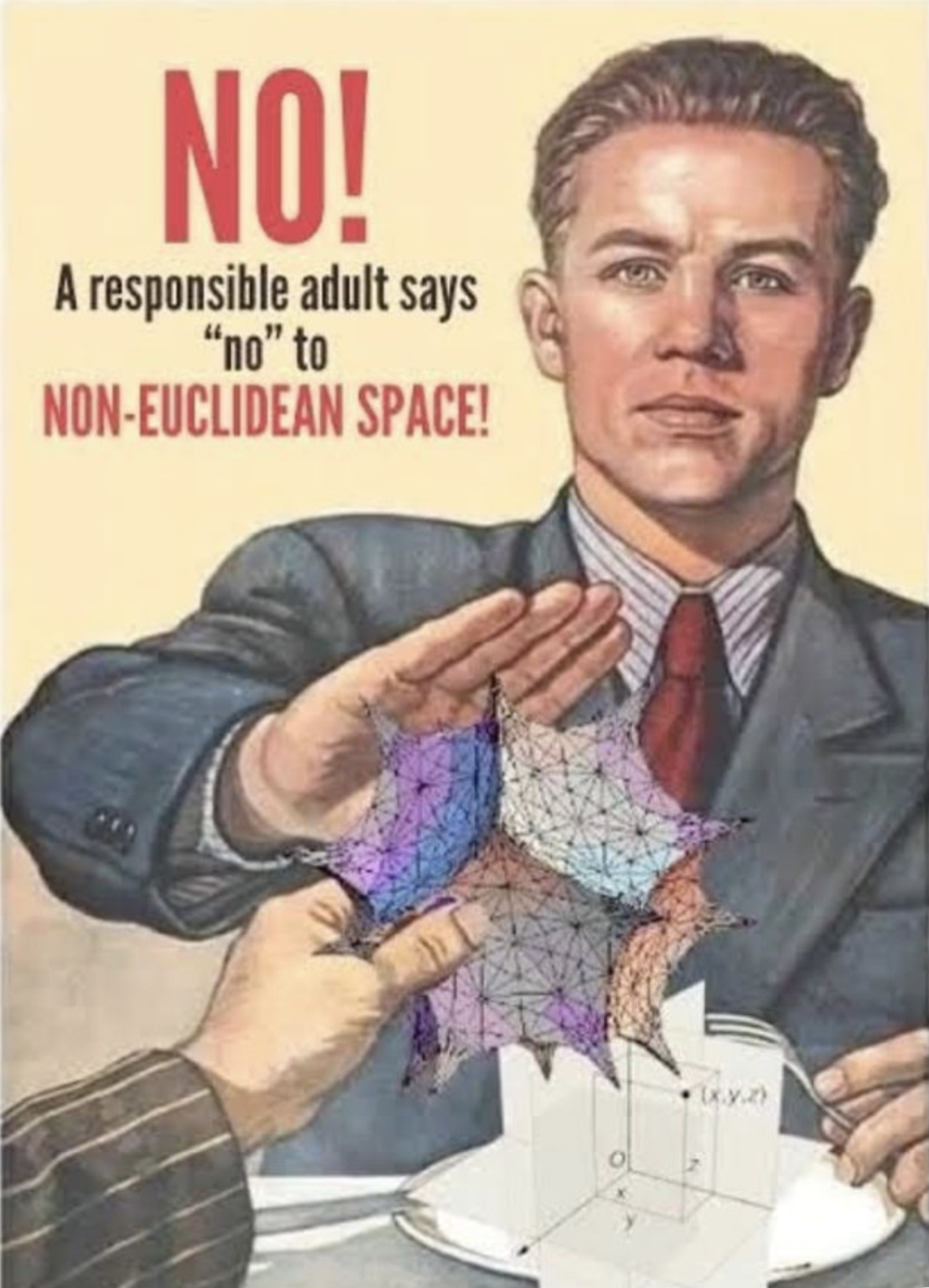}}

\end{document}